\RequirePackage{ifpdf}
\documentclass[hyper,letterpaper]{JHEP3}
\usepackage{amsmath,amssymb,amsfonts,bm,amscd}
\usepackage{cite}
\usepackage{graphicx, wrapfig}
\usepackage{multirow}
\usepackage{verbatim}
\usepackage{appendix}
\usepackage{fancybox}
\usepackage{slashed}

\usepackage{url}
\usepackage{float}

\newcommand{\longrightleftarrows}{\mathrel{\substack{\xrightarrow{\hspace{1.6cm}} \\[-.6ex] \xleftarrow{\hspace{1.6cm}}}}}

\newcommand{\llongrightleftarrows}{\mathrel{\substack{\xrightarrow{\hspace{4cm}} \\[-.6ex] \xleftarrow{\hspace{4cm}}}}}

\newcommand{\bea}{\begin{eqnarray}}
\newcommand{\eea}{\end{eqnarray}}
\newcommand{\be}{\begin{equation}}
\newcommand{\ee}{\end{equation}}
\newcommand{\eq}[1]{(\ref{#1})}

\newcommand{\unknot}{{\,\raisebox{-.08cm}{\includegraphics[width=.8cm]{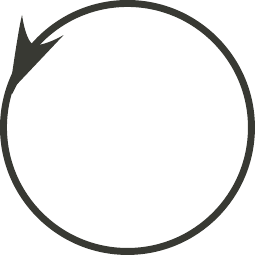}}\,}}

\newcommand{\longmapsfrom}{\mathrel{\reflectbox{\ensuremath{\longmapsto}}}}

\newcommand{\Id}{{\rm Id}}
\newcommand{\id}{{\rm id}}
\newcommand{\p}{^{\prime}}

\newcommand{\qE}{{\mathcal E}}
\newcommand{\qF}{{\mathcal F}}

\def \I{\mathcal{I}}

\newcommand{\Z}{{\mathbb Z}}
\newcommand{\R}{{\mathbb R}}
\newcommand{\C}{{\mathbb C}}

\def\eg{{\it e.g.~}}
\def\ie{{\it i.e.~}}
\def\cf{{\it c.f.~}}

\def\frak{\mathfrak}

\def\tilde{\widetilde}
\def\hat{\widehat}
\def\bar{\overline}

\def\CA{{\mathcal A}}

\def\CC{{\mathcal C}}

\def\CH{{\mathcal H}}

\def\CJ{{\mathcal J}}

\def\CL{{\mathcal L}}
\def\CM{{\mathcal M}}
\def\CN{{\mathcal N}}

\def\CS{{\mathcal S}}

\def\CU{{\mathcal U}}

\def\CX{{\mathcal X}}

\def\unknot{{\,\raisebox{-.08cm}{\includegraphics[width=.4cm]{unknot}}\,}}

\renewcommand{\P}{{\cal P}}
\newcommand{\cp}{{\mathbb{C}}{\mathbf{P}}}

\renewcommand{\bar}{\overline}
\renewcommand{\hat}{\widehat}

\title{Junctions of surface operators  and \\ categorification of quantum groups}

\author{Sungbong Chun$^{1}$, Sergei Gukov$^{1}$, Daniel Roggenkamp$^{2}$
\\
$^{1}$ Walter Burke Institute for Theoretical Physics, California Institute of Technology, Pasadena, CA 91125 USA\\
$^{2}$ Institut f\"ur Theoretische Physik, Universit\"at Heidelberg, Philosophenweg 19, 69120 Heidelberg, Germany}

\abstract{We show how networks of Wilson lines realize quantum groups ${\mathbf{U}}_q(\mathfrak{sl}_m)$, for arbitrary $m$, in 3d $SU(N)$ Chern-Simons theory.
Lifting this construction to foams of surface operators in 4d theory we find that
rich structure of junctions is encoded in combinatorics of planar diagrams.
For a particular choice of surface operators we 
make a connection to 
known mathematical
constructions of categorical representations and categorified quantum groups.
\\
\\
\\
\\
\\
\\
\\
{\tt CALT-TH 2015-040}}

\begin{document}
\cornersize{1}

\section{Introduction}

In this paper we develop a physical framework for categorification of quantum groups,
which is consistent with (and extends) the physical realization of homological knot invariants
as $Q$-cohomology of a certain brane system in M-theory (that we review in section \ref{sec:LG}).

Our starting point will be the celebrated relation between
quantum groups and Chern-Simons TQFT in three dimensions.
Even though it has a long history, many aspects of this relation remain mysterious,
even with respect to some of the most basic questions.
For instance, quantum groups are usually defined via generators and relations,
which are not easy to ``see'' directly in Chern-Simons gauge theory.
Instead, one can see certain combinations of the generators that implement braiding
of Wilson lines \cite{MR1030450}.

The realization of quantum groups we discuss in this paper is also based on Chern-Simons gauge theory, but in constrast to the conventional one, it allows for gauge group and quantum group to be of completely different rank! Moreover, the quantum group generators have an immediate and very concrete interpretation. 
This approach is based on networks of Wilson lines and gives a physical realization of the {\it skew Howe duality}, which was introduced in the context of knot homology in \cite{MR2593278}:
As we will show in section \ref{sec:CS}, upon concatenation, networks of Wilson lines exhibit quantum group relations, which for the Lie algebra $\frak{g} = \mathfrak{sl}_2$ take a very simple form:
\bea
&&KK^{-1}=1=K^{-1}K\,,\nonumber\\
&&KE=q^2EK\,,\qquad
KF=q^{-2}FK \label{Uqsltwo} \\
&&\left[E,F\right]={K-K^{-1}\over q-q^{-1}}\,.\nonumber
\eea
Here, the quantum variable $q$ is related to the Chern-Simons coupling constant in the usual way, \cf \eq{qvariable}. What is not usual is that configurations of $m$ Wilson lines in totally antisymmetric representations
\be
R_1 \otimes \ldots \otimes R_m \; = \; \Lambda^{k_1}\C^N\otimes\ldots\otimes\Lambda^{k_m}\C^N
\label{RRRweight}
\ee
are interpreted as weight spaces of quantum $\mathfrak{sl}_m$, on which the Chevalley generators act by adding extra Wilson line segments. This idea is due to \cite{MR3263166}.
So, generators and relations of the quantum group have a very concrete realization (shown in Figures~\ref{fig:SHfunctor} and \ref{fig:rels}, respectively); and the rank $m-1$ of the quantum group has nothing to do with the rank $N-1$ of Chern-Simons gauge theory. Instead it is determined by the number of incoming (equally, outgoing) Wilson lines.

\begin{figure}[htb]
\centering
\includegraphics[width=6.0in]{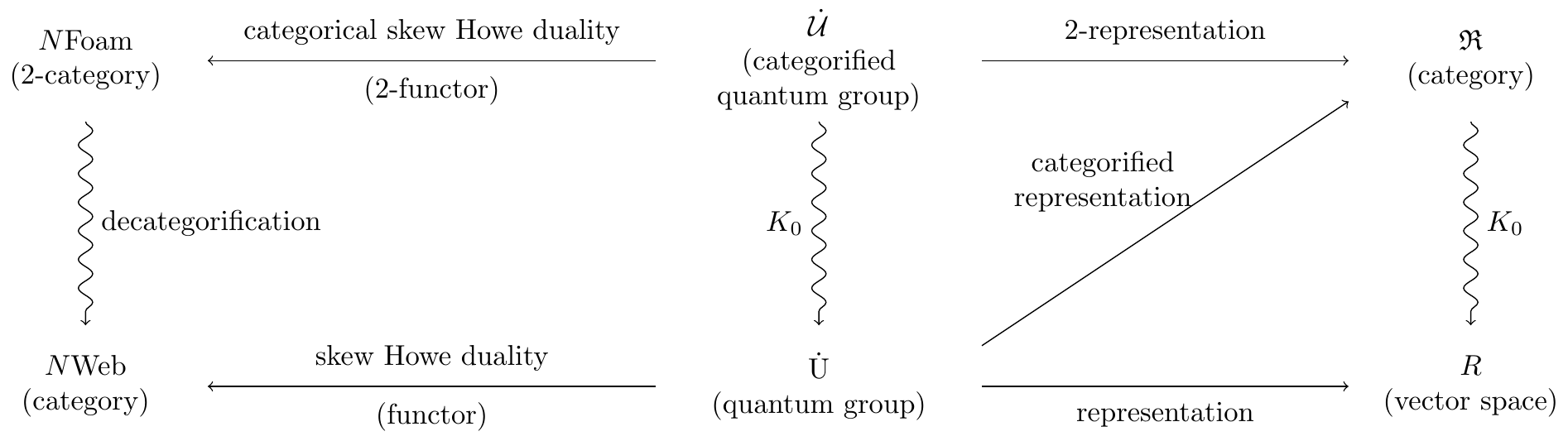}
\caption{Categorified representation theory and categorified skew Howe duality.}
\label{fig:BIGpicture}
\end{figure}

By categorifying representation theory, one usually means replacing weight spaces, such as \eqref{RRRweight},
by graded categories $\mathfrak{R}_{\lambda}$ on which raising and lowering operators act as functors,
\cf \cite{MR1074310,MR1197834,MR1714141,MR2305608,MR2503405}.
The quantum group itself is promoted to a 2-category $\dot \CU$,
for which $\mathfrak{R}_{\lambda}$ is a 2-representation, as illustrated in Figure~\ref{fig:BIGpicture}.
Sounds a bit scary, doesn't it?

Luckily, the rich and abstract structure of categorical representations and categorified quantum groups
can be made very ``user-friendly'' and intuitive in the {\it diagrammatic approach}
recently developed by Khovanov and Lauda \cite{MR2729010,MR2525917,MR2763732,KLIII}
(see \cite{MR2747932,Lauda,KamnitzerRev} for excellent expositions and \cite{Rouquier} for a related algebraic construction).
One of our main goals is to provide a physical realization of this diagrammatic approach.

In the context of quantum field theories, categorification can be achieved by adding a dimension, and networks of Wilson line operators used to realize quantum groups in Chern-Simons theory become {\it foams} of surface operators in 4d. 
Indeed, foams have been related to Khovanov-Lauda diagrams  in the math literature in \cite{MR3426687,QR14}.  The physical realizations of foams discussed here are built by sewing surface operators along 1d junctions, which are interesting in their own right and so far did not even appear in the physics literature.
 In section \ref{sec:surface},  we will not only present
evidence that BPS junctions of surface operators exist, but we will also discuss various applications, ranging from math to physics.

In fact, the relevant surface operators \cite{Gukov:2006jk,Gukov:2007ck}
have already been used to construct group actions on categories
(notably, in the context of the geometric Langlands program \cite{Gukov:2008sn})
and to realize many elements of geometric representation theory \cite{Gukov:2008ve}.
In these realizations, as well as in many other similar problems, groups acting on categories
are generated by codimension-$1$ walls (interfaces) acting on categories of boundary conditions. The group law comes from the ``fusion'' product of interfaces.
This is an instance of the well known fact that the algebraic structure of various operators and defects in $(n+1)$-dimensional topological quantum field theories is governed by $n$-categories (see {\it e.g.} \cite{Gukov:2007ck} for a review).

\begin{figure}[htb]
\centering
\includegraphics[width=5.0in]{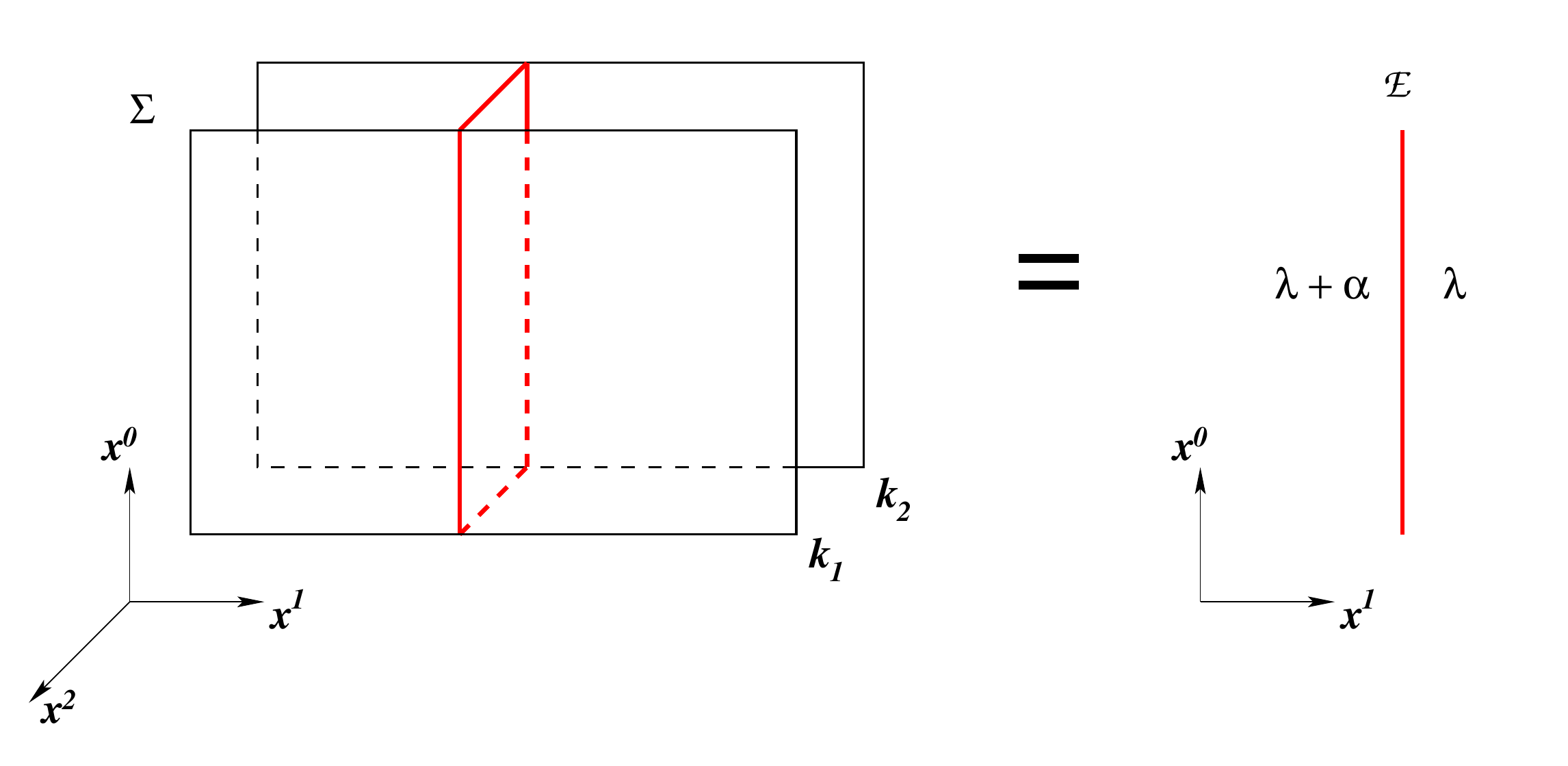}
\caption{Projecting $\Sigma$ onto $(x^0,x^1)$ plane gives a product of Landau-Ginzburg theories,
in which junctions (singular edges of $\Sigma$) are represented as interfaces.}
\label{fig:junction14}
\end{figure}

Our physical realization of categorified quantum groups is conceptually similar, but with several twists. One way to describe it is to consider the world-volume theory on the foam of surface operators. By moving the $m$ parallel surface defects representing the $\mathfrak{sl}_m$ weight spaces close together we can describe the world-volume theory of the combined system 
as tensor product of the theories on the individual surface operators. Junctions of surface operators introduce interactions between tensor factors along $1$-dimensional loci, and are incorporated by interfaces between generally different 2d theories. For instance, collapsing the $x^2$-direction of the configuration of surface operators on the left of Figure~\ref{fig:junction14}, the combined world-volume theory can be described by a 2d theory in the $(x^0,x^1)$-plane. The surface operator supsended along the $x^2$-direction introduces a 1d interface.

Now, while in the spririt of the remark above the structure of a 2d TQFT is encoded in a $1$-category, different 2d TQFTs together with interfaces between them form a $2$-category, whose objects are not boundary conditions, but rather the different 2d theories. $1$-morphisms are interfaces, and $2$-morphisms are interface changing fields. As a bonus, these $2$-categories come with $2$-representations on the categories of boundary conditions of the 2d TQFTs.

It is in this way that we extract 
the building blocks of 
categorified quantum groups $\dot \CU$ out of 
foams of surface operators: $\mathfrak{sl}_m$ weight spaces correspond to tensor products of the world-volume theories of $m$ surface operators, and the generators of the quantum group are realized as interfaces between them. 

In fact, this 2d perspective on surface operators and their junctions provides a physical realization of the planar diagrams of
Khovanov-Lauda \cite{MR2729010,MR2525917,MR2763732,KLIII} and vast generalizations,
for more general types of surface operators.\footnote{After embedding in eleven-dimensional M-theory,
this will also provide an answer to the following question:
Which two dimensions of space-time, relative to the fivebranes, compose the plane
where the diagrams of \cite{MR2729010,MR2525917,MR2763732,KLIII} are usually drawn?}

There are various ways to describe relevant 2d TQFTs, and in this paper we mainly consider two variants:
one is based on topological Landau-Ginzburg (LG) models,
while the second involves the UV topological twist of sigma-models whose targets are  certain flag-like
subspaces of affine Grassmannians which play an important role in the geometric Satake correspondence.
We generally favor the former, where interfaces and their compositions are easier to analyze using matrix factorizations \cite{Brunner:2007qu}.

In the Landau-Ginzburg approach (described in detail in section~\ref{sec:LG})
the 2d space-time is precisely the plane on which the planar diagrams (with dots on the lines) of Khovanov-Lauda are drawn.
Each 2d region of the plane colored by the highest weigh $\lambda$ defines a LG model or, to be more accurate, a product of LG models.
It is basically a projection of $\Sigma$ from the three-dimensional space --- that later in the text we parameterize with $(x^0,x^1,x^2)$ ---
to a two-dimensional plane $(x^0,x^1)$.
Then, LG interfaces describe transitions between different sheets of $\Sigma$, regarded as a multi-cover
of the $(x^0,x^1)$-plane, \cf Figure~\ref{fig:junction14}.
These 1-dimensional interfaces (defects) are precisely the arcs in the planar diagrams of \cite{MR2729010,MR2525917,MR2763732,KLIII}:
\be\nonumber
\begin{array}{ccccc}
\vspace{0.1cm}
\text{region on a plane} & \quad \leftrightarrow \quad & \text{weight space \eqref{RRRweight}} & \quad \leftrightarrow \quad & \text{LG model } LG_{k_1}\otimes\ldots\otimes LG_{k_m} \\
\vspace{0.1cm}
\text{lines} & \leftrightarrow & \text{1-morphisms} & \leftrightarrow & \text{interfaces between LG models} \\
\text{dots}  & \leftrightarrow &  \text{2-morphisms}  & \leftrightarrow &  \text{interface changing fields}
\end{array}
\ee
The reader is invited to use this dictionary to translate virtually any question in one subject to a question in another.
For example, it would be interesting to study 2-categories that one finds for more general types of surface operators
in various gauge theories, and interfaces in Landau-Ginzburg models, other than examples studied in this paper.
Mathematically, they should lead to interesting generalizations of the Khovanov-Lauda-Rouquier (KLR) algebras \cite{MR2525917,MR2763732,Rouquier}.
Conversely, it would be interesting to identify surface operators and LG interfaces for mathematical constructions
generalizing $\dot \CU$.

In summary, if you are studying junctions
of surface operators or interfaces in Landau-Ginzburg models,
most likely you are secretly using the same mathematical
structure --- and, possibly, even the same diagrams --- as those categorifying representations and quantum groups,
or interesting generalizations thereof.


\section{Junctions of Wilson lines and quantum groups}
\label{sec:CS}

In this section, we will discuss Wilson lines in $SU(N)$ Chern-Simons theory and their junctions. We will show that (upon concatenation) certain networks of Wilson lines satisfy the defining relations of (Lusztig's idempotent form \cite{Lusztig93} of) the quantum groups $\dot{\mathbf{U}}_{q}(\frak{sl}_{m})$.

We start by reviewing the necessary techniques from \cite{Witten:1988hf, Witten:1989wf, Witten:1989rw} in section \ref{subsec:JWL}.
In section \ref{subsec:rels}, we explore various relations satisfied by networks of Wilson lines and their junctions,
which surprisingly include the quantum group relations.
We give an explicit derivation of one such relation, relegating the rest to Appendix~\ref{appendix:derivations}.
Then, by playing with $m$ strands, in section \ref{subsec:SH} we (re)discover the representation category
of the quantum group of $\mathfrak{sl}_{m}$.
This gives a physical realization of the diagrammatic approach of Cautis-Kamnitzer-Morrison \cite{MR3263166}
to quantum skew Howe duality.
Finally, in section \ref{sec:whysurf}, we give an {\it a priori} explanation why categorification of these beautiful
facts should involve surface operators and their junctions, leading us into sections \ref{sec:surface} and \ref{sec:LG}.

\subsection{Junctions of line operators}
\label{subsec:JWL}
Our starting point is Chern-Simons theory in three dimensions with gauge group $G=SU(N)$.
The action on a closed $3$-manifold $M_3$ is given by
\be
S_{CS} = \frac{\kappa}{4\pi}\int_{M_3} (A \wedge dA + \frac{2}{3}A \wedge A \wedge A).
\ee
where $A$ is an $SU(N)$-gauge connection on $M_3$, and the coupling constant $\kappa\in{\mathbb N}$ is usually called the ``level.''

The most important (topological) observables of this theory are derived from the parallel transport with respect to $A$: To a path $\Gamma\subset M_3$ and a representation $R$ of the gauge group one associates the Wilson line
\be
U_R(\Gamma) = {\mathcal P}\exp\left(\int_\Gamma \rho_R(A)\right):~~~R\longrightarrow R\,.
\ee
Here ${\mathcal P}$ denotes path-ordering, and $\rho_R$ is the associated representation on the Lie algebra. 
For an {\it open} path $\Gamma$ such $U_R(\Gamma)$ is not gauge invariant,
but one can form gauge invariant combinations, such as the familiar Wilson loops
\be
W_R(\Gamma)={\rm tr}_R\,U_R(\Gamma)
\ee
by closing the path $\Gamma$.
More general gauge invariant combinations can be associated to networks of Wilson lines \cite{Witten:1989wf}
by forming junctions of Wilson lines in representations $R_1,\ldots,R_n$ and contracting
the corresponding observables $U_{R_i}$ with invariant tensors 
\be
\epsilon \; \in \; \text{Hom} (R_1\otimes\ldots\otimes R_n,\C) \,.
\label{invtensors}
\ee
The latter correspond to additional junction fields,
which need to be specified along with representations of Wilson lines in order define the observable.

\FIGURE[thb]{
\includegraphics {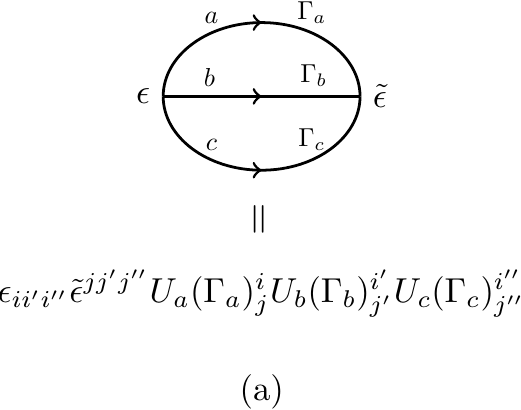}
\hspace{1.5cm}
\includegraphics {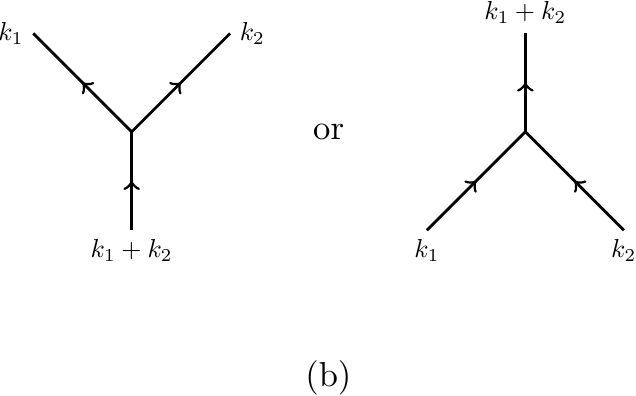}
\caption{(a) A trivalent network of Wilson lines and the corresponding gauge invariant observable.
(b) Trivalent junctions of Wilson lines in antisymmetric representations: Labels $k$ refer to  representations $\Lambda^k\square$.\label{fig:theta}}
}
For instance, to the graph in Figure \ref{fig:theta}(a), one can associate an observable by contracting the Wilson lines $U_{a}(\Gamma_a)$, $U_b(\Gamma_b)$ and $U_c(\Gamma_c)$ with invariant tensors $\epsilon \in \bar{a} \otimes \bar{b} \otimes \bar{c}$ in the ``incoming" junction and $\tilde{\epsilon} \in a \otimes b \otimes c$ in the ``outgoing'' junction of the graph. Here $\bar{R}$ denotes the dual representation.
(The Wilson lines carry gauge indices $i,j$ of representation $a$, $i',j'$ of $b$, and $i'',j''$ of $c$, respectively. After contracting the gauge indices of Wilson lines via $\epsilon$ and $\tilde{\epsilon}$, we obtain a gauge invariant observable.)

In the following we will mostly consider Wilson lines in totally antisymmetric representations $\wedge^{k}\square$, of the fundamental representation $\square$
of $SU(N)$, which for ease of notation we will just label by $k$. These Wilson lines admit two possible trivalent junctions, depicted in Figure
\ref{fig:theta}(b), which are dual to each other. Since $Hom(\Lambda^{k_1}\square\otimes\Lambda^{k_2}\square,\Lambda^{k_1+k_2}\square)$ is one-dimensional,
there is just one possible junction field, whose normalization we will fix shortly. We can then omit the labels of the junctions.
As it turns out, junctions of more than three such Wilson lines can all be factorized into trivalent junctions, \cf Figure \ref{fig:rels}(a).

In order to unambiguously define Wilson line invariants in the quantum theory,
a {\it framing}, \ie the choice of orthogonal vector fields along the Wilson lines is required, which fit together in the junctions.
This is in particular needed to regularize the self-linking number of Wilson loops. In our discussion, we will always consider two-dimensional projections of graphs of Wilson lines, and choose the corresponding {\it vertical} framing.

Moreover, Wilson lines in the quantum theory are labeled only by those representations of $SU(N)$ which correspond to integrable highest weight representations
of $\widehat{su}(N)_{\kappa}$; all other Wilson lines decouple in the quantum theory. The totally antisymmetric representations considered here are integrable for all $\kappa\geq 1$.

Now that we have defined the gauge invariant observables we are interested in, let us proceed to summarize some relevant machinery from \cite{Witten:1988hf, Witten:1989wf, Witten:1989rw} to compute their expectation values.

\subsubsection*{Hilbert space and the connected sum formula}

Quantization of Chern-Simons theory on a product $\Sigma\times\R$
associates to any surface $\Sigma$ a Hilbert space ${\mathcal H}_\Sigma$.
The Chern-Simons path integral on an open 3-manifold $M_3$ with boundary $\partial M_3=\Sigma$ gives rise to a vector ${\mathcal Z}(M_3)=|M_3\rangle_\Sigma\in{\mathcal H}_\Sigma$.
Moreover, if a closed 3-manifold $M_3$ can be cut along a surface $\Sigma$ into two
disconnected components $M_3=M_3^{(1)}\#_\Sigma M_3^{(2)}$, then the path integral on $M_3$ evaluates to the scalar product ${\mathcal Z}(M_3)={}_\Sigma\langle M_3^{(1)}|M_3^{(2)}\rangle_\Sigma$ of the vectors associated to $M_3^{(1)}$ and $M_3^{(2)}$. This also holds in the presence of Wilson lines, as long as they intersect the surface $\Sigma$ transversely. The corresponding Hilbert space then also depends on the punctures, at which the surface is pierced by the Wilson lines and the associated representations.

It turns out that the Hilbert spaces ${\mathcal H}_\Sigma$ are isomorphic to the spaces of conformal blocks of the $\widehat{su}(N)_\kappa$-WZW models on $\Sigma$, where, at the intersection points, primary fields with the respective integrable highest weight representations of $\widehat{su}(N)_\kappa$ are inserted \cite{Witten:1988hf}.
These spaces are well studied. For instance, the Hilbert spaces associated to the 2-sphere with two punctures
labeled by irreducible representations $R_i$ and $R_j$ is one-dimensional if $R_i\cong \bar{R}_j$ and zero-dimensional otherwise:
\be\label{dimargument}
{\rm dim}{\mathcal H}_{\Sigma\{R_i,R_j\}}=\delta_{R_i,\bar{R}_j}\,.
\ee
As before $\bar{R}$ is the dual of the representation $R$.
This in particular implies the following. Consider a configuration of Wilson lines on a 3-sphere $M_3=S^3$,
which intersects a given great sphere $\Sigma=S^2$ at two points colored by representations $R$ and $R^\prime$, \cf Figure~\ref{fig:direct sum}. Denote the hemispheres (with Wilson lines) obtained by cutting $M_3$ along $\Sigma$ by $M_3^{(1)}$ and $M_3^{(2)}$.
Now, according to \eq{dimargument}, we have ${\mathcal Z}(M_3^{(1)})=|M_3^{(1)}\rangle_{S^2\{R,R^\prime\}}=0$
if $R\ncong\bar{R}^\prime$.
If, on the other hand, $R\cong \bar{R}^\prime$, then $|M_3^{(1)}\rangle_{S^2\{R,\bar{R}\}}$ is proportional
to the vector $|N_3 \rangle_{S^2\{R,\bar{R}\}}$ that corresponds to a hemisphere $N_3$ with a single strand
of $R$ Wilson line connecting two points in the boundary $S^2$. The proportionality constant is given by
\be
|M_3^{(1)}\rangle_{S^2\{R,\bar{R}\}}=|N_3\rangle_{S^2\{R,\bar{R}\}}\,
{{}_{S^2\{R,\bar{R}\}}\langle N_3|M_3^{(1)}\rangle_{S^2\{R,\bar{R}\}}\over
{}_{S^2\{R,\bar{R}\}}\langle N_3|N_3\rangle_{S^2\{R,\bar{R}\}}}
\,,
\ee
which in particular implies the connected sum formula
\be
{}_{S^2\{R,\bar{R}\}}\langle M_3^{(2)}|M_3^{(1)}\rangle_{S^2\{R,\bar{R}\}}={{}_{S^2\{R,\bar{R}\}}\langle M_3^{(2)}|N_3\rangle_{S^2\{R,\bar{R}\}}\;
{}_{S^2\{R,\bar{R}\}}\langle N_3|M_3^{(1)}\rangle_{S^2\{R,\bar{R}\}}\over
{}_{S^2\{R,\bar{R}\}}\langle N_3|N_3\rangle_{S^2\{R,\bar{R}\}}}
\,.
\ee
Here, the inner products on the right are the $S^3$ partition function with three configurations of Wilson lines: the two in the numerator are obtained by joining the $R$ and $\bar{R}$ ends of the Wilson line configurations in $M_3^{(2)}$ and $M_3^{(1)}$, respectively, while the one in the denominator is a single Wilson loop (the unknot) in the representation $R$. A pictorial representation of this formula is given in Figure \ref{fig:direct sum}.

\FIGURE[t]{
\includegraphics {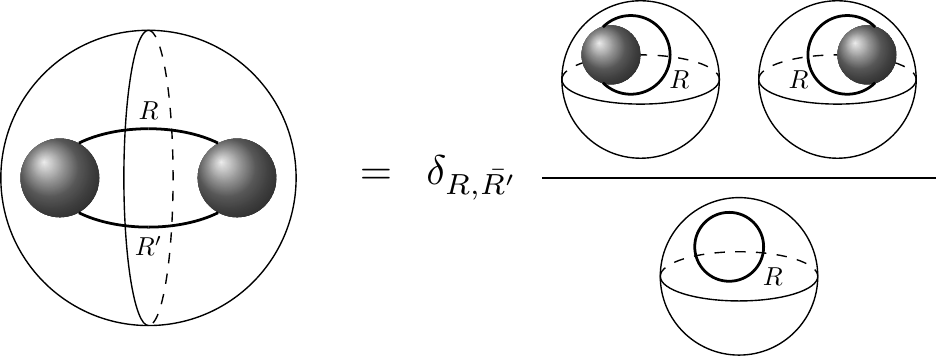}
\caption{An illustration of the connected sum formula, applied to a network of Wilson lines in $S^3$ and a separating 2-sphere $S^2$ that cuts the network at two Wilson lines in representations $R$ and $R^\prime$. When $R\cong\bar{R}^\prime$, its partition function factorizes into contributions from the Wilson line networks in the left and right hemisphere, respectively, where the $R$ and $R^\prime=\bar{R}$ Wilson lines are joined.}
\label{fig:direct sum}}

\subsubsection*{Framing and skein relations}

Surgeries other than a simple connected sum enable us to study braiding of Wilson lines, skein relations, and to compute the expectation values of Wilson loops colored by various representations. In general, given a surgery presentation of $M_3$, we can compute the expectation values of Wilson line operators in it by the surgery formula \cite{Witten:1988hf}.

\FIGURE[htb]{\centering
\includegraphics {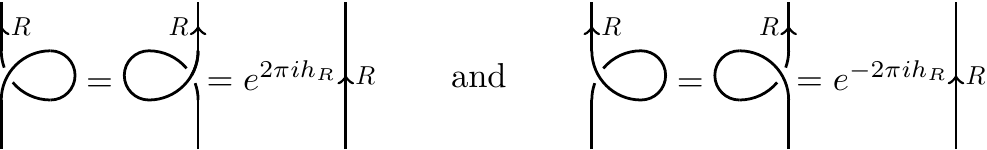}
\caption{Wilson lines with extra twists.}
\label{fig:line braiding}}
The simplest example is the relation among twisted Wilson lines. The path integral on a $3$-ball containing a straight Wilson line in some representation $R$ ending on the boundary $2$-sphere determines a vector in the one-dimensional Hilbert space $\CH_{S^2\{R,\bar{R}\}}$. This is also true when the Wilson line has extra twists, as in Figure~\ref{fig:line braiding}. In particular, the vectors associated to Wilson lines with different number of twists are proportional. Assuming vertical framing, the Wilson lines in Figure~\ref{fig:line braiding} are related by $(\pm1)$-Dehn twists on the boundary $2$-sphere. The proportionality constant in this case is $e^{2 \pi i h_{R}}$ where $h_{R}$ is the conformal weight of the primary field of the WZW model transforming in  representation $R$.

\FIGURE[t]{
\includegraphics {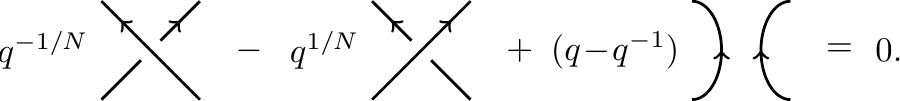}
\caption{Skein relation of vertically framed Wilson lines in the fundamental representation $\square$.}
\label{fig:skein}}

In a similar fashion one obtains skein relations. Consider a $3$-ball with two Wilson lines labeled by the fundamental representation $\square$, ending on the boundary $2$-sphere. The associated Hilbert space $\CH_{S^{2},\{\square, \square, \bar{\square}, \bar{\square} \}}$ is two-dimensional, so the three configurations of Wilson lines in Figure~\ref{fig:skein} have to satisfy a linear relation.
The coefficients of this relation can be determined from the fact that the configurations are related by half-twists on the boundary $2$-sphere, \cf \cite{Witten:1988hf,Witten:1989wf}. In Figure~\ref{fig:skein} we expressed them in terms of the variable
\be\label{qvariable}
q = e^{\pi i/(N+\kappa)}\,.
\ee
Note that the exact form of the relation depends on the choice of framing. We use the vertical framing throughout this paper, which is why the skein relation here looks different from its more familiar form in the canonical framing\footnote{in which self-linking numbers of knots are zero}, in which the coefficients of the first two terms are $q^{-N}$ and $-q^{N}$, respectively. A thorough discussion of this can be found in \cite{Witten:1989wf}.

Next, let us show how to obtain the expectation value of 
an unknotted
 Wilson loop in $S^3$ labeled by an irreducible representation $R_i$. The idea is that such a configuration can be obtained by Dehn surgery on $S^2\times S^1$, with the Wilson line running along the $S^1$. More precisely, let $M_3$ be a tubular neighborhood of the Wilson line in $S^2\times S^1$, and $M_3^\prime$ its complement. Then $S^3$ can be obtained by gluing $M_3$ and $M_3^\prime$ along the boundary torus $T^2$, with a non-trivial identification by the global diffeomorphism $S \in SL(2,\Z)$:
\be
{\mathcal Z}(S^3,R_i)={}_{T^2}\langle M_3^\prime|S|M_3,R_i\rangle_{T^2}\,.
\ee
Here we use the crucial fact that the mapping class group of $T^2$ acts on
$\CH_{T^2}$ by the modular transformation on the characters of the respective Kac-Moody algebra. In particular,
\be
S|M_3,R_i\rangle_{T^2}=\sum_j S_{ij}|M_3,R_j\rangle_{T^2}\,,
\ee
where $S_{ij}$ is the modular $S$-matrix.
Hence,
\be
{\mathcal Z}(S^3,R_i)=
\sum_j S_{ij}\;{}_{T^2}\langle M_3^\prime|M_3,R_j\rangle_{T^2}
=\sum_jS_{ij}\;{\mathcal Z}(S^2\times S^1,R_j)
\,,
\ee
but
${\mathcal Z}(S^2\times S^1,R_j)={\rm tr}_{\CH_{S^2\{R_j\}}}(1)={\rm dim}({\CH_{S^2\{R_j\}}})$, which is $1$ if $R_j$ is the trivial representation and $0$ otherwise.
It immediately follows that the expectation value of a Wilson loop in $S^3$ is given by the {\it quantum dimension}
\be
\langle W_{R_j}\rangle_{S^3}={{\mathcal Z}(S^3,R_i)\over {\mathcal Z}(S^3,R_0={\rm triv})}
={S_{i0}\over S_{00}}\,.
\ee
For Wilson lines in the totally antisymmetric representations in $SU(N)$ Chern-Simons theory at level $\kappa$, this gives
\be \label{eqn:loop exp value}
\langle W_{\wedge^{m}\square} \rangle = \dfrac{S_{0,\wedge^{m}\square}}{S_{0,0}} = {N \brack m}\, ,
\ee
where
\be
  {N \brack m} = \dfrac{[N] \cdots [N-m+1]}{[m] \cdots [1]}\, ,
 \quad
\left[N\right] = \dfrac{q^{N} - q^{-N}}{q-q^{-1}}\,
\ee
is the quantum binomial coefficient.

\subsubsection*{Tetrahedral network and normalization of trivalent vertices}

\FIGURE[r]{
\includegraphics{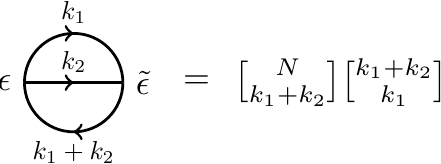}
\caption{The evaluation of a ``$\theta$-web'' that determines the normalization of the vertices $\epsilon$ and $\tilde{\epsilon}$. Here $k$ stands for a totally antisymmetric representation $\wedge^{k}\square$.}
\label{fig:normalization}
}
As we explained earlier, there are only two possible 
trivalent
junctions between Wilson lines in totally antisymmetric representations, \cf Figure~\ref{fig:theta}(b). Moreover, the spaces of junction fields
\bea
\epsilon & \in & \text{Hom} (\Lambda^{k_1+k_2}\square,\Lambda^{k_1}\square\otimes\Lambda^{k_2}\square) \,, \label{invtensant} \\
\tilde{\epsilon} & \in & \text{Hom} (\Lambda^{k_1}\square\otimes\Lambda^{k_2}\square,\Lambda^{k_1+k_2}\square)\nonumber
\eea
are each one-dimensional.
We choose $\epsilon$ and $\tilde{\epsilon}$ as positive multiples of the respective antisymmetrizations of the identity maps $\square^{k_1}\otimes\square^{k_2}\longleftrightarrow \square^{k_1+k_2}$. Note that this choice depends on the ordering of $(k_1,k_2)$, where a change of the ordering leads to a sign factor $(-1)^{k_1k_2}$.
The normalization is fixed by requiring the relation in Figure \ref{fig:normalization} to hold.
Note that this differs from the normalization used in  \cite{Witten:1989wf}, where the ``$\theta$-web'' would evaluate to $\sqrt{{N\brack k_1}{N\brack k_2}{N\brack k_1+k_2}}$.
\begin{figure} [b] \centering
\includegraphics {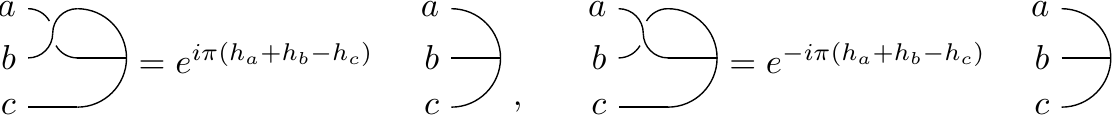} \\[1.5ex]
\includegraphics {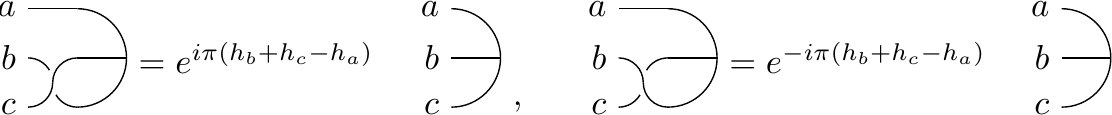}
\caption{Relations between vertices in the conventions of \cite{Witten:1989wf}. Our choice for the vertices between totally antisymmetric representations $\Lambda^{k_1}\square$, $\Lambda^{k_2}\square$ and $\Lambda^{k_1+k_2}\square$ involves an extra sign factor $(-1)^{k_1k_2}$.}
\label{fig:vertexBraiding}
\end{figure}
Also, our choice of signs leads to an additional factor of $(-1)^{k_1k_2}$ in the vertex relations depicted in Figure
\ref{fig:vertexBraiding}, which are valid in the conventions of \cite{Witten:1989wf}, when $a,b,c$ are totally antisymmetric representations $\Lambda^{k_1}\square$, $\Lambda^{k_2}\square$ and $\Lambda^{k_1+k_2}\square$.

\FIGURE[h]{\centering
\includegraphics {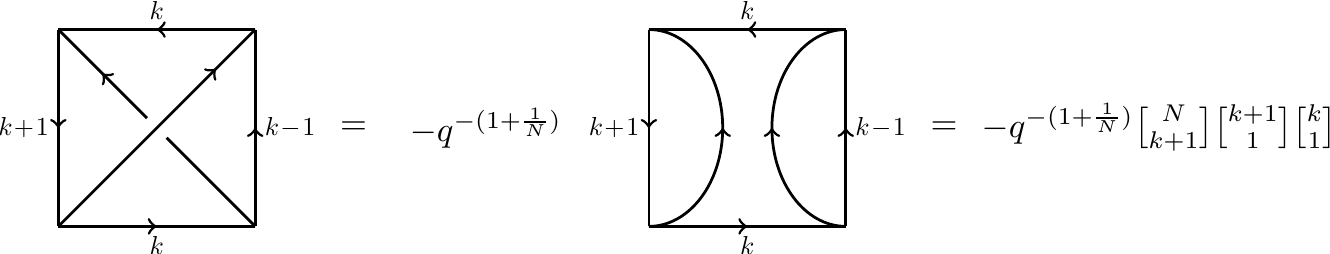}
\caption{Evaluation of the tetrahedral web.}
\label{fig:tetrahedron}}

We conclude this review by computing the expectation value of the tetrahedral network in Figure~\ref{fig:tetrahedron}. It has four Wilson lines in antisymmetric representations $\wedge^{k}\square, \wedge^{k}\square, \wedge^{k+1}\square, \wedge^{k-1}\square$ and two diagonal Wilson lines in the fundamental $\square$, which are positively crossed. The braiding relations for vertices (\cf Figure \ref{fig:vertexBraiding}) on the two junctions at the end of a diagonal line, imply that the expectation value of this tetrahedron is proportional to the expectation value of the one, in which the fundamental Wilson lines are negatively crossed.
The constant of proportionality can be easily calculated to be $q^{-2-\frac{2}{N}}$. (The relevant conformal weights satisfy $\exp(2\pi i\, h(\Lambda^k\square))=q^{(1+\frac{1}{N})(N-k)k}$.)
Using the skein relation of Figure~\ref{fig:skein} one arrives at the first equation in Figure~\ref{fig:tetrahedron}.
The second equality follows from the connected sum formula (Figure~\ref{fig:direct sum}).

\subsection{Web relations}
\label{subsec:rels}

\begin{figure}[tbp] \centering
\includegraphics {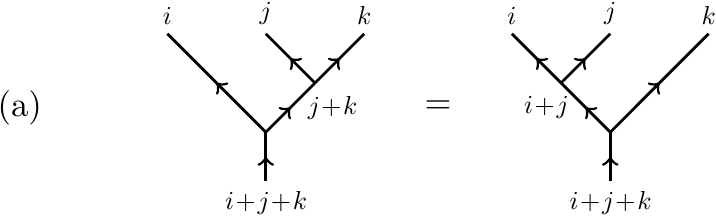} \\[1.5ex]
\includegraphics {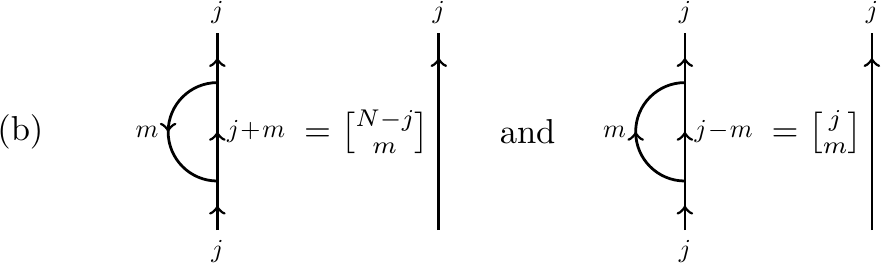} \\[1.5ex]
\includegraphics {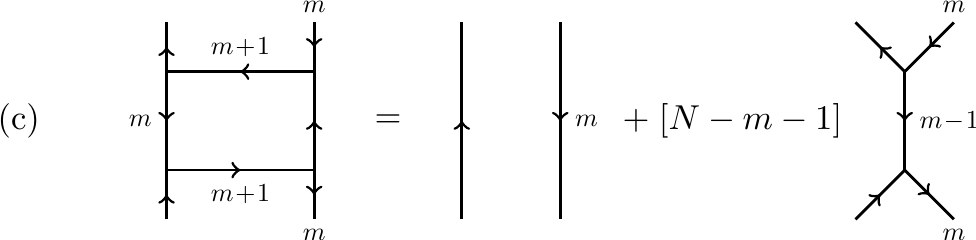} \\[1.5ex]
\includegraphics {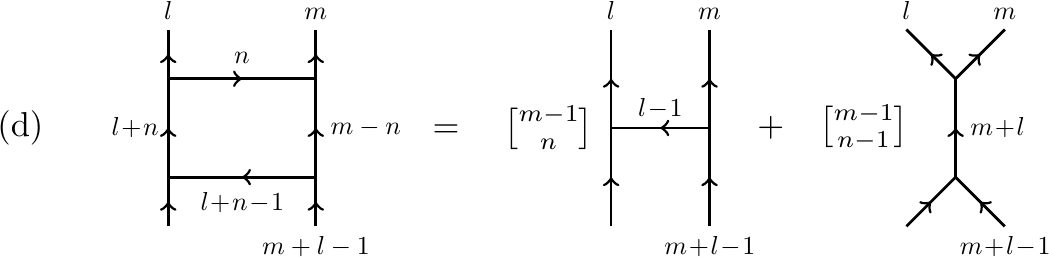} \\[1.5ex]
\includegraphics {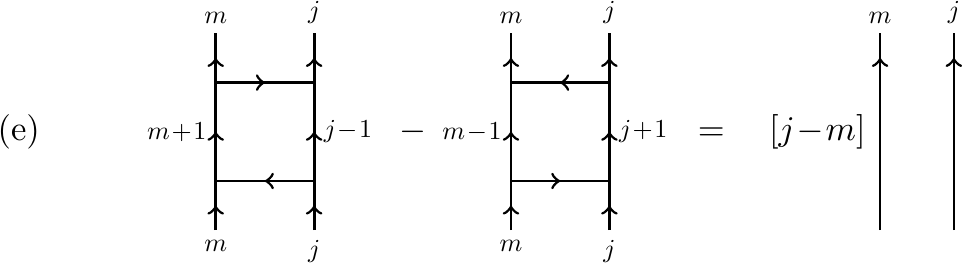} \\[1.5ex]
\includegraphics {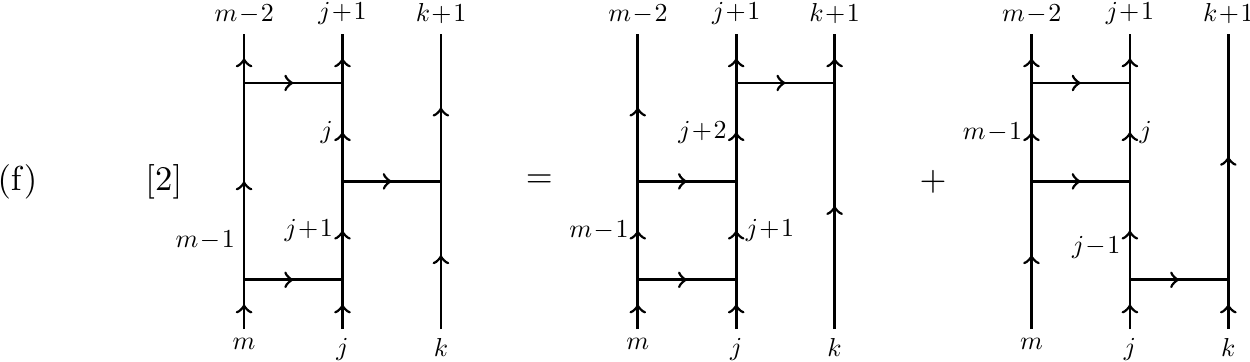} \\[1.5ex]
\caption{Relations among networks of Wilson lines.}
\label{fig:rels}
\end{figure}
Networks of Wilson lines in totally antisymmetric representations satisfy the linear relations depicted in Figure \ref{fig:rels}.
These relations have interesting implications. For instance, identities (e) and (f)
are part of the defining relations of the quantum group $\dot{\mathbf{U}}_{q}(\frak{sl}_{m})$.
The connection between Wilson lines and quantum groups is the subject of section \ref{subsec:SH} below.
Here, we will demonstrate how to derive such relations using the one in Figure~\ref{fig:rels}(e) as an example.
The proofs of the other relations are deferred to Appendix~\ref{appendix:derivations}.

\begin{figure} [htb]\centering
\includegraphics {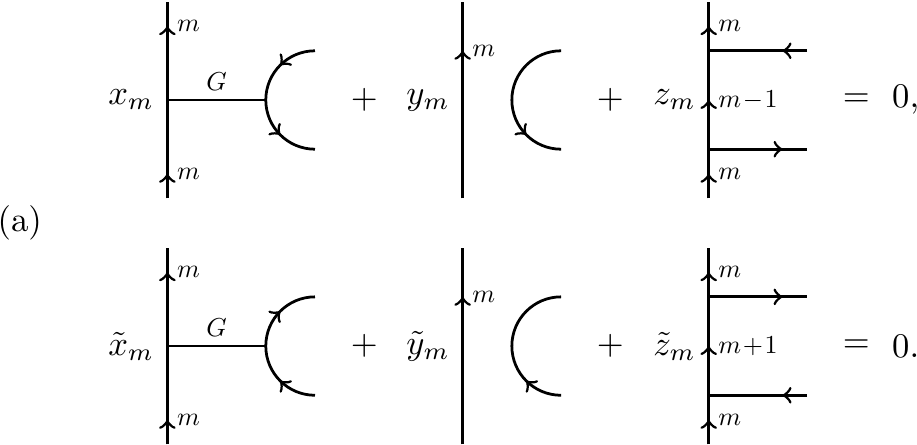}  \\[1.5ex]
\includegraphics {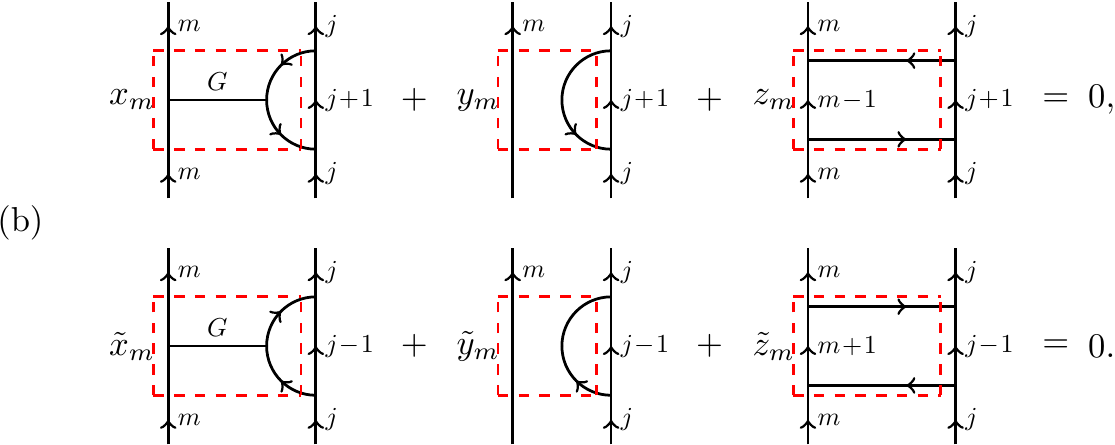} \\[1.5ex]
\includegraphics {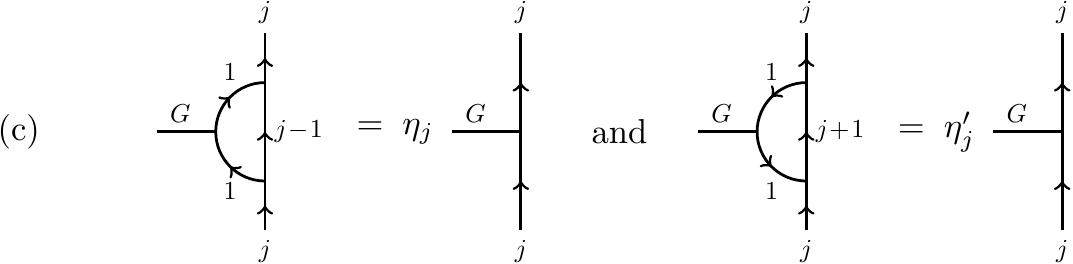}
\caption{(a) Two linear relations among three Wilson lines in $\CH_{S^{2},\{1,m,\bar{1},\bar{m}\}}$, (b) two linear relations among three Wilson lines containing those of (a) in the red dashed box, and (c) proportionality relations between two vectors in $\CH_{S^{2},\{G,j,\bar{j} \}}$.}
\label{fig:xyzs}
\end{figure}

Consider the three configurations of Wilson lines in Figure~\ref{fig:rels}(e) (all lying in $3$-balls with ends on the boundary $2$-sphere). The path integral in these systems gives three vectors in $\CH_{S^{2}, \{m,j,\bar{m},\bar{j} \}}$. Since the dimensions of this space is greater than $2$ for general $m$ and $j$, it is not a priori clear that a relation we are seeking exists. In order to obtain it, we start with the relations among networks in Figure~\ref{fig:xyzs}(a).
That such relations (with some coefficients) have to hold follows from the fact that $\CH_{S^{2},\{1,m,\bar{1},\bar{m}\}}$ is $2$-dimensional.
Here, $G$ stands for the adjoint representation, and we have chosen non-trivial vertices at the end-points of the Wilson lines colored by $G$.
%
(In fact, $\CH_{S^{2},\{G,m,\bar{m}\}}$ is one-dimensional, so the gauge invariant tensor lying at the junction of $G, m, \bar{m}$-colored Wilson lines is proportional to the multiplication $G \otimes \wedge^{m}\square \rightarrow \wedge^{m}\square$. Nevertheless, the precise normalization need not be specified, since as long as $x_{m}$ and $\tilde{x}_{m}$ in Figure \ref{fig:xyzs}(a) are nonzero, we will eventually get Figure \ref{fig:almostEF}. That $x_{m}$ and $\tilde{x}_{m}$ are nonzero can be easily shown by closing off the Wilson lines of Figure \ref{fig:xyzs}(a) in two inequivalent ways.)
Inserting these relations into larger networks of Wilson lines leads to relations of Figure~\ref{fig:xyzs}(b).

Next, from $1$-dimensionality of $\CH_{S^{2},\{G,j,\bar{j} \}}$ one derives the relations in Figure \ref{fig:xyzs}(c).
(Again, we have chosen non-trivial junction fields.) This allows us to relate the first two terms in the identities of Figure \ref{fig:xyzs}(b), and hence to eliminate them from the relations. One arrives at the new relation depicted in Figure \ref{fig:almostEF}, which is a linear relation of the type we are after. To deduce (e) of Figure \ref{fig:rels}, it remains to determine the coefficients $\alpha$ and $\beta$.
\FIGURE[h]{\centering
\includegraphics {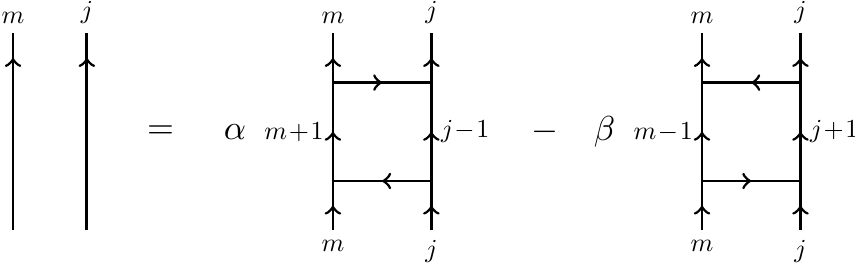}
\caption{A linear relation among three vectors in $\mathcal{H}_{S^{2}, \{ j,m,\bar{j},\bar{m} \}}$. The coefficients $\alpha$ and $\beta$ are functions of $x_m,y_m,z_m,\tilde{x}_m,\tilde{y}_m,\tilde{z}_m,\eta_j,\eta^\prime_j$.}
\label{fig:almostEF}}
We will do this in two steps. First we close off the $m$- and $j$-colored Wilson lines in Figure~\ref{fig:almostEF}.
Using the expectation values of all the resulting networks of Wilson lines that have been determined earlier we obtain the following relation
\be\label{ab1}
\left[N\right]=\alpha\left[N-j\right]\left[m\right]-\beta\left[N-m\right]\left[j\right]\,.
\ee

\begin{figure} [h]
\includegraphics{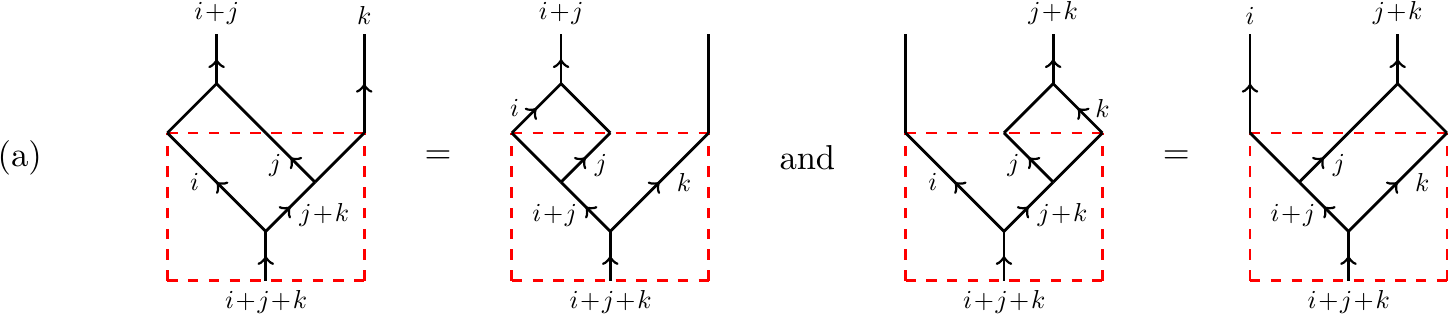}\\[1.5ex]
\includegraphics{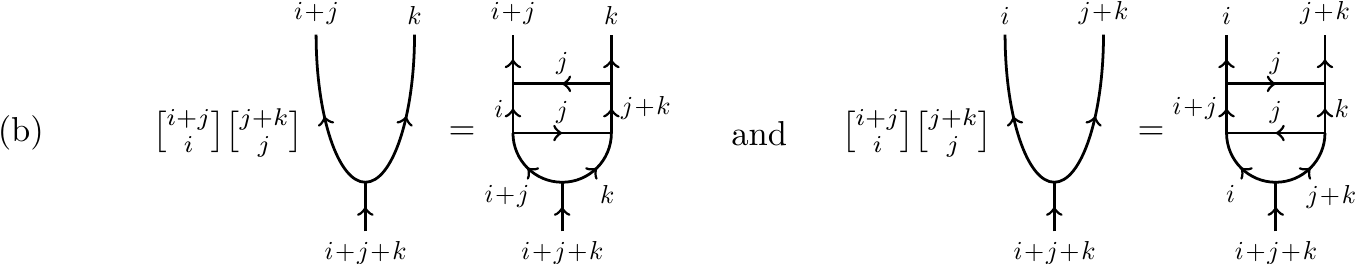}
\caption{(a) Capping off the ``associativity identity''.
(b) Relations in $\CH_{S^2,\{j,m,\overline{j+m}\}}$.}
\label{fig:vertexrel}
\end{figure}

Another relation can be found by connecting the incoming $m$- and $j$-colored Wilson lines in Figure~\ref{fig:almostEF} with a junction to an incoming $j+m$ Wilson line. Since $\CH_{S^2,\{j,m,\overline{j+m}\}}$ is one-dimensional, the vectors associated to all three configurations of Wilson lines are proportional to one another. The constants of proportionality can be easily found: close off Wilson lines in relation of Figure \ref{fig:rels}(a) in a way shown in Figure~\ref{fig:vertexrel}(a), then insert the identity of Figure \ref{fig:rels}(b), and finally apply the resulting identity twice.
The result is depicted in Figure~\ref{fig:vertexrel}(b), from which we obtain another relation on the coefficients $\alpha$ and $\beta$:
\be\label{ab2}
1=\alpha\left[m\right]\left[j+1\right]-\beta\left[m+1\right]\left[j\right]\,.
\ee
Together with \eq{ab1} this fixes the sought-after coefficients to be
\be
\frac{1}{\alpha}=\left[m-j\right]=\frac{1}{\beta}\,,
\ee
finally proving the relation of Figure~\ref{fig:rels}(e).

Let us conclude this subsection by adding a remark on the expectation values of closed and planar trivalent graphs $\Gamma$ of Wilson lines. The relations in Figure \ref{fig:rels}, together with the expectation value of the Wilson loops (equation \ref{eqn:loop exp value}), form the complete set of Murakami-Ohtsuki-Yamada (MOY) graph polynomial relations which uniquely determines the MOY graph polynomials $P_N (\Gamma;q) \in \Z [q,q^{-1}]$, see \cite{Wu}. As the MOY graphs are closed, oriented graphs generated by the junctions of Figure \ref{fig:theta}(b), the uniqueness theorem implies that the expectation value of any planar Wilson lines in antisymmetric representations (and their junctions) can be computed from the definition of MOY graph polynomials. This not only provides us a consistency check for our methods, but also a combinatorial way to compute the expectation value of networks of Wilson lines, as the MOY graph polynomials are combinatorially defined.

\subsection{Skew Howe duality and the quantum group}
\label{subsec:SH}

The quantum group ${\mathbf{U}}_q(\mathfrak{sl}_2)$ is usually defined by means of generators $E,F,K,K^{-1}$ and relations \eqref{Uqsltwo}.
In finite dimensional representations of 
${\mathbf{U}}_q(\mathfrak{sl}_2)$,
 $K$ can be diagonalized with eigenvalues $q^n$, here $n\in\Z$ is the respective $\mathfrak{sl}_2$-weight. $E$ and $F$ raise, respectively lower the weights by $2$. For such representations, one can trade the generators $K,K^{-1}$ for idempotents
$1_n$, $n\in\Z$, projecting on weight spaces of weight $n$
\be\label{qsl2rel1}
1_n1_m=\delta_{n,m}1_n\,.
\ee
This leads \cite{MR1074310} to a modified quantum group $\dot{\mathbf{U}}_q(\mathfrak{sl}_2)$ generated by $E,F$ and the $1_n$.
Since $K1_n=q^{n}1_n$, the algebra relations become
\bea\label{qsl2rel2}
&&E1_n=1_{n+2}E=1_{n+2}E1_n\,,\qquad
F1_n=1_{n-2}F=1_{n-2}F1_n\\
&&\left[E,F\right]1_n=\left[n\right]1_n\,.\nonumber
\eea
It is now very easy to see that if we identify
\be
1_{k_2-k_1}\;\mapsto\;\,\begin{array}{c}
\\
\begin{picture}(30,55)
\put(0,10){\vector(0,1){35}}
\put(0,45){\line(0,1){10}}
\put(-3,0){${\scriptstyle k_1}$}
\put(25,10){\vector(0,1){35}}
\put(25,45){\line(0,1){10}}
\put(22,0){${\scriptstyle k_2}$}
\end{picture}
\end{array}\,,
\qquad
E\;\mapsto\;\,
\begin{array}{c}
\\
\begin{picture}(30,55)
\put(0,10){\vector(0,1){15}}
\put(0,25){\vector(0,1){20}}
\put(0,45){\line(0,1){10}}
\put(25,10){\vector(0,1){15}}
\put(25,25){\vector(0,1){20}}
\put(25,45){\line(0,1){10}}
\put(0,32){\vector(1,0){15}}
\put(15,32){\line(1,0){10}}
\put(11,36){${\scriptstyle 1}$}
\end{picture}
\end{array}\,,\qquad
F\;\mapsto\;\,
\begin{array}{c}
\\
\begin{picture}(30,55)
\put(0,10){\vector(0,1){15}}
\put(0,25){\vector(0,1){20}}
\put(0,45){\line(0,1){10}}
\put(25,10){\vector(0,1){15}}
\put(25,25){\vector(0,1){20}}
\put(25,45){\line(0,1){10}}
\put(25,32){\vector(-1,0){15}}
\put(0,32){\line(1,0){10}}
\put(11,36){${\scriptstyle 1}$}
\end{picture}\end{array}
\,,
\ee
as in \cite{MR3263166},
then, upon concatenation, these configurations of Wilson lines satisfy the quantum group relations, \eq{qsl2rel1} and \eq{qsl2rel2}. In particular, the commutation relation of $E$ and $F$ is nothing but the identity of Figure~\ref{fig:rels}(e), which was explained in the previous subsection. Here, composition of webs (networks) is drawn from bottom to top, and Wilson lines with different labels cannot be joined, \ie their concatenation vanishes. Note, that $E$ and $F$ do not change $k=k_1+k_2$, which characterizes the quantum group representation. For fixed $k$ the identification of strands and idempotents is unambiguous.

Interestingly, while it is well known that the quantum group
${\mathbf{U}}_{q}(\mathfrak{sl}_N)$ appears in the description of $SU(N)$ Chern-Simons theories,
here we realize the quantum group $\mathfrak{sl}_2$ in Chern-Simons theory with gauge group $SU(N)$ for any $N$. The choice of gauge group merely restricts the possible $\mathfrak{sl}_2$ weights which can appear. After all, the labels $k_i$ stand for totally antisymmetric representations $\Lambda^{k_i}\square$ in $SU(N)$ Chern-Simons theory. Hence, the $\mathfrak{sl}_2$ weights can only lie between $-N$ and $N$. The networks of Wilson lines in $SU(N)$ Chern-Simons theory therefore only realize quantum group representations with highest weights $\leq N$.

\begin{figure} [htb] \centering
\includegraphics {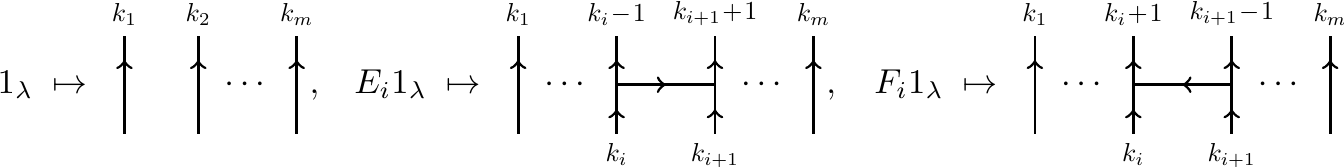}
\caption{$\dot{\mathbf{U}}_{q}(\mathfrak{sl}_{m})$ idempotents and generators in terms of Wilson lines and their junctions.}
\label{fig:SHfunctor}
\end{figure}
This construction can easily be generalized by increasing the number of strands of Wilson lines to $m>2$.
One defines idempotents $1_\lambda$, $\lambda=(k_{2}\!-\!k_{1}, \cdots, k_{m}\!-\!k_{m-1}) \in \mathbb{Z}^{m-1}$
by $m$ parallel strands of Wilson line with labels $k_1,\ldots,k_m$.
For $1\leq i\leq m$, the generators $E_i$ (resp. $F_i$) are defined by
suspending a Wilson line in the fundamental representation between the $i$th and $(i+1)$st strand
(resp. between the $(i+1)$st and the $i$th strand).
These definitions are illustrated in Figure~\ref{fig:SHfunctor}.
Then, using in particular identities (e) and (f) of Figure \ref{fig:rels},
one finds that these webs satisfy the defining relations
of the higher rank quantum group $\dot{\mathbf{U}}_{q}(\mathfrak{sl}_{m})$:
\be
\label{eq:uqslm}
\begin{split}
1_{\lambda}1_{\lambda '} = \delta_{\lambda, \lambda ' }1_{\lambda}, \quad  E_{i}1_{\lambda} =  1_{\lambda+l_i} E_{i}, \quad F_{i}1_{\lambda}=1_{\lambda-l+i}F_{i}, \\[1.5ex]
[E_{i},F_{j}]1_{\lambda} = \delta_{i,j}[\lambda_{i}]1_{\lambda}, \quad [E_{i},E_{j}]1_{\lambda} = 0 \quad \text{for} \quad |i-j|>1, \\[1.5ex]
\text{and} \quad E_{i}E_{j}E_{i}1_{\lambda} = E_{i}^{(2)}E_{j}1_{\lambda}+E_{j}E_{i}^{(2)}1_{\lambda} \quad \text{for} \quad |i-j|=1.
\end{split}
\ee
Here $\lambda$ denotes an $\mathfrak{sl}_m$ weight.
The generators $E_i$ (resp. $F_i$) associated to the simple roots of $\mathfrak{sl}_m$ raise (resp. lower) $\lambda$
by $l_i=(0,\ldots,0,-1,2,-1,0,\ldots,0)$ where $2$ appears in position $i$.

To summarize, in $SU(N)$ Chern-Simons theory with arbitrary $N$, we obtain a realization of $\dot{\mathbf{U}}_{q}(\mathfrak{sl}_{m})$ on a configurations of Wilson lines with $m$ strands. The rank of the Chern-Simons gauge group only restricts the possible representations of the quantum group which can be obtained in this way. What we described is a physical realization of the skew Howe 
duality of \cite{MR2593278}
\be
\Lambda^k(\C^m\otimes\C^N)\cong
\bigoplus_{k_1+\ldots+k_m=k}
\Lambda^{k_1}\C^N\otimes\ldots\otimes\Lambda^{k_m}\C^N\,.
\label{skewHowekmn}
\ee
The actions of $\mathfrak{sl}_m$ and $\mathfrak{sl}_N$, respectively, on $\C^m$ and $\C^N$ commute,
and the direct sum on the RHS of this equation is indeed the weight decomposition with respect to the $\mathfrak{sl}_m$ action.
This duality has a direct generalization to quantum groups $\dot{\mathbf{U}}_{q}(\mathfrak{sl}_{m})$.

In Chern-Simons theory we interpret each summand on the RHS of \eqref{skewHowekmn}
as the Hilbert space of $m$ Wilson lines in representations, $R_1 = \Lambda^{k_1}\C^N$, {\it etc.},
\be
R_1 \otimes \ldots \otimes R_m \; = \; \Lambda^{k_1}\C^N\otimes\ldots\otimes\Lambda^{k_m}\C^N \,,
\ee
so that raising and lowering operators, $E_i$ and $F_i$, which relate different weight spaces are realized
as configurations of $SU(N)$ Wilson lines in totally antisymmetric representations with $m$ incoming and $m$ outgoing strands.

\FIGURE[h!]{\centering
\includegraphics {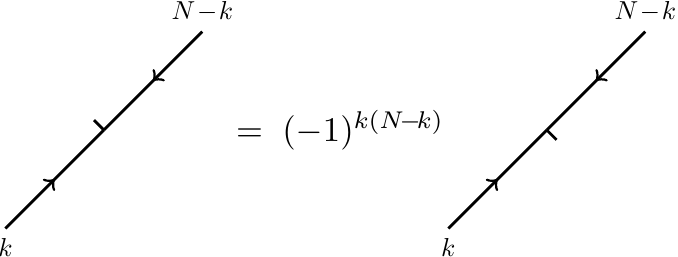}
\caption{``Tag'' morphism relation of the $N\rm{Web}$ category.}
\label{fig:tag}}
In fact, the identification of Figure~\ref{fig:SHfunctor} exactly corresponds to the skew Howe duality functor from
the quantum group $\dot{\mathbf{U}}_{q}(\mathfrak{sl}_{m})$ to the category $N\rm{Web}_{m}$ \cite{MR3263166},
whose objects are tuples $(k_1,\ldots,k_m)$, $0\leq k_i\leq N$, and whose morphisms are ``$\mathfrak{sl}_N$-webs''.
The latter are generated by the morphisms depicted in Figure~\ref{fig:SHfunctor} modulo the relations of Figure~\ref{fig:rels}.

Note that the category $N\rm{Web}_m$ contains a ``tag'' morphism satisfying the relation in Figure \ref{fig:tag}.
This tag corresponds to a junction with the $N$th antisymmetric representation. Since this is a trivial representation, the respective Wilson line is trivial and does not have to be drawn. The junction however still requires a specification of the junction field, \ie an invariant tensor $\Lambda^k\square \otimes\Lambda^{N-k}\square\longrightarrow\Lambda^N\square$, which was determined by the ordering of incoming Wilson lines. This ordering is specified by the direction of the outgoing Wilson line, hence the tag, and the change of ordering leads to a sign factor $(-1)^{k(N-k)}$ as in the tag relation.


\subsection{Why ``categorification = surface operators''}
\label{sec:whysurf}

In general, a $d$-dimensional TQFT assigns a number (= partition function) to every closed $d$-manifold $M_d$,
a vector space $\CH (M_{d-1})$ to every $(d-1)$-manifold, a category $\frak{C} (M_{d-2})$ to every $(d-2)$-manifold, and so on.
Therefore, the process of ``categorification'' that promotes each of these gadgets to the higher categorical level
has an elegant interpretation as ``dimensional oxidation'' in TQFT, which promotes a $d$-dimensional TQFT to
a $(d+1)$-dimensional one \cite{Crane:1994ty} (see also \cite{Gukov:2007ck}).

\begin{figure}[htb]
\centering
\includegraphics[width=4.5in]{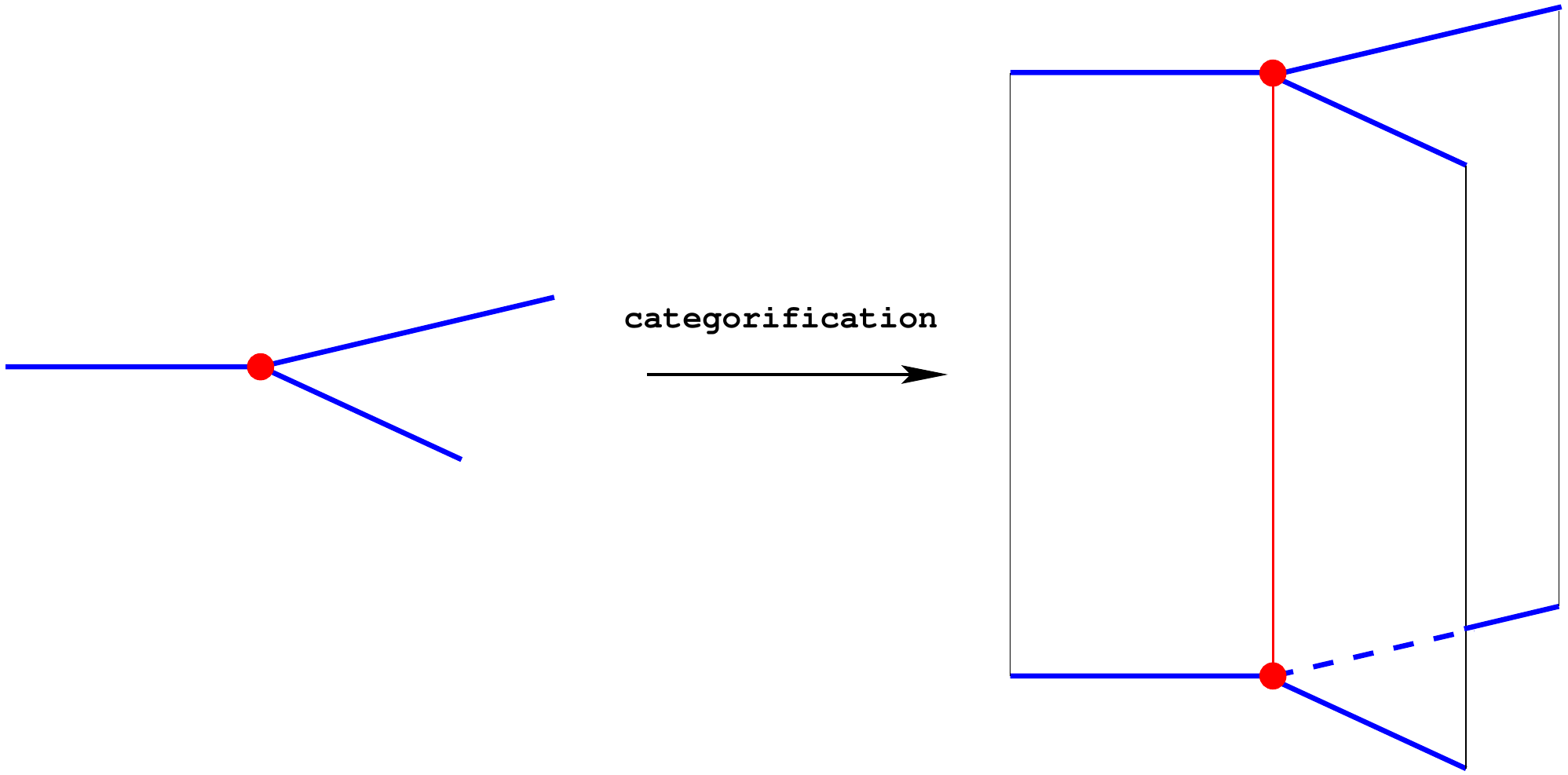}
\caption{Categorification means lifting a given theory to one dimension higher. This operation of adding an extra dimension
turns line operators into surface operators \cite{Gukov:2007ck} and graphs (networks) into foams (surfaces with singular edges).}
\label{fig:junction11}
\end{figure}

Many interesting theories come equipped with non-local operators (often called ``defects'')
that preserve and frequently ameliorate the essential structure of the theory.
Categorification has a natural extension to theories with such non-local operators which also gain an extra dimension,
much like the theory itself.
Prominent examples of such operators are codimension-$2$ line operators in 3d Chern-Simons TQFT that we
encountered earlier and that do not spoil the topological invariance of the theory.
One dimension lower, such codimension-$2$ operators would actually be local operators supported at points $p_i$ on a 2-manifold $M_2$.
And, one dimension higher, in a 4d TQFT such codimension-$2$ operators would be supported on surfaces $\Sigma \subset M_4$.
Therefore, on general grounds a categorification of 3d Chern-Simons invariants should be a functor (4d TQFT) that assigns
\bea
{\mbox{\rm surface}} ~\Sigma \subset M_4 \quad & \leadsto & \quad {\mbox{\rm number}} ~Z(M_4;\Sigma) \nonumber\\
{\mbox{\rm knot/web}} ~K \subset M_3 \quad & \leadsto & \quad {\mbox{\rm vector space}} ~\CH (M_3;K) \label{categorlore} \\
{\mbox{\rm points}} ~p_1, \ldots, p_n \in M_2 \quad & \leadsto & \quad {\mbox{\rm category}} ~\frak{C} (M_2;p_1, \ldots, p_n) \nonumber \\
& \vdots & \nonumber
\eea
In particular, as illustrated in Figure \ref{fig:junction11},
categorification of graphs and networks is achieved by studying surface operators with singular edges (junctions) in four dimensions.
This naturally leads us to the study of junctions of surface operators and the effective 2d theory on their world-sheet $\Sigma$
which, respectively, will be the subjects of sections \ref{sec:surface} and \ref{sec:LG}.

\FIGURE[thb]{\includegraphics[width=5.5cm]{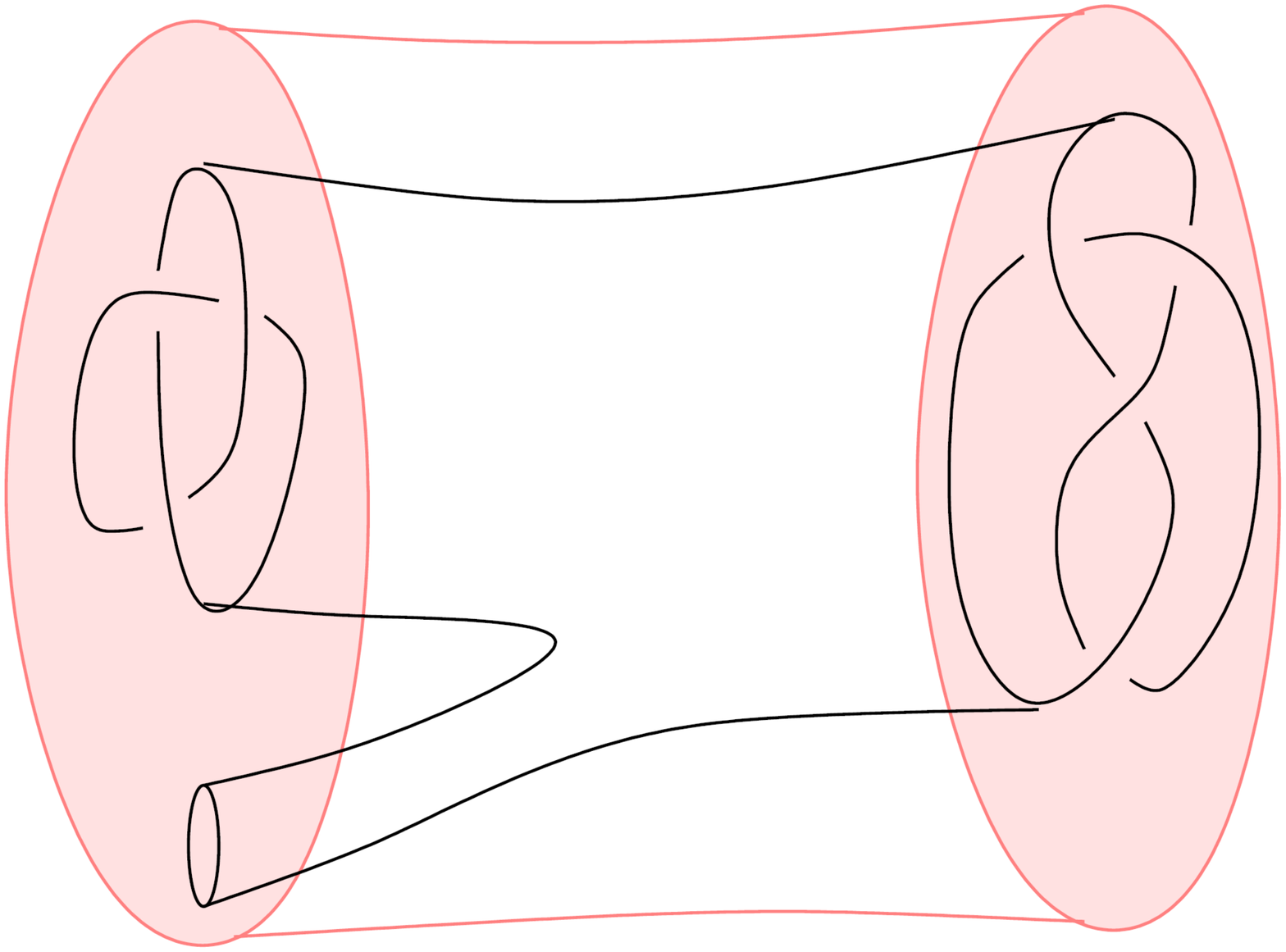}
\caption{\label{fig:cobordism2}
A link cobordism $\Sigma$ defines a map between the corresponding homology groups.}}
Moreover, functoriality implies that the structure at each level should be compatible with cobordisms as well as operations of cutting and gluing.
It turns out to be very rich even before one tackles the very interesting question of studying
tangle cobordisms in non-trivial ambient manifolds. In other words, even when $M_2 = \R^2$, $M_3 = \R^3$ and $M_4 = \R^4$ in \eqref{categorlore}
functoriality can be highly non-trivial due to interesting topologies of $K$ and $\Sigma$.
In particular, a surface cobordism $\Sigma$ between links $K_1$ and $K_2$ gives rise to a linear map (see Figure~\ref{fig:cobordism2}):
\be
Z(\Sigma) ~: \quad \CH (K_1) \longrightarrow \CH (K_2)
\label{cobordmapHH}
\ee
where, to avoid clutter, we tacitly assumed that all components of $K_1$ and $K_2$ have the same color.
(Otherwise, $\Sigma$ can not be a smooth surface and must be a {\it foam}, {\it i.e.} have singular edges.)
In fact, all our defects --- line operators in 3d, surface operators in 4d, as well as points in 2d --- carry
certain labels or colors which we denote by $R_i$.
For example, making this part of our notation explicit and, on the other hand, suppressing $M_2 = \R^2$,
the third line in \eqref{categorlore} should read:
\be
(p_1, R_1), \ldots , (p_n, R_n) \quad \leadsto \quad {\mbox{\rm category}} ~\frak{C}_{R_1, \ldots, R_n}
\label{unknotcategory}
\ee
The Hochschild homology of this category must be the cohomology of $n$ copies of the unknot, colored by $R_1, \ldots, R_n$, respectively:
\be
HH^* (\frak{C}_{R_1, \ldots, R_n}) \; = \; \CH^{R_1} \big( \unknot \big) \otimes \ldots \otimes \CH^{R_n} \big( \unknot \big)
\label{Hochschunknot}
\ee
This is a part of the reverse process, called ``decategorification,'' which in TQFT corresponds to dimensional reduction (of the ``time'' direction). It will prove very useful later, in section \ref{sec:LGlocal},
where it will help us to better understand the cohomology of the colored unknot as well as the structure of the category \eqref{unknotcategory}. Notably, applying \eqref{cobordmapHH} to the cobordism between on the one hand the disjoint union $K_1 = K \bigsqcup \unknot$ of a knot $K$ with the unknot and on the other hand their
connected sum $K_2 = K \# \unknot \cong K$ which gives back $K$, yields a map
\be
\CA \otimes \CH^{R} (K) \; \to \; \CH^{R} (K)\,,
\label{HHHunknot}
\ee
where
\be
\CA \; := \; \CH^{R} (\unknot)\,.
\label{Aunknot}
\ee
In fact, for the special case
$K=\unknot$ this defines an algebra structure 
\be
{\raisebox{-.4cm}{\includegraphics[width=1.0cm]{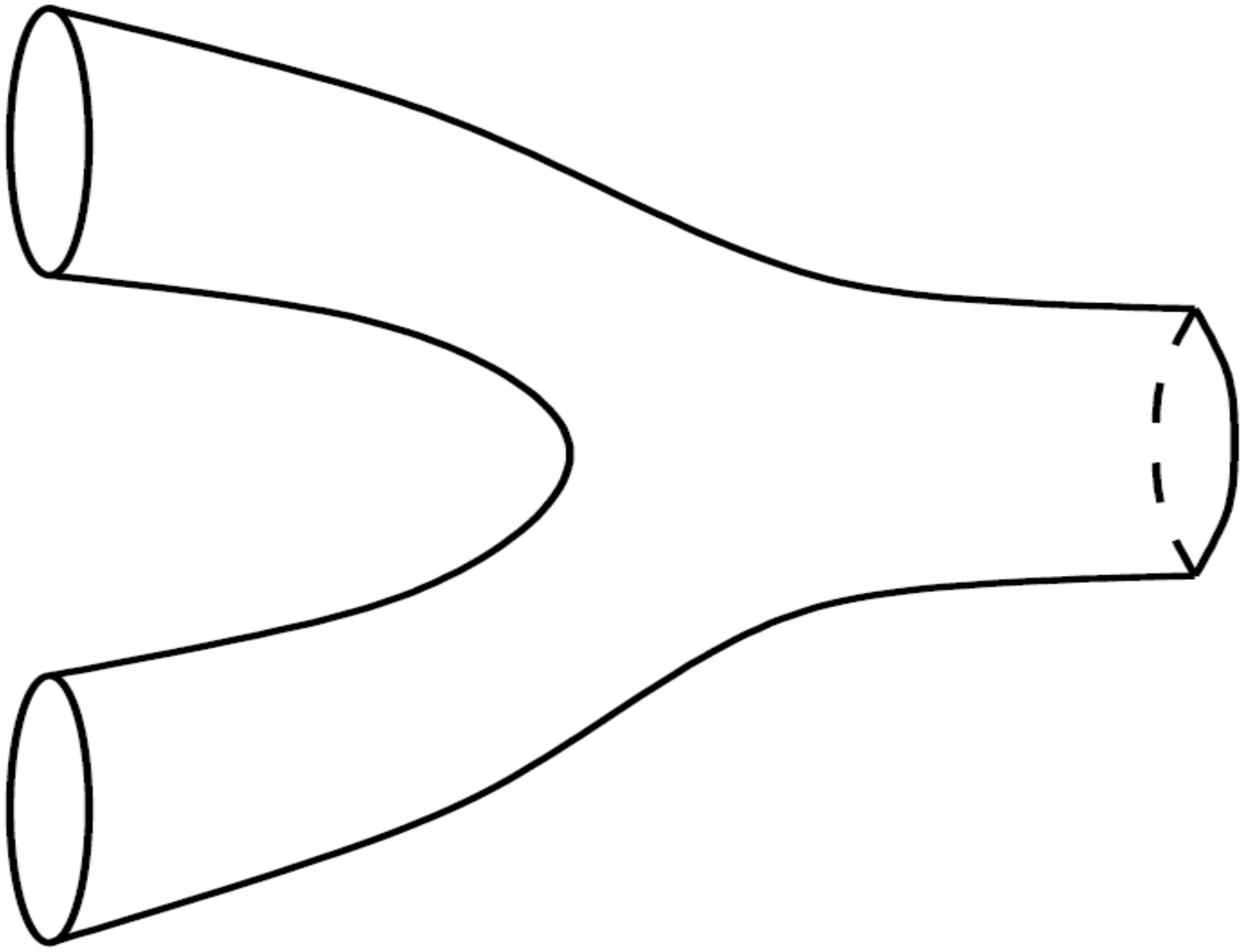}}\,}
~: \qquad
\CA \otimes \CA \; \to \; \CA\,,
\label{unknotAAA}
\ee
on the homology of the unknot, and \eq{HHHunknot} promotes the homology $\CH^{R}(K)$ of any knot $K$ to an $\CA$-module.

It turns out that the algebra structure of the unknot homology \eqref{unknotAAA} contains a lot of useful information
about how the theory behaves under cobordisms \cite{MR2174270,MR2253455,MR2336253}
and has been used to construct various deformations of Khovanov-Rozansky
homology \cite{MR2173845,MR2232858,Gornik,MR2826932,MR2916277,Rose:2015pla}.
Much of this algebraic structure extends even to colored HOMFLY-PT homology \cite{Gorsky:2013jxa}.

A 2d TQFT is basically determined by a Frobenius algebra, defined by a ``pair-of-pants'' product as in \eqref{unknotAAA}. In case the theory is obtained from a supersymmetric one by topological twisting, than  this algebra is the chiral ring of the untwisted theory. 
In our present context this implies that the chiral ring of the world-volume theory of a surface operator labeled by representation $R$ has to agree with the algebra \eqref{unknotAAA} associated to the unknot colored by $R$,
\be
\CA \; = \; \CH^{R} (\unknot) \; = \; \text{chiral ring of 2d TQFT on the surface operator}\,.
\label{unknotchiralA}
\ee
This provides a useful clue for identifying the surface operators (and their world-volume theories) which are relevant for the categorification of quantum groups. 

For example, the original Khovanov homology \cite{Khovanov} is based on the nilpotent Frobenius algebra
\be
\CA \; = \; \C [X] / \langle X^2 \rangle
\ee
A semi-simple deformation of this algebra studied by Lee \cite{MR2173845}
led to a number of important developments, including Rasmussen's proof of the Milnor conjecture \cite{MR2729272}.


\section{Junctions of surface operators}
\label{sec:surface}

Surface operators are non-local operators which,
in four-dimensional QFT, are supported on two-dimensional surfaces embedded in four-dimensional space-time $M_4$,
\be
\Sigma \; \subset \; M_4
\label{4dspacetime}
\ee
Intuitively, surface operators can be thought of as non-dynamical flux tubes (or vortices)
much like Wilson and 't Hooft line operators can be thought of as static electric and magnetic sources, respectively.
Compared to line operators, however, surface operators have a number of peculiar features.
See \cite{Gukov:2014gja} for a recent review on surface operators.

Thus, in gauge theory with gauge group $G$, line operators are labeled by discrete parameters,
namely electric and magnetic charges of the static source, while surface operators in general
are labeled by {\it both} discrete and continuous parameters.
The former are somewhat analogous to discrete labels of line operators,
but the latter are a novel feature of surface operators.

In practice, there are two basic ways of constructing surface operators:

\begin{itemize}

\item
as a singularity along $\Sigma$, {\it e.g.}
\be
F = 2\pi \alpha \delta_{\Sigma} + \ldots
\label{alphafirst}
\ee

\item
as a coupled 2d-4d system, {\it e.g.} described by the action
\be
S_{\text{tot}}
= \int_{M_4} d^4 x \; \left( \CL_{\text{4d}} + \delta_{\Sigma} \cdot \CL_{\text{2d}} \right)
= \int_{M_4} d^4 x \; \CL_{\text{4d}} + \int_{\Sigma} d^2 x \; \CL_{\text{2d}}
\label{S2d4d}
\ee
where the global symmetry $G$ of the 2d Lagrangian $\CL_{\text{2d}}$ is gauged when coupled to the 4d Lagrangian $\CL_{\text{4d}}$.

\end{itemize}

\noindent
Both descriptions can be very useful for establishing the existence and analyzing junctions of surface operators,
which locally look like a product of ``time'' direction $\R_t$ with a Y-shaped graph where three (or more)
surface operators meet along a singular edge of the surface $\Sigma$,
thus providing a physics home and a microscopic realization of the ideas advocated in \cite{Alford:1992yx,Rozansky:2003hz,MR2322554}.
A surface $\Sigma$ with such singular edges is often called a ``foam'' or a ``seamed surface''.

Before one can start exploring properties of junctions and their applications,
it is important to establish their existence at two basic levels:

$a)$ at the level of ``kinematics'' as well as

$b)$ at the level of field equations (or BPS equations, if one wants to preserve some supersymmetry).

\noindent
The former means that junctions of surface operators must obey certain charge conservation conditions
analogous to \eqref{invtensors},
while the latter means they must be a solution to field equations or BPS equations, at least classically.
Addressing both of these questions will be the main goal of the present section, where we treat cases with
different amount of supersymmetry in parallel.
In particular, sections \ref{sec:N4}--\ref{sec:N1} will be devoted to the latter question,
whereas kinematics will be largely the subject of section \ref{sec:Horn}.
Before we proceed to technicalities, however, let us give a general idea of what the answer to each of these questions looks like.

When a configuration of surface operator junctions is static, we have $\Sigma = \R_t \times \Gamma$,
where $\R_t$ is the ``time'' direction and $\Gamma$ is a planar trivalent graph or, more generally,
a trivalent oriented graph in a 3-manifold $M_3$ which, for most of what we need in this paper, will be simply $\R^3$ (or $S^3$).
Then, as we shall see, in many cases the BPS equations on $M_3$ will reduce to (or, at least, contain solutions of)
the simple flatness equations for the gauge field $A$:
\be
F_A \; = \; 0
\label{flatA}
\ee
For example, the famous instanton equations $F_A^+ = 0$ in Donaldson-Witten theory reduce to \eqref{flatA}
on a 4-manifold of the form $\R_t \times M_3$ when one requires solutions to be invariant under translations along $\R_t$.
Since \eqref{flatA} is a ``universal'' part of BPS equations for many different systems,
with different amounts of supersymmetry, it makes sense to illustrate how the questions
of ``kinematics'' and the ``existence of solutions'' look in the case of \eqref{flatA}.
By doing so, we also focus our attention on the most interesting ingredient, namely the gauge dynamics.
Then, adding more fields and interactions to the system is a relatively straightforward exercise and we comment on it in each case.

\FIGURE[htb]{\includegraphics[width=4.5cm]{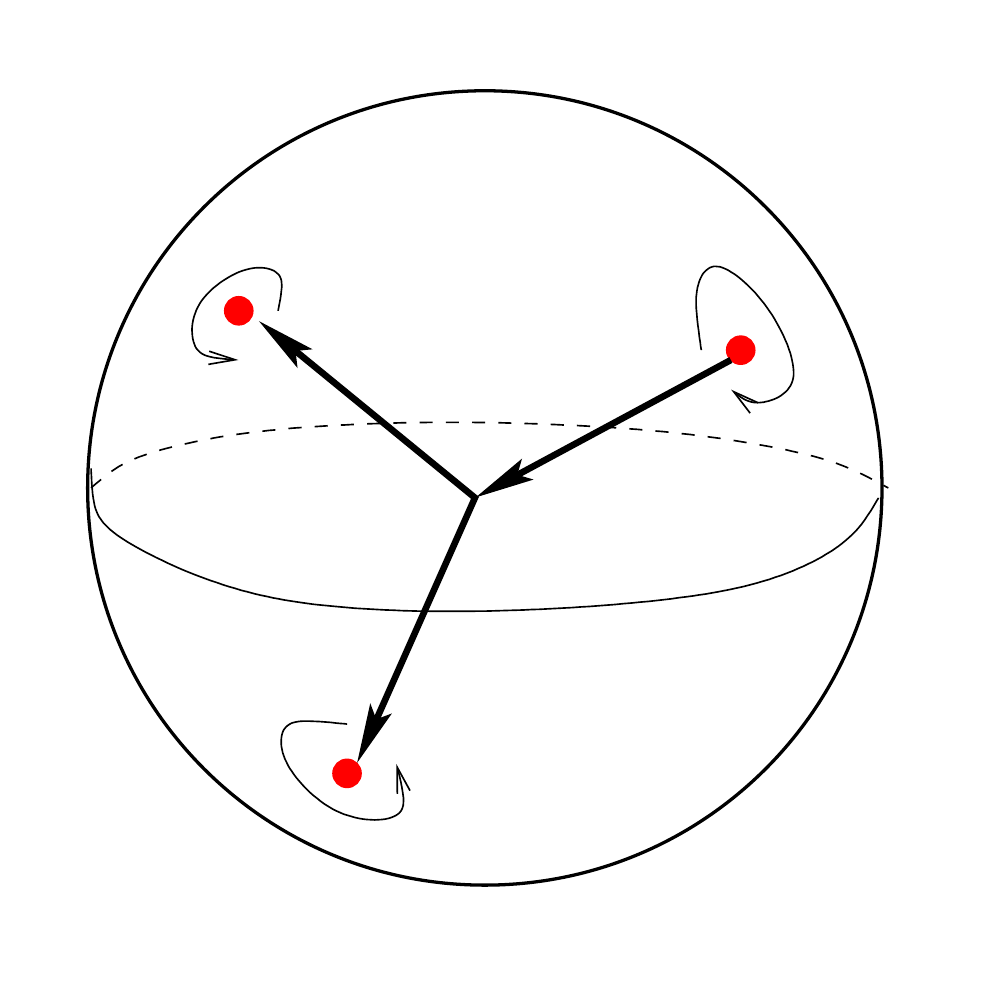}
\caption{\label{fig:junction13}
A small ball around a trivalent junction.}}
As explained in \cite{Gukov:2006jk}, a surface operator with non-zero $\alpha$ in \eqref{alphafirst}
can be thought of as a Dirac string of a magnetic monopole with improperly quantized magnetic charge $\alpha$.
Magnetic charges of monopoles that obey Dirac quantization condition take values in the root lattice $\Lambda_{\text{rt}}$
of the gauge group $G$. When this condition is not obeyed, the world-sheet of a Dirac string becomes
visible to physics and this is precisely what a surface operator is. In this way of describing surface operator
--- as a singularity for the gauge field (and, possibly, other fields) --- it should be clear that $\alpha$ in \eqref{alphafirst}
takes values in the Lie algebra of the maximal torus of the gauge group, $\frak t = \text{Lie} (\mathbb{T})$,
modulo the lattice of magnetic charges $\Lambda_{\text{rt}}$.
For example, when $G = U(1)$ the parameter $\alpha$ is a circle-valued variable.

In this description of a surface operator --- as a singularity \eqref{alphafirst} or, equivalently, as a Dirac string
of an improperly quantized magnetic charge --- the orientation of the edges of the graph $\Gamma$ has a natural interpretation:
it represents the magnetic flux. Of course, one can change the orientation of each edge and simultaneously invert
the holonomy $U = \exp (2\pi i \alpha) \in G$ of the gauge field around it, without affecting the physics:
\be
\xleftarrow[~~~~~~~~]{~U} \quad = \quad \xrightarrow[~~~~~~~~]{~U^{-1} }
\label{holorient}
\ee
Moreover, if the gauge field satisfies \eqref{flatA} away from the singularity locus $\Sigma = \R_t \times \Gamma$,
then the magnetic flux must be conserved at every junction.
In the abelian theory with $G = U(1)$ it simply means that the signed sum (signs determined by the orientation)
of the parameters $\alpha$ for all incoming and outgoing edges is equal to zero at every vertex of $\Gamma$, {\it e.g.}
\be
\alpha = \alpha' + \alpha'' \quad (\text{mod}~1)
\label{afluxcons}
\ee
for a basic trivalent junction as depicted in Figure \ref{fig:junction13}.
A non-abelian version of flux conservation is more involved and will be discussed in section \ref{sec:Horn}.
This preliminary discussion, however, should give a reasonably good idea of what a junctions looks like
in the description of a surface operator as a singularity (or, ramification) along $\Sigma$.

\FIGURE[htb]{\includegraphics[width=5.4cm]{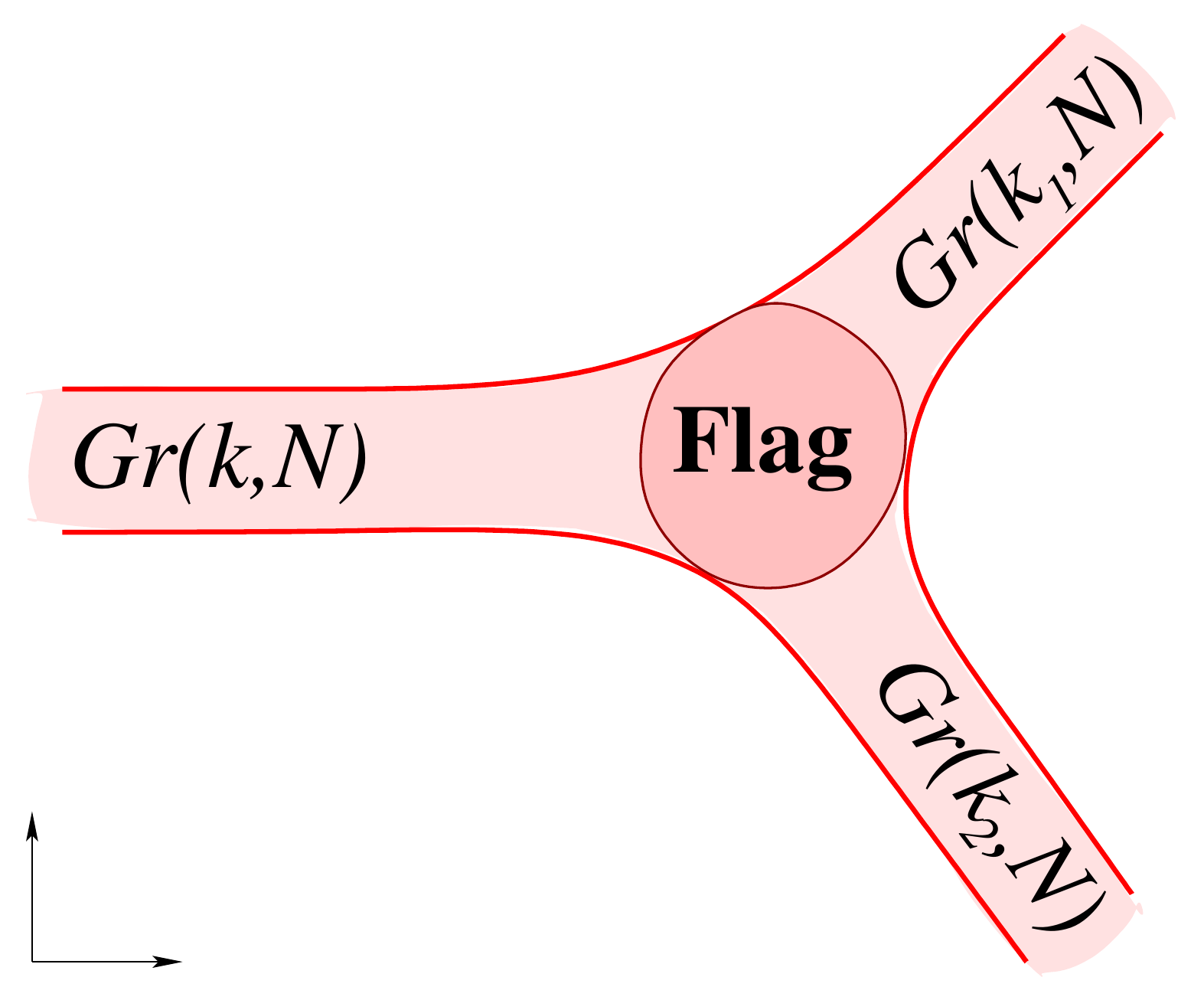}
\caption{\label{fig:junction5}
The core of a junction between surface operators \eqref{Grassmannian} carries a partial flag variety $Fl(k_1,N)$, $k=k_1+k_2$.
Shown here is the $(x^1,x^2)$ plane.}}Now, let us have a similar preliminary look at the same question from the viewpoint of a 2d-4d coupled system {\it \`a la} \eqref{S2d4d}.
In this approach, each face of the seamed surface $\Sigma$ is decorated with a 2d theory that lives on it,
in many examples simply a 2d sigma-model with a target space $\CC$ that enjoys an action of the symmetry group $G$,
{\it e.g.} a conjugacy class\footnote{Notice, this is similar to labeling of surface operators defined as singularities,
where conjugacy classes characterize possible values of the holonomies $U \in \CC$.} or a flag variety $\CC = G / \mathbb{L}$.
In the approach based on 2d-4d coupled system,
a junction of surface operators (or, a more general codimension-1 defect along $\Sigma$)
determines a ``correspondence'' $\mathcal{M} \subset \CC \times \CC' \times \CC''$:
\be
\begin{matrix}
 & & \mathcal{M} & & \\
 & \swarrow & & \searrow & \\
\CC & & & & \CC' \times \CC''
\end{matrix}
\label{correspondences}
\ee
where $\CC'$ and $\CC''$ are target spaces of 2d sigma-models on ``incoming'' edges of $\Gamma$
that join into an edge that carries sigma-model with target space $\CC$.
The description of such interfaces (or line defects) in the dual\footnote{in the sense of LG / sigma-model duality \cite{Lerche:1989uy,Witten:1993yc}}
Landau-Ginzburg model will be the main subject of section \ref{sec:LG}.

Surface operators that will be relevant to skew Howe duality and categorification of quantum groups
are labeled by ``Levi types" $\mathbb{L} = S(U(k)\times U(N-k))$
in a theory with gauge group $G = SU(N)$, or simply by $k \in \{ 1, \ldots, N-1 \}$.
In the 2d-4d description of such ``$k$-colored'' surface operators the target space $\CC = G / \mathbb{L}$
is the Grassmannian of $k$-planes in $\C^N$,
\be
\CC \; = \; Gr(k,N) \; = \; SU(N) / S(U(k)\times U(N-k))
\label{Grassmannian}
\ee
As we proceed, we will encounter Grassmannian varieties\footnote{In case of a more general irreducible representation of highest weight $\lambda$,
the space $Gr(k,N)$ would be replaced by a finite-dimensional subspace $\bar{Gr}^{\lambda}$ of the affine Grassmannian
that plays an important role in the geometric Satake correspondence.
Since such spaces are generally singular, we leave detailed study of the corresponding surface operators
to future work and focus here on the ones relevant to skew Howe duality and categorification of quantum groups.}
more and more often, as moduli spaces of solutions in gauge theory (later in this section)
as well as in the description of interfaces and chiral rings in Landau-Ginzburg models (in section \ref{sec:LG}).

What about junctions of surface operators that carry Grassmannian sigma-model on their world-sheet?
They can be conveniently described as correspondences \eqref{correspondences}, where $\CM$ is a partial flag
variety $Fl(k_1,k,N)$ of $k_1$-planes in $k=(k_1+k_2)$-planes in $\C^N$, as illustrated in Figure~\ref{fig:junction5}:
\be\label{correspondence}
\begin{array}{ccccc}
 Gr(k,N) & \longleftarrow & Fl(k_1,k,N) & \longrightarrow &Gr(k_1,N)\times Gr(k_2,N)\\
V & \longmapsfrom
& (V_1\subset V) & \longmapsto& (V_1,V_1^\perp\subset V)
\end{array}
\ee

Now, once we had a preliminary discussion of how the decorations of surface operators flow across junctions,
both from the singularity perspective and in the approach of 2d-4d coupled system, it is natural to tackle
the question of their existence at the level of field equations.
As we already noted earlier, it involves solving PDEs in three dimensions with prescribed boundary conditions
around $\Gamma \subset M_3$, which in general is not an easy task, even for the basic version of the equations \eqref{flatA}.
While it would be extremely interesting to pursue the construction of explicit solutions, we will only need to know
whether they exist and what their moduli space looks like. Luckily, this latter question can be addressed without
constructing explicit solutions and an illuminating way to do that is via embedding the gauge theory into string / M-theory.

Many interesting SUSY field theories on $M_4$ can be realized on stacks of M-theory
fivebranes, such as configurations of  
\be
N~\text{M5-branes}\;{\rm on}\; M_4 \times C
\label{M5branes}
\ee
Indeed, for different choices of $2$-manifolds $C$ and their embeddings in the eleven-dimensional space-time
one can realize 4d field theories on $M_4$ with $\CN=1$, $\CN=2$, or $\CN=4$ supersymmetry.
Below we consider each of these choices in turn.

In such constructions, one can ``engineer'' surface operators by introducing
additional M5-branes or M2-branes which share only the two-dimensional part of the 4d geometry, $\Sigma \subset M_4$,
with the original set of fivebranes \eqref{M5branes}.
In particular, in such brane constructions, the junctions of surface operators
relevant to skew Howe duality and categorification of quantum groups
have a simple interpretation where one stack of $k$ branes splits up into two stacks of $k_1$ and $k_2=k-k_1$ branes,
as illustrated in Figure~\ref{fig:junction4}.

\FIGURE[htb]{\includegraphics[width=5.0cm]{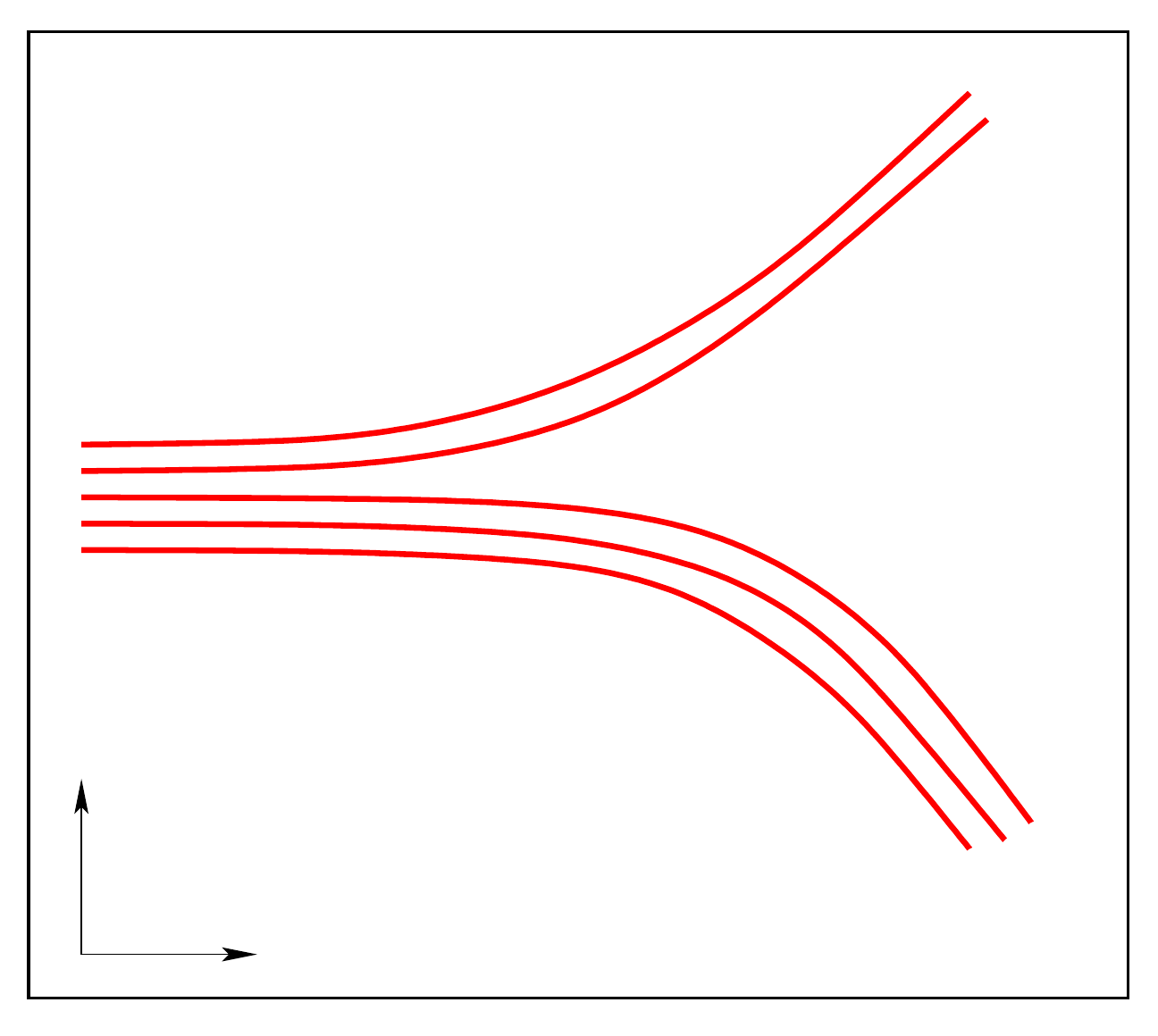}
\caption{\label{fig:junction4}
Brane realization of the junction of surface operators in Figure~\ref{fig:junction5}.}}

\subsection{Junctions in 4d $\CN=4$ theory}
\label{sec:N4}

When $C=T^2$ or $\R^2$, the fivebrane configuration \eqref{M5branes} preserves maximal ($\CN=4$) supersymmetry in 4d space-time $M_4$.
Then, depending on the geometry of $\Sigma$ and $M_4$ in \eqref{4dspacetime}, one finds different fractions of unbroken supersymmetry.

If both $M_4$ and $\Sigma$ are flat, {\it i.e.} $M_4 = \R^4$ and $\Sigma = \R^2$, then surface operators can be half-BPS
and admit the following brane construction
\be
\begin{array}{l@{\;}|@{\;}ccccccccccc}
& ~0~ & ~1~ & ~2~ & ~3~ & ~4~ & ~5~ & ~6~ & ~7~ & ~8~ & ~9~ & ~10~ \\
\hline
\text{M5}~ & \times & \times & \times & \times &  &  & \times &  &  &  & \times \\
\text{M5}'~ & \times & \times &  &  &  &  & \times &  & \times & \times & \times
\end{array}
\label{M5N4}
\ee
where, following conventions of \cite{Witten:1997sc}, we assume that $M_4$ is parametrized by $(x^0,x^1,x^2,x^3)$
and $C$ is parametrized by $x^6$ and $x^{10}$.

The half-BPS surface operators in 4d $\CN=4$ theory can form $\frac{1}{4}$-BPS junctions of the form $\Sigma = \R_t \times \Gamma$,
where $\R_t$ is the $x^0$ ``time'' direction and $\Gamma$ is an arbitrary trivalent graph in the $(x^1,x^2)$-plane.
In fact, without breaking supersymmetry further, we can take the 4d space-time to be $M_4 = \R_t \times M_3$,
with an arbitrary 3-manifold $M_3$ locally parametrized by $(x^1,x^2,x^3)$, and a knotted trivalent graph $\Gamma \subset M_3$.
This requires partial topological twist of the 4d $\CN=4$ theory along $M_3$, which in the brane construction \eqref{M5N4}
is realized by replacing the space $\R^3$ parametrized by $(x^7,x^8,x^9)$ with the cotangent bundle to $M_3$.

Then, M5-branes are supported on $\R^3 \times M_3$, while M5$'$-branes are supported on $\R^3 \times L_{\Gamma}$,
where $M_3$ and $L_{\Gamma}$ are special Lagrangian submanifolds in a local Calabi-Yau space $T^* M_3$, such that
\be
L_{\Gamma} \cap M_3 = \Gamma
\label{LMGamma}
\ee
This configuration of fivebranes preserves 4 real supercharges, {\it i.e.} a quarter of the original $\CN=4$ SUSY.

When $C=T^2$ or $\R^2$, we can consider a further generalization in which junctions of surface operators
are not necessarily static.
Such configurations preserve only 2 real supercharges --- namely, $\CN=(0,2)$ supersymmetry along $C$ --- and
recently have been studied \cite{Gorsky:2013jxa,Gadde:2013sca} in a closely related context.
In particular, one can combine the time direction with $M_3$ into a general $4$-manifold $M_4$
and take $\Sigma$ to be an arbitrary surface \eqref{4dspacetime} with singular trivalent edges, \ie a foam or a seamed surface \cite{Rozansky:2003hz,MR2322554}.

For generic $\Sigma$, such foams or non-static junctions of surface operators will break supersymmetry
unless we extend the partial topological twist along $M_3$ to all of $M_4$.
In the fivebrane system \eqref{M5N4} this twist is realized by replacing $(x^7,x^8,x^9)$-directions
with a non-trivial bundle over $M_4$, namely the bundle of self-dual 2-forms:
\bea
{\mbox{\rm space-time:}} && \qquad  \Lambda^{2}_{+} (M_4) \times \R^4 \nonumber\\
{\mbox{\rm $N$ M5-branes:}} && \qquad  ~~~~~~M_4 \times \R^2 \label{5branesnonstatic} \\
{\mbox{\rm $k$ M5$'$-branes:}} && \qquad  ~~~~~~L_{\Sigma} \times \R^2  \nonumber
\eea
where, much as in \eqref{LMGamma}, $L_{\Sigma}$ and $M_4$ are coassociative submanifolds in
a local $G_2$-holonomy manifold $\Lambda^{2}_{+} (M_4)$, such that
\be
L_{\Sigma} \cap M_4 = \Sigma
\label{LMSigma}
\ee
As we already stated earlier, this generic configuration of surface operators preserves only
2 real supercharges (which, moreover, are chiral from the 2d perspective of $x^6$ and $x^{10}$).

\subsection{Line-changing operators in class $\CS$ and network cobordisms}
\label{sec:linechanging}

\begin{figure}[htb]
\centering
\includegraphics[width=5.0in]{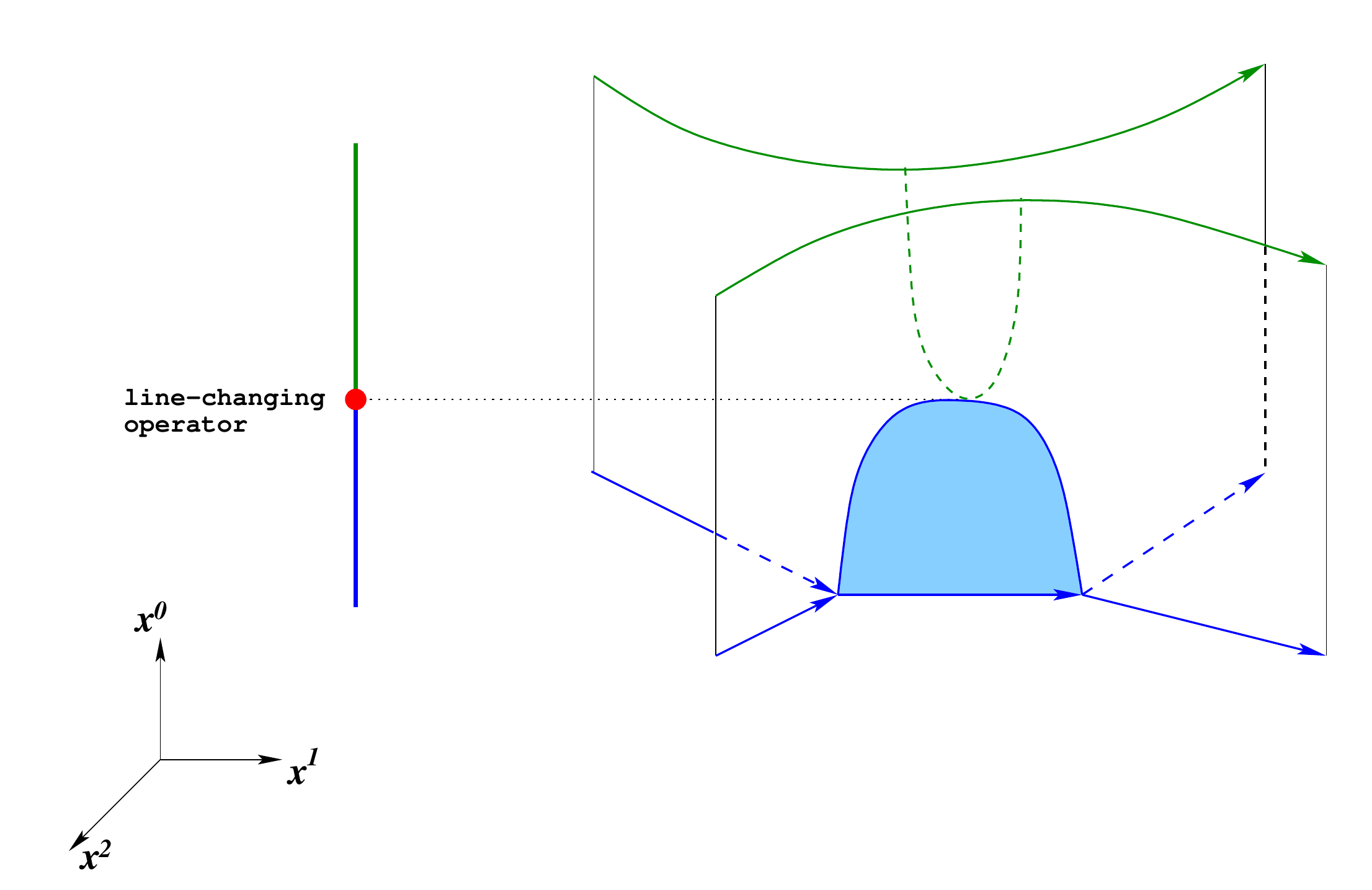}
\caption{6d fivebrane theory on a $2$-manifold $M_2$ parametrized by $x^1$ and $x^2$ yields 4d $\CN=2$ theory $T[M_2]$
in the remaining four dimensions. In this theory, a network of line defects on $M_2$ determines a BPS line operator,
whereas a ``time evolution'' of such network corresponds to segments of different line operators joined together
at those values of ``time'' $x^0$ where the topology of network changes.}
\label{fig:junction12}
\end{figure}

The intersection of M5-branes with M5$'$-branes produces the so-called codimension-2 defect in
the 6d $(0,2)$ theory on the fivebrane world-volume \eqref{M5branes}.
The latter also admits codimension-4 defects which too can be used to produce surface operators in
4d theory on $M_4$ and which in M-theory are realized by M2-branes ending on M5-branes:
\be
\begin{array}{l@{\;}|@{\;}ccccccccccc}
& ~0~ & ~1~ & ~2~ & ~3~ & ~4~ & ~5~ & ~6~ & ~7~ & ~8~ & ~9~ & ~10~ \\
\hline
\text{M5}~ & \times & \times & \times & \times &  &  & \times &  &  &  & \times \\
\text{M2}~ & \times & \times &  &  &  &  &  & \times &  &  &
\end{array}
\label{M5M2}
\ee
where the M2-brane world-volume is $\Sigma \times \R_+$, with $\R_+ = \{ x^7 \ge 0 \}$.
In this realization of surface 
operators, 
BPS junctions can be obtained by considering codimension-4 defect supported on $\Sigma = \R_t \times \Gamma$.
In fact, let $M_4 = \R_t \times M_2 \times \R$, where the $2$-manifold $M_2$ is a Riemann surface (possibly with punctures);
this requires a partial topological twist of the fivebrane theory which, similar to \eqref{5branesnonstatic},
is realized by embedding $M_2$ in the space-time $T^* M_2$.
Also, let $\Gamma \subset M_2$ be a colored trivalent graph on $M_2$ without self-intersections:
\be
\begin{matrix}
{\mbox{\rm 6d (0,2) theory:}} & \qquad & \R_t & \times & M_2 & \times & \R & \times & C  \\
& \qquad & \| &  & \cup \\
{\mbox{\rm surface operator:}} & \qquad & \R_t & \times & \Gamma && &&
\end{matrix}
\label{line-networks}
\ee
If we apply this to $C=\R^2$ (or $C=T^2$), as in section \ref{sec:N4}, it is natural to ask what this configuration
looks like from the viewpoint of $\R_t \times \R \times C \cong \R_t \times \R^3$, after compactification on the Riemann surface $M_2$.
Due to partial topological twist along $M_2$, the resulting 4d theory $T[M_2]$ has only $\CN=2$ supersymmetry
and the codimension-$4$ defect \eqref{line-networks} produces a line operator in this theory labeled by
a colored trivalent network $\Gamma \subset M_2$.

This is precisely the configuration recently used in \cite{Xie:2013lca,Xie:2013vfa,Bullimore:2013xsa,Coman:2015lna,Tachikawa:2015iba} to study line
operators in 4d $\CN=2$ theories $T[M_2]$.
In our application here, we simply wish to interpret the Riemann surface $M_2$ as part of the 4d space-time.
Then, the same configuration \eqref{line-networks} describes ``static'' surface operators in 4d $\CN=4$ theory on $M_4 = \R_t \times M_2 \times \R$.
Non-static surface operators correspond to networks $\Gamma (t)$ which vary with $t \in \R_t$; from the viewpoint
of 4d $\CN=2$ theory $T[M_2]$ they correspond to segments of different line operators glued together in a single line
via local operators supported at those points $t_i \in \R_t$ where the network $\Gamma (t)$ changes topology.
(Recall, that due to partial topological twist along $M_2$ only topology of the network $\Gamma$ matters.)

In other words, basic topological moves on $\Gamma$ realized via cobordisms correspond to line-changing operators
in 4d $\CN=2$ theory $T[M_2]$:
\be
\text{network cobordism} \qquad \Leftrightarrow \qquad \text{line-changing operator}
\ee
Turning on Omega-background in 4d $\CN=2$ theory $T[M_2]$ corresponds to replacing $C = \R^2$ with $C = \R^2_{\epsilon}$.
This is precisely the setup we will use in application to knot homologies and categorification of quantum groups, {\it cf.} \eqref{thetwothys}.

\subsection{Junctions in 4d $\CN=2$ theory}

When $C$ is an arbitrary Riemann surface (possibly with boundaries and punctures) of genus $g \ne 1$,
in order for the fivebrane configuration \eqref{M5branes} to preserve supersymmetry its world-volume
theory must be (partially) twisted along $C$.
For the embedding of the fivebrane world-volume in the ambient space-time the partial topological
twist means that $C$ must be either a holomorphic Lagrangian submanifold in a Calabi-Yau $2$-fold
(that locally, near $C$ always looks like $T^* C$) or a holomorphic curve in a Calabi-Yau $3$-fold.
The first choice preserves $\CN=2$ supersymmetry in the 4d theory on $M_4$, while the second option
preserves only $\CN=1$ SUSY and will be considered next.

Consistent with \eqref{M5N4}, our choice of coordinates is
\be
\label{xxxconventions}
\begin{array}{c|c|c}
M_4 & T^* C & \R^3 \\
\hline
~x^0,~x^1,~x^2,~x^3~ & ~x^4,~x^5,~x^6,~x^{10}~ & ~x^7, ~x^8,~x^9~ \\
\end{array}
\ee
In particular, the $U(1)_r \times SU(2)_R$ R-symmetry group in this setup is identified with
the $U(1)_{45} \times SU(2)_{789}$ symmetry of the eleven-dimensional geometry.
When both $M_4 = \R^4$ and $\Sigma = \R^2$ are flat, we can still use brane configuration \eqref{M5N4}
to describe half-BPS surface operators, this time in 4d $\CN=2$ theory on $M_4$.

Also, as in the $\CN=4$ case, we can describe junctions by taking $M_4 = \R_t \times M_3$
and replacing $(x^7,x^8,x^9)$ with the cotangent bundle to $M_3$:
\bea
{\mbox{\rm space-time:}} && \qquad  \R_t \times T^* M_3 \times  T^* C \nonumber\\
{\mbox{\rm $N$ M5-branes:}} && \qquad  \R_t \times ~M_3~ \times ~C \label{5branesstatic} \\
{\mbox{\rm $k$ M5$'$-branes:}} && \qquad  \R_t \times ~L_{\Gamma}~ \times  ~C  \nonumber
\eea
where, as in \eqref{LMGamma}, $L_{\Gamma}$ and $M_3$ are special Lagrangian submanifolds in $T^* M_3$
which intersect over a knotted trivalent graph $\Gamma \subset M_3$.
Then, the resulting brane configuration \eqref{5branesstatic} describes junctions of half-BPS surface
operators supported on a surface $\Sigma = \R_t \times \Gamma$.

Due to the partial topological twist along $M_3$, the latter can be an arbitrary $3$-manifold and $\Gamma$ can be
an arbitrary knotted graph. The price to pay for it is that junctions \eqref{5branesstatic} preserve only
$2$ real supercharges, {\it i.e.} again are $\frac{1}{4}$-BPS in the original 4d theory on $M_4$.

\subsection{Junctions in 4d $\CN=1$ theory}
\label{sec:N1}

Even though 4d $\CN=1$ theories are not directly related to the main subject of our paper,
here for completeness we consider possible junctions of surface operators in such theories.

First, in a typical brane construction \cite{Witten:1997ep} of 4d $\CN=1$ theories, $C$ is embedded as a holomorphic curve in
a local Calabi-Yau $3$-fold:
\be
M_4 \times CY_3 \times \R
\ee
This leaves little room for surface operators. In fact, even when $M_4 = \R^4$ this system admits
a half-BPS surface operator that breaks supersymmetry down to $\CN=(0,2)$ along $\Sigma = \R^2$.

In the fivebrane construction \eqref{M5branes} of the 4d $\CN=1$ theory, this half-BPS surface operator
can be realized by an additional system of M5$'$-branes supported on a holomorphic $4$-cycle $D \subset CY_3$:
\bea
{\mbox{\rm space-time:}} && \qquad  M_4 \times CY_3 \times  \R \nonumber\\
{\mbox{\rm $N$ M5-branes:}} && \qquad  M_4 \times ~C  \\
{\mbox{\rm M5$'$-branes:}} && \qquad  ~\Sigma~ \times ~D  \nonumber
\eea
Equivalently, via a ``brane creation'' phase transition \cite{Frenkel:2015rda}, one can represent half-BPS surface operators
in 4d $\CN=1$ theory by M2-branes with a semi-infinite extent in the $x^7$-direction ($x^7 \ge 0$):
\bea
{\mbox{\rm space-time:}} && \qquad  M_4 \times CY_3 \times  \R \nonumber\\
{\mbox{\rm $N$ M5-branes:}} && \qquad  M_4 \times ~C  \\
{\mbox{\rm M2-branes:}} && \qquad  ~\Sigma~ \times \{ \text{pt} \} \times \R_+  \nonumber
\eea
It is easy to verify that $\CN=1$ supersymmetry on $M_4$ is not sufficient to allow for non-trivial BPS junctions of surface operators.

\subsection{OPE of surface operators and the Horn problem}
\label{sec:Horn}

In non-abelian theory, the analogue of the flux conservation \eqref{afluxcons} is known as the multiplicative Horn problem:
\be
\text{What conjugacy classes are contained in}~ \CC' \cdot \CC'' ?
\label{Hornquestion}
\ee
Recall, that when a surface operator is described as a singularity (ramification) for the gauge field,
it is naturally labeled by the conjugacy class $\CC \subset G$ where the holonomy $U$ --- defined only modulo gauge
transformations, {\it i.e.} modulo conjugaction --- takes values. 

If away from $\Sigma = \R_t \times \Gamma$, that is away from $\Gamma \subset M_3$, the gauge field satisfies \eqref{flatA},
then the product of holonomies around edges of each vertex in the graph $\Gamma$ should be trivial.
This is a non-abelian version of the flux conservation \eqref{afluxcons} carried by surface operators,
which for a basic trivalent junction as in Figure~\ref{fig:junction13} takes the form
\be
U = U' \cdot U''\,.
\label{holprod}
\ee
The same condition describes junctions of non-abelian vortex strings. 
When written in terms of the respective conjugacy classes 
\be
\CC \subset \CC'\cdot\CC''
\ee
it indeed turns into the multiplicative Horn problem \eqref{Hornquestion}.

By reversing the orientation of the surface operator characterized by holonomy $U$ using \eq{holorient}, we can equivalently describe the more symmetric situation, in which all surface operators are outgoing.
This replaces $U\mapsto U^{-1}$ in \eq{holprod} and leads to
\be
1 \; \in \; \CC \cdot \CC' \cdot \CC''
\label{conjprod}
\ee
where $1 \in G$ is the identity. 

In gauge theory, \eqref{Hornquestion} and \eqref{conjprod} can be interpreted as operator product expansion (OPE) of surface operators.
Then, depending on the context, the ``OPE coefficient'' is either the moduli space $\CM_{0,3}^G (\CC,\CC',\CC'')$
of flat $G$-bundles on the punctures 2-sphere $S^2 \setminus \{ p,p',p'' \}$ with holonomies around the punctures lying
in $\CC$, $\CC'$, and $\CC''$, or its suitable cohomology or K-theory.
Indeed, this moduli space describes the space of gauge fields which satisfy \eqref{flatA} in the vicinity of
a trivalent junction, which we can surround with a small ball, as illustrated in Figure~\ref{fig:junction13}.
The boundary of such small ball is a trinion $S^2 \setminus \{ p,p',p'' \}$.
Since its fundamental group is generated by loops around the punctures,
with a single relation $abc = 1$,
\be
\pi_1 (S^2 \setminus \{ p,p',p'' \}) = \langle a,b,c \rangle / abc = 1\,,
\ee
the moduli space of its representations into $G$ is precisely the set of triples \eqref{holprod},
and each such triple determines a flat $G$-bundle on the trinion:
\be
\CM_{0,3}^G (\CC,\CC',\CC'') \; = \; \{ (U,U',U'') \in \CC \times \CC' \times \CC'' ~\vert~ U \cdot U' \cdot U'' =1 \} / G
\ee
In particular, this moduli space is non-empty if and only if \eqref{conjprod} holds,
\be
\CM_{0,3}^G (\CC,\CC',\CC'') \ne \emptyset \quad \Leftrightarrow \quad \CC \cdot \CC' \cdot \CC'' \ni 1
\label{mspaceconjprod}
\ee
This moduli space is a complex projective variety and is also symplectic.
For instance, when $G=SU(2)$ and the conjugacy classes satisfy \eqref{conjprod}
we have\footnote{The ``complex case'' of $G_{\C} = SL(2,\C)$ is even easier since eigenvalues of the holonomies
obey algebraic relations, analogous to the familiar A-polynomials for knots, see {\it e.g.} \cite{Gukov:2007ck}.}
\be
\CM_{0,3}^{SU(2)} (\CC,\CC',\CC'') \; = \; \cp^3 / \!\! / U(1)^3 \; = \; \text{point}\,.
\label{sutwomccc}
\ee
Note, that the problem considered here has an obvious analogue for a 2-sphere with an arbitrary number of punctures; it characterizes junctions where more than three surface operators meet along a singular edge of $\Sigma$. Since this generalization is rather straightforward, we shall focus on trivalent junctions, which are relevant in the context of the skew Howe duality.

Our next goal is to describe the selection rules of the ``OPE" of surface operators,  \ie to find solutions to the multiplicative Horn problem \eqref{Hornquestion} or its more symmetric form \eqref{mspaceconjprod}.

To do so, it is useful to parametrize conjugacy classes by the logarithm $\alpha$ of the eigenvalues of $U = \exp (2 \pi i \alpha)$,
that take values in the fundamental alcove of $G$,
\be
\alpha \; \in \; \frak{U}\,.
\label{Weylalcove}
\ee
For example, for $G=SU(N)$ parameters $\alpha$ take values in the simplex
\be
\frak{U} \; = \; \{ \alpha_1 \ge \alpha_2 \ge \ldots \ge \alpha_N \ge \alpha_1 -1 \vert \sum_i \alpha_i =0 \}\,.
\ee

Since in a theory with gauge group $G$ surface operators are parametrized by points in the Weyl alcove \eqref{Weylalcove} and, possibly,
other data, it is natural to study the image of \eqref{mspaceconjprod} in $\frak{U}^3$, which defines
a convex polytope of maximal dimension:
\be
\Delta_G \; := \; \{ (\alpha, \alpha', \alpha'') \in \frak{U}^3 :~ \CC \cdot \CC' \cdot \CC'' \ni 1 \}
\ee
In other words, the polyhedron $\Delta_G \subset \frak{U}^3$ describes the set of triples of conjugacy classes such that the moduli space \eqref{mspaceconjprod} is non-empty.
For instance, for $G=SU(2)$, $\frak{U}^3 = [0,\tfrac{1}{2}]^3$ is a cube and $\Delta_G$
is a regular tetrahedron inscribed in it \cite{MR1204322}, {\it cf.} \eqref{afluxcons}:
\be
|\alpha' - \alpha''| \; \le \; \alpha \; \le \; \min \{ \alpha' + \alpha'', 1 - \alpha' - \alpha'' \}
\label{tetrsutwo}
\ee
This tetrahedron is precisely the image of the moment map under $U(1)^3$ action on the ``master space'' $\cp^3$ in \eqref{sutwomccc}.

More generally, for $G = SU(N)$ the facets of $\Delta_G$ are defined by linear inequalities \cite{MR1671192,MR1856023}:
\be
\sum_{i \in I} \alpha_i + \sum_{j \in J} \alpha_j' + \sum_{k \in K} \alpha_k'' \le d
\label{SUNpolyhedron}
\ee
for each $d\geq 0$, $1\leq r \leq N$ and all triples $I$, $J$, $K$ of $r$-element subsets of $\{ 1, \ldots, N \}$, such that the 
degree-$d$ Gromov-Witten invariant of the Grassmannian $Gr(r,N)$ satisfies 
\be
GW_d (\sigma_I, \sigma_J, \sigma_K) = 1\,,
\label{SUNpolyGW}
\ee
where the {\it Schubert cycles} $\sigma_I$ are the cohomology classes associated to the {\it Schubert subvarieties}
\be
\CX_J= \{ W \in Gr (r,N) ~\vert~ \dim (W \cap F_{i_j}) \ge j, \; j =1, \ldots, r \}
\label{Schubertfirst}
\ee
for $I=\{i_1,\ldots,i_r\}$, 
which are defined 
with respect to a complete flag
\be
F_0 = \{ 0 \} \subset F_1 \subset \ldots \subset F_{N-1} \subset F_N = \C^N\,.
\label{completeflag}
\ee
The $\sigma_I$ form a basis of $H^* (Gr(r,N),\Z)$ and will be the subject of section \ref{sec:Schubert}.

A similar list of inequalities that defines $\Delta_{G}$ for any compact group $G$
was explicitly described by Teleman and Woodward \cite{MR2008438} and then further optimized in \cite{Ressayre,BelkaleKumar}.
The inequalities of Teleman and Woodward also have an elegant interpretation in terms of the small quantum cohomology,
except that instead of the Grassmannian they involve the flag variety $Fl = G_{\C} / \frak{P}$,
where $G_{\C}$ is the complexification of $G$ and $\frak{P}$ is a maximal parabolic subgroup.

For example, conjugacy classes of order-2 elements $U^2 = (U')^2 = (U'')^2 = 1 \in SU(N)$
such that $U \cdot U' \cdot U'' = 1$ have the moduli space
\be
\CM_{0,3}^{SU(N)} (\alpha,\alpha',\alpha'') = \text{point}
\ee
and therefore belong to the boundary of the polyhedron $\Delta_G$, $(\alpha,\alpha',\alpha'') \in \partial \Delta_G$.
The dimensions $i$, $j$, $k$ of their $-1$ eigenspaces satisfy the Clebsch-Gordan rules: $|i-j| \le k \le i+j$, {\it etc.}

Multiplying such order-2 conjugacy classes by $\zeta^k$,
where $k$ is the dimension of $(-1)$-eigenspace
and $\zeta = \exp (i \pi / N)$ is a primitive $N^{\text{th}}$ root of $-1$,
we obtain conjugacy classes that correspond to $R = \Lambda^k (\C^N)$ and play an important
role in MOY invariants of colored trivalent graphs and in the skew Howe duality:
\be
U_k \; = \; \text{diag} ( \, \underbrace{\zeta^k, \ldots, \zeta^k}_{N-k}, \, \underbrace{-\zeta^k, \ldots, -\zeta^k}_{k} \, )\,.
\label{UkslN}
\ee
Namely, as in section \ref{sec:CS}, let $\Gamma$ be a planar oriented trivalent graph,
whose edges are colored by elements in $\{ 1, 2, \ldots, N-1 \}$
with the signed sum (signs given by the orientation) of colorings around each vertex equal to 0.

With every such graph we can associate a configuration of surface operators supported on $\Sigma = \R_t \times \Gamma$,
such that an edge colored by $k$ is represented by a surface operator with holonomy in the conjugacy class of \eqref{UkslN}:
\be
\text{color}~k \quad \Leftrightarrow \quad \text{representation}~\Lambda^k (\C^N) \quad \Leftrightarrow \quad \text{holonomy}~U_k\,.
\ee
Note, this way of associating a particular type of surface operator to a colored edge of $\Gamma$
is consistent with the BPS equation \eqref{flatA} which, in turn, leads to flux conservation condition \eqref{conjprod}.
In particular, one can verify \eqref{SUNpolyhedron} using the parameters \eqref{Weylalcove}
for the conjugacy class of \eqref{UkslN}:
\be
\alpha \; = \; \Big( \, \underbrace{\frac{k}{2N}, \ldots, \frac{k}{2N}}_{N-k}, \, \underbrace{\frac{k-N}{2N}, \ldots, \frac{k-N}{2N}}_{k} \, \Big)\,.
\ee
In this case, the moduli space of solutions to the BPS equations with a ``foam'' of surface operators
on $\Sigma = \R_t \times \Gamma$ is the representation variety of the fundamental group of the graph complement into $SU(N)$
with meridional conditions on the edges of the graph
\be
\CM (\Gamma) \; = \; \text{Rep} \left( \pi_1 (S^3 \setminus \Gamma) ; SU(N) \right)\,.
\ee
Lobb and Zentner \cite{MR3190356} (see also \cite{MR3125899})
point out that the Poincar\'e polynomial of this space is closely
related to the MOY polynomial $P_N (\Gamma) \in \Z [q,q^{-1}]$ of the colored graph $\Gamma$.
In fact,
\be
\chi \left( \CM (\Gamma) \right) \; = \; P_N (\Gamma) \vert_{q=1}\,.
\label{euMP}
\ee
Moreover, the authors of \cite{MR3190356,MR3125899} provide a simple model for $\CM (\Gamma)$
by decorating each edge colored with $k$ by a point in the Grassmannian $Gr (k,N)$,
with the condition that at every vertex with the three edges colored by $k_1$, $k_2$, and $k=k_1+k_2$
the corresponding decorations consist of two orthogonal $k_1$- and $k_2$-planes in $\C^N$ and the $k$-plane that they span.
Such decorations are called {\it admissible} and the space of all admissible decorations
of the colored graph $\Gamma$ is homeomorphic to $\CM (\Gamma)$.
For example, the moduli space for the $k$-colored unknot is
\be
\CM \big( \unknot k \big) \; \cong \; Gr (k,N)\,,
\ee
in agreement with the fact that its cohomology indeed gives the colored homology of the unknot \eqref{Aunknot}
for $R = \Lambda^k (\C^N)$:
\be
\CA \; = \; \CH^{R} (\unknot) \; \cong \; H^* (Gr (k,N))\,.
\label{AHGr}
\ee
Also note that the moduli space associated to the trivalent junction with the edges colored by $k_1$, $k_2$, and $k=k_1+k_2$
is precisely the partial flag variety that appears in the correspondence \eqref{correspondence}.
In section \ref{sec:LG} we describe the corresponding interface in the product of Landau-Ginzburg models
with chiral rings $\CA = H^* (Gr (k,N))$, which provide an equivalent (and more user-friendly) description
of the topologically twisted theory. But before that, in the next section, we will briefly comment on how Schubert calculus can be employed to approach the ``OPE of surface operators".

\subsection{OPE and Schubert calculus}
\label{sec:Schubert}

Note, that in degree $d=0$, the condition \eq{SUNpolyGW} in the description \eqref{SUNpolyhedron} of the OPE selection rules corresponds to the cup product in 
the classical cohomology ring of the Grassmannian $Gr(r,N)$
\be
\sigma_{\lambda} \cdot \sigma_{\mu} \; = \; \sum_{\nu} c_{\lambda, \mu}^{\nu} \sigma_{\nu}\,,
\label{HGRclass}
\ee
where the structure constants $c_{\lambda, \mu}^{\nu}$ are the Littlewood-Richardson coefficients.
Here, our goal will be to explain \eqref{HGRclass} and its ``quantum deformation''
\be
\sigma_{\lambda} \star \sigma_{\mu} \; = \; \sum_{d \ge 0} \sum_{\nu} c_{\lambda, \mu}^{\nu, d} q^d \cdot \sigma_{\nu}
\label{QHGRquant}
\ee
that conveniently packages all Gromov-Witten invariants that appear in \eqref{SUNpolyhedron}, see \eg \cite{MR1454400}.
A careful reader will notice that, compared to \eqref{SUNpolyhedron}, we have labeled the cohomology classes differently here.
Instead of {\it Schubert symbols} $J = \{ j_1, \ldots, j_r \}$ (which are  sequences $1 \le j_1 < j_2 < \ldots j_r \le N$), we used partitions $\lambda$: Indeed, to each  Schubert symbol $J$, we can associate 
a partition
\be
\lambda_J = (\lambda_1 \ge \ldots \ge \lambda_r \ge 0)\,,\qquad
\lambda_i=N-r-j_i+i\,,
\ee
whose associated Young diagram fits into a rectangle of size $r \times (N-r)$, {\it i.e.} $\lambda_1 \le N-r$. 
Note, that such Young diagrams are in one-to-one correspondence with integrable highest weight representations
of $\hat{\mathfrak{su}}(N-r+1)_r$ (equivalently, of $\hat{\mathfrak{u}}(r)_{N-r+1}$).
Moreover, there is a partial ordering of the labels: $I \le J$ if $i_n\le j_n$ for all $n$, which expressed in partitions becomes $\lambda_I \subseteq \lambda_J$.

To explain this more slowly, let us recall a few basic facts about the geometry of $Gr(r,N)$,
the space of $r$-dimensional linear subspaces $V$ of a fixed $N$-dimensional 
complex
vector space $E$.
Thinking of these subspaces as 
spans of 
row vectors,  to each $V \in Gr (r,N)$ we can associate
a matrix $M \in GL_r \backslash \text{Mat}^*_{r,N}$, which leads to the Pl\"ucker embedding
\be
Gr (r,N) \; \subset \; \cp^{{N \choose r}-1}
\ee
and defines $Gr(r,N)$ as an algebraic variety, cut out by homogeneous polynomial equations.
For example, the Grassmannian $Gr (2,4)$, which we will use for illustrations, is defined by
a single equation in $\cp^5$:
\be
X_{12} X_{34} - X_{13} X_{24} + X_{14} X_{23} = 0
\ee

We are interested in the cohomology ring of $Gr(r,N)$ and its quantum deformation \eqref{QHGRquant}.
The classical cohomology $H^*(Gr(r,N))$ has total rank ${N \choose r}$ and is non-trivial
only in even degrees ranging from zero to the dimension of the Grassmannian,
\be
\dim_{\R} Gr (r,N) = 2 r(N-r)
\ee
In fact, the Grassmannian has a decomposition into a disjoint union 
\be
Gr (r,N) \; = \; \bigsqcup_{J} \CS_J
\ee
of {\it Schubert cells}, which gives it the structure of a CW complex. The Schubert cells can be defined with respect to a any fixed complete flag $E_1\subset E_2\subset\ldots\subset E_{N-1}\subset E_N$ in $E$ as follows:
\be
\CS_J=\{V\in Gr(r,N)\,|\, \dim(V\cap E_i)=\#\{a|j_a\leq i\}\,{\rm for\, all}\, 1\leq i\leq N\}.
\ee 
Their complex dimensions can be expressed as
\be
\dim \CS_J = \sum_n j_{n} - n = |\lambda_J|
\ee
where $|\lambda| = \lambda_1 + \ldots + \lambda_r$.
Thus, our favorite example $Gr (2,4)$ has six Schubert cells labeled by Schubert symbols
$J=(1,2)$, $(1,3)$, $(1,4)$, $(2,3)$, $(2,4)$, and $(3,4)$, of complex dimensions 0, 1, 2, 2, 3, and 4, respectively.

The closure $\bar \CS_J$ of a Schubert cell is a union of Schubert cells, {\it e.g.} $\bar \CS_{13} = \CS_{13} \cup \CS_{12}$ in $Gr (2,4)$.
This inclusion of Schubert cells defines the so-called Bruhat order which indeed is compatible with the partial order on the Schubert symbols mentioned above:
\be\label{partialorder}
I \le J \quad \Leftrightarrow \quad \CS_{I} \subseteq \bar \CS_{J}\,.
\ee
For $Gr (2,4)$ this partial order can be conveniently described by the following diagram
\be
\begin{matrix}
& (3,4) & \\
& (2,4) & \\
(2,3) &  & (1,4) \\
& (1,3) & \\
& (1,2) &
\end{matrix}
\label{Gr24Hasse}
\ee
A similar partial order on the set of conjugacy classes defined by closure
plays an important role in the gauge theory approach to the ramified case of the geometric Langlands correspondence \cite{Gukov:2008sn}
and the geometric construction of Harish-Chandra modules \cite{Gukov:2008ve}.

The {\it Schubert variety} $\CX_J$ is the Zariski closure of $\CS_J$:
\be
\CX_J := \bigsqcup_{I \le J} \CS_I
\ee
Clearly, $\dim \CX_J = |\lambda_J|$; and the Schubert cycles 
  $\sigma_{\lambda} \in H^{2 |\lambda|} (Gr (r,N))$ form an integral basis
for the cohomology ring of the Grassmannian,
\be
H^* (Gr (r,N); \Z) = \bigoplus_{\lambda \subseteq (N-r)^t} \Z \cdot \sigma_{\lambda}
\ee
For example, the Poincar\'e polynomial of $Gr (2,4)$ is ({\it cf.} \eqref{Gr24Hasse})
\be
P (Gr (2,4)) \; = \; 1 + t^2 + 2t^4 + t^6 + t^8
\ee

As alluded to above, in the ``classical'' Schubert calculus \eqref{HGRclass}, the non-negative integers
$c_{\lambda, \mu}^{\nu}$ are the Littlewood-Richardson coefficients; they vanish unless $|\lambda| + |\mu| = |\nu|$.
Note that, $\sigma_{(0)} \equiv \sigma_{(0, \ldots, 0)}$ associated to the partition $\lambda = (0, \ldots, 0)$ is the unit in the cohomology ring of the Grassmannian.
The classical cohomology ring has two important generalizations (deformations), one of which we already mentioned earlier.
Namely, much like its classical counterpart, the (small) quantum cohomology of the Grassmannian $Gr(r,N)$
has a $\Z [q]$-basis formed by Schubert classes $\{ \sigma_{\lambda} \}$ labeled by partitions
$\lambda = (\lambda_1, \ldots, \lambda_r)$ that fit into a rectangle of size $r \times (N-r)$,
{\it i.e.} $N-r \ge \lambda_1 \ge \lambda_2 \ge \ldots \ge \lambda_r\ge 0$.
The structure constants $c_{\lambda, \mu}^{\nu, d}$ of the quantum multiplication \eqref{QHGRquant}
are often called {\it quantum Littlewood-Richardson coefficients}. They vanish,
$c_{\lambda, \mu}^{\nu, d} = 0$, unless $|\lambda| + |\mu| = |\nu| + Nd$,
which means that $q$ carries homological degree $2N$.

For example, the quantum cohomology ring $QH^* (Gr(2,4))$ can be described by the following relations:
\be
\begin{array}{lll}
\sigma_{(1,0)} \star \sigma_{(1,0)} = \sigma_{(2,0)} + \sigma_{(1,1)}
& \quad &
\sigma_{(1,0)} \star \sigma_{(2,0)} = \sigma_{(2,1)} \\
\sigma_{(1,0)} \star \sigma_{(1,1)} = \sigma_{(2,1)}
& \quad &
\sigma_{(1,0)} \star \sigma_{(2,1)} = \sigma_{(2,2)} + q \\
\sigma_{(1,0)} \star \sigma_{(2,2)} = q \sigma_{(1,0)}
& \quad &
\sigma_{(2,0)} \star \sigma_{(2,0)} = \sigma_{(2,2)} \\
\sigma_{(2,0)} \star \sigma_{(1,1)} = q
& \quad &
\sigma_{(2,0)} \star \sigma_{(2,1)} = q \sigma_{(1,0)} \\
\sigma_{(2,0)} \star \sigma_{(2,2)} = q \sigma_{(1,1)}
& \quad &
\sigma_{(1,1)} \star \sigma_{(1,1)} = \sigma_{(2,2)} \\
\sigma_{(1,1)} \star \sigma_{(2,1)} = q \sigma_{(1,0)}
& \quad &
\sigma_{(1,1)} \star \sigma_{(2,2)} = q \sigma_{(2,0)} \\
\sigma_{(2,1)} \star \sigma_{(2,1)} = q \sigma_{(2,0)} + q \sigma_{(1,1)}
& \quad &
\sigma_{(2,1)} \star \sigma_{(2,2)} = q \sigma_{(2,1)} \\
\sigma_{(2,2)} \star \sigma_{(2,2)} = q^2
& \quad &
\end{array}
\ee
Setting $q=0$ we obtain the ordinary cohomology ring \eqref{HGRclass},
while specializing to $q=1$ we get the Verlinde algebra \cite{Witten:1993xi}
(namely, the algebra of Wilson loops in $U(r)$ Chern-Simons theory at level $N-r$ in our conventions).
A categorification of this algebra was recently constructed in \cite{Gukov:2015sna}.

\begin{figure}[htb] \centering
\includegraphics[width=4.0in]{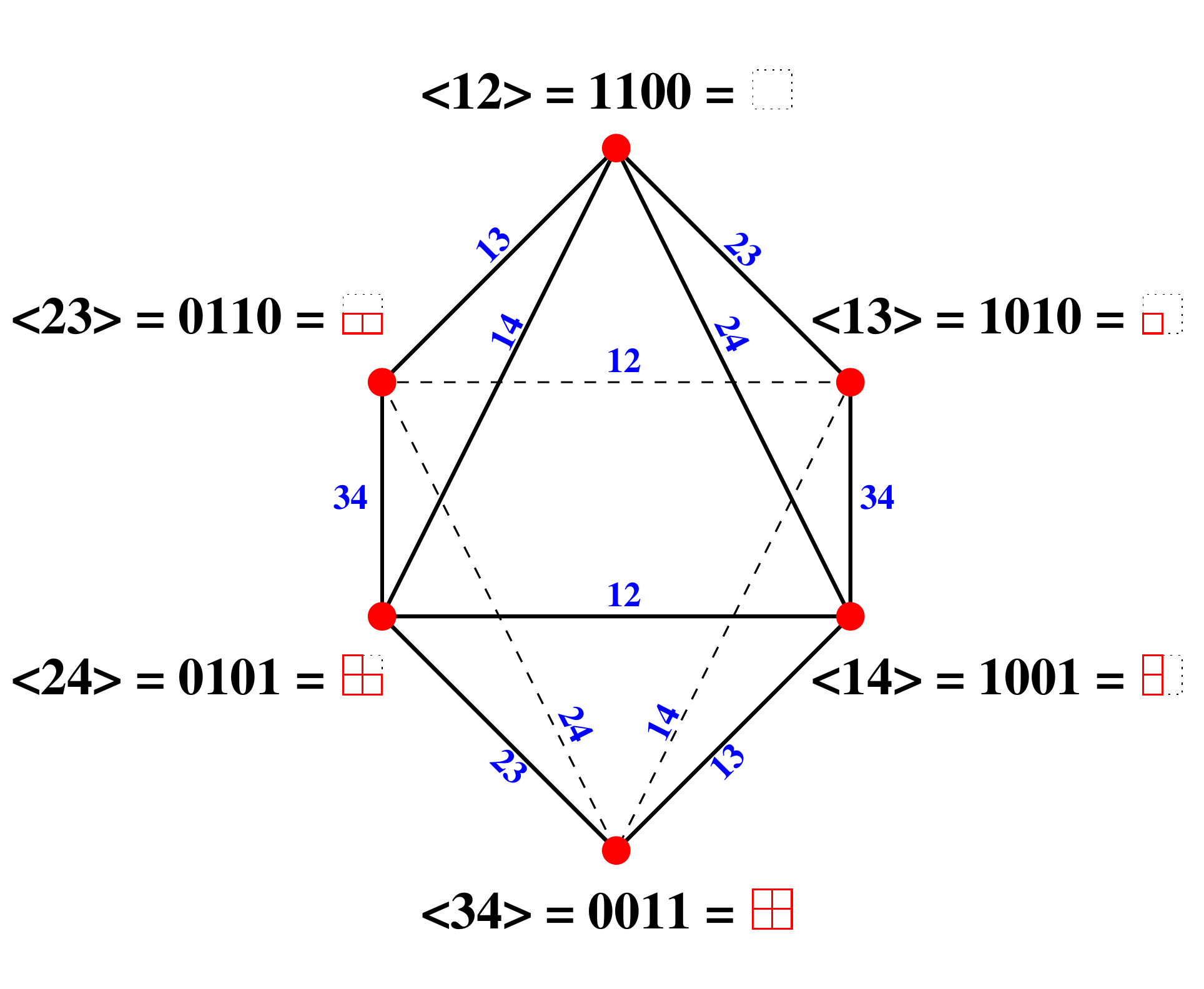}
\caption{\label{fig:junction2}
The Schubert cells of the Grassmannian (as well as vacua of 4d $\CN=2$ SQCD) can be labeled by Young diagrams
that fit into $r \times (N-r)$ rectangle, or, equivalently, by a string that contains $r$ ones and $N-r$ zeros,
or, yet another way, by length-$r$ subsets $I \subset \{ 1, \ldots, N \}$.
Shown here is the example with $r=2$ and $N=4$.}
\end{figure}

The second important deformation of the classical Schubert calculus \eqref{HGRclass}
is based on the fact that $Gr(r,N)$ admits a torus action of $T=U(1)^N$.
(Note that since the diagonal $U(1)$ acts trivially, we effectively have an action of $U(1)^{N-1}$.)
There is a fixed point for each subset $J = \{ j_1 < \ldots < j_r \} \subset \{ 1, \ldots, N \}$, so that
\be\label{Tequiv}
\dim H^* (Gr(r,N)^T) = \dim H^* (Gr(r,N)) = {N \choose r}\,.
\ee
This is the general property of spaces with vanishing odd cohomology: all such spaces are equivariantly formal.
In general, the $T$-equivariant cohomology of such a space $X$ is conveniently described by the {\it moment graph} (a.k.a. the Johnson graph),
whose vertices are in bijection with $X^T$, edges correspond to one-dimensional orbits, {\it etc.}
For example, the moment graph of the Grassmannian $Gr(2,4)$ is illustrated in Figure~\ref{fig:junction2},
whereas for $r=1$ and $N=4$ we get precisely the regular tetrahedron $\Delta_{SU(2)}$
obtained in \eqref{tetrsutwo} as the image of the moment map under the $U(1)^3$ action on $\cp^3$.
In terms of the moment graph, the $T$-equivariant cohomology $H_T^* (X)$ has a simple explicit description as the ring
\be
(f_x) \in \bigoplus_{x \in \text{vertices}} H_T^* (\text{pt}) ~\text{s.t.}~~ {\chi_E \text{ divides } f_x - f_y  \text{ whenever} \atop x \text{ and } y \text{ lie in a common edge } E }
\ee
where
\be
H_T^* (\text{pt}) = \bigotimes_{i=1}^N \C[m_i] = \C [m_1, \ldots, m_N]\,,
\label{equivring}
\ee
and $\chi_E=m_i-m_j$ if the vertices of the edge $E$ correspond to Schubert symbols $I=(I\cap J)\cup\{i\}$ and $J=(I \cap J)\cup\{j\}$.
%

The quantum and equivariant deformations of the cohomology ring \eqref{HGRclass} can be combined into the
$T$-equivariant quantum cohomology of $Gr(r,N)$ which has multiplication of the form \eqref{QHGRquant},
where the {\it equivariant quantum Littlewood-Richardson coefficients} $c_{\lambda, \mu}^{\nu, d}$
are homogeneous polynomials in the ring \eqref{equivring} of degree $|\lambda| + |\mu| - |\nu| - Nd$.
The explicit form of these polynomials for $Gr(2,4)$ can be found in \cite[sec.~8]{MR2231042}.
In particular, $c_{\lambda, \mu}^{\nu, d}$ is equal to the quantum Littlewood-Richardson coefficient
when $|\lambda| + |\mu| = |\nu| + Nd$, while for $d=0$ it is equal to the ordinary
equivariant Littlewood-Richardson coefficient $c_{\lambda, \mu}^{\nu}$.
The remarkable fact about the equivariant quantum Littlewood-Richardson coefficients
is that they are homogeneous polynomials in variables $m_1 - m_2$, $\ldots$, $m_{N-1} - m_N$
with positive coefficients. Hence, they can be categorified!

\subsection*{Domain walls in 4d $\CN=2$ SQCD}

The geometry of Schubert varieties described here has a simple interpretation in terms of
moduli spaces of domain walls in 4d $\CN=2$ super-QCD with $U(r)$ gauge group and $N_f=N$ flavors
in the fundamental representation of the gauge group, see \eg \cite{Eto:2005mx,Eto:2006mz,Eto:2008dm}.
Since this physical manifestation is somewhat tangential to the subject of the present paper
we relegate it to Appendix \ref{app:domainwalls}.


\section{Categorification and the Landau-Ginzburg perspective}
\label{sec:LG}

As we already explained in section \ref{sec:whysurf}, surface operators supported on knot and graph cobordisms
provide a categorification of the corresponding knot/graph invariants, thus making the results of section \ref{sec:surface} directly relevant.
Moreover, the choice of the Riemann surface $C$ in section \ref{sec:surface} determines the type of knot homology that one finds.
Thus, a compact Riemann surface $C$ typically leads to a singly-graded homology theory (see \cite{Gukov:2007ck} for concrete examples).
In this paper, we are mostly interested in applications to doubly-graded knot homologies categorifying quantum group invariants;
in the language of section \ref{sec:surface},
they correspond to a somewhat peculiar choice $C \cong \R^2_{\epsilon}$ (= ``cigar'' in the Taub-NUT space)
that we already encountered in section \ref{sec:linechanging} and will describe in more detail below.

In particular, we describe junctions of surface operators from the perspective of the 2d TQFT on $\Sigma$.
We restrict the discussion to surface operators associated with antisymmetric tensor products
of the fundamental representation of $\mathfrak{sl}_N$.
Generalization of the discussion below to $\mathfrak{sl}_{N|M}$ and to symmetric representations of $\mathfrak{sl}_N$
should provide categorification of the super $q$-Howe duality 
%
\cite{QS15}
and symmetric $q$-Howe duality \cite{SymmetricHowe}, respectively.
We leave these generalizations to future work.


\subsection{Physics perspectives on categorification}
\label{sec:bpssummary}

Soon after the advent of the first homological knot invariants \cite{Khovanov,OShfk,RasmussenHFK}
it was proposed \cite{Gukov:2004hz,Gukov:2007ck}
that, in general, knot homology should admit a physical interpretation as a $Q$-cohomology of 
a
suitable physical system,
\be
\text{knot homology} \; = \; Q\text{-cohomology} \; \equiv \; \CH_{\text{BPS}}
\ee
where $Q$ is a supercharge.
Since then, many concrete realizations of this general scenario have been proposed for different knot homologies.

In most of the physics approaches to knot homologies, the starting point is one of the two closely related fivebrane systems:
\be
\begin{array}{rcl}
\multicolumn{3}{c}{~~\text{doubly-graded}~} \\[.1cm]
\hline
{\mbox{\rm space-time:}} && ~~  \R_t \times T^* S^3 \times TN_4 \\
{\mbox{\rm $N$ M5-branes:}} && ~~  \R_t \times ~S^3~ \times ~\R^2_\epsilon \\
{\mbox{\rm $k$ M5$'$-branes:}} && ~~  \R_t \times ~L_{\Gamma}~ \times  ~\R^2_\epsilon
\end{array}
\quad \xleftarrow[~\text{transition}]{~\text{phase}} \!\!\! \to \quad
\begin{array}{rcl}
\multicolumn{3}{c}{~~~\text{triply-graded}~~} \\[.1cm]
\hline
&& \R_t \times CY_3 \times  TN_4 \\
&& \R_t \times ~L_{\Gamma} \times  \R^2_\epsilon
\end{array}
\label{thetwothys}
\ee
which is precisely a variant of \eqref{5branesstatic}, with $M_3 = S^3$ and $C = \R_{\epsilon}^2$.
(Other variants, with $C=S^2$ and $C=T^2$, have been studied in \cite{Gukov:2007ck}.)
Keeping track of the $U(1) \times U(1)$ quantum numbers associated with the rotation symmetry of $C=\R^2_\epsilon$ 
and its normal bundle in the Taub-NUT space where it is naturally embedded, $TN_4 \cong T^* C$,
leads to two gradings, namely the $q$-grading and the homological $t$-grading.

Much like the familiar phases of water, the two fivebrane configurations in \eqref{thetwothys}
are conjecturally related by a phase transition \cite{GV-I,GV-II}
and describe the same physical system in different regimes of parameters, one of which has the fixed rank $N$
and hence is more suited for doubly-graded knot homologies, while the other has an additional, third grading on the space
of refined BPS states (= $Q$-cohomology) and provides a physical realization of colored HOMFLY-PT homologies.
Studying the two fivebrane systems \eqref{thetwothys} from different vantage points led to different physical
perspectives on doubly-graded and triply-graded knot homologies:

\begin{itemize}

\item
{\bf Enumerative invariants:} Looking at the system \eqref{thetwothys} from the vantage point of the Calabi-Yau space $CY_3$ leads to the original 
%
physical
description \cite{Gukov:2004hz} in terms of enumerative invariants.

\item
{\bf Refined Chern-Simons:} Looking from the vantage point of $M_3 = S^3$ on the doubly-graded side of \eqref{thetwothys} leads
to $\mathfrak{sl}_N$ double affine Hecke algebra (DAHA) and refined Chern-Simons theory
%
 \cite{AS,IK11,Cherednik,Nakajima:2012gx,Samuelson:2014}.

\item
{\bf 5d gauge theory:} Also on the doubly-graded side, analyzing the system \eqref{thetwothys} from the vantage point
of the world-volume theory on $N$ M5-branes leads to a formulation of knot homology via counting solutions
to the Haydys-Witten equations in five dimensions \cite{Witten:2011zz,Haydys:2010dv}.

\item
{\bf Landau-Ginzburg} model is a two-dimensional theory that ``lives'' on $(\text{time}) \times (\text{knot})$ or, to be more precise,
on the cylinder $\R_t \times S^1_{\sigma}$, where $S^1_{\sigma} \subset S^3$ is the great circle and $K \to S^1_{\sigma}$
is a multiple cover
%
\cite{Rozansky:2003hz, MR2322554, Gukov:2005qp,Carqueville:2011zea}.
 In this description of doubly-graded $\mathfrak{sl}_N$ homology, each braiding (crossing) in a two-dimensional projection of a link $K$ is represented by the corresponding line defect $\mathcal{D}_i$ in the two-dimensional Landau-Ginzburg model, as illustrated in Figure \ref{fig:LGdefects1}. 

\item
{\bf 3d $\CN=2$ theory} labeled by the knot $K$ that ``lives'' on a $3$-dimensional part, $\R_t \times \R^2_\epsilon$,
of the fivebrane world-volume provides yet another way to compute the $Q$-cohomology on either side of \eqref{thetwothys}.
This approach is a certain variant of the so-called ``3d-3d correspondence'' \cite{Fuji:2012pi,Chung:2014qpa}.

\item
{\bf Vortex counting:} Compactification of M-theory on the Calabi-Yau $3$-fold $CY_3$
is known to engineer 5d $\CN=2$ gauge theory \cite{engineering},
in which fivebranes \eqref{thetwothys} produce a codimension-$2$ defect, a ``surface operator''. Thus, from the vantage point
of 5d gauge theory on $\R_t \times TN_4$ with a ramification (determined by the knot $K$) along $\R_t \times \R^2_\epsilon$,
the problem of counting BPS states can be formulated in terms of K-theoretic instanton-vortex counting \cite{DGH,Gorsky:2013jxa}
that involves Hilbert schemes of points, {\it etc.}

\end{itemize}

\noindent
These physical perspectives stimulated development of many new structures in knot homologies,
including the formulation of triply-graded homology categorifying the HOMFLY-PT polynomial \cite{KR2},
which came as a bit of surprise \cite{MR2100691},
led to many new differentials ({\it canceling}, {\it universal}, {\it colored}, {\it exceptional}),
to new connections with knot contact homology \cite{Aganagic:2013jpa},
recursion relations with respect to color-dependence \cite{Fuji:2013rra}, {\it etc.}

The setup \eqref{thetwothys} has a natural extension to knot cobordisms
(on both sides, although the two extensions are not obviously related by a phase transition), {\it cf.} \eqref{5branesnonstatic}:
\be
\begin{array}{rcl}
\multicolumn{3}{c}{~~\text{doubly-graded}~} \\[.1cm]
\hline
{\mbox{\rm space-time:}} && ~~  \Lambda^{2}_{+} (M_4) \times TN_4 \\
{\mbox{\rm $N$ M5-branes:}} && ~~  ~M_4~ \times ~R_\epsilon^2 \\
{\mbox{\rm M5$'$-branes:}} && ~~  ~L_{\Sigma}~ \times  ~R_\epsilon^2
\end{array}
\quad \qquad \quad
\begin{array}{rcl}
\multicolumn{3}{c}{~~~\text{triply-graded}~~} \\[.1cm]
\hline
{\mbox{\rm space-time:}} &&~~ X_{G2} \times  TN_4 \\
{\mbox{\rm M5$'$-branes:}} && ~L_{\Sigma} \times  R_\epsilon^2
\end{array}
\label{thetwocobord}
\ee
where the cobordism $\Sigma$ combines the ``time'' direction and the direction along the knot
into a non-trivial 2-manifold, $X_{G2}$ is a 7-manifold with $G_2$ holonomy (= $\R_t \times CY_3$ in simple cases),
and $L_{\Sigma}$ is a coassociative submanifold described in \eqref{LMSigma}.

A variant of the transition \cite{Frenkel:2015rda} relates the doubly-graded side of \eqref{thetwothys} and \eqref{thetwocobord}
with surface operators produced from codimension-2 defects on the fivebrane world-volume to similar brane configurations
with M2-branes (codimension-4 surface operators). For instance, a suitable analogue of \eqref{thetwocobord} is
\bea
{\mbox{\rm space-time:}} \qquad  \Lambda^{2}_{+} (M_4) & \times & TN_4 \nonumber\\
{\mbox{\rm $N$ M5-branes:}} \qquad  ~~~M_4~~~ & \times & R_\epsilon^2\label{2branesnonstatic} \\
{\mbox{\rm M2-branes:}} \qquad  \Sigma \; \times \!\!\!\!\! \tilde{~}~ \; \R_+ & \times & ~\{ 0 \}  \nonumber
\eea
where the twisted product $\Sigma \; \times \!\!\!\!\! \tilde{~}~ \; \R_+$ is an associative submanifold in $\Lambda^{2}_{+} (M_4)$.
In the special case $M_4 = \R_t \times S^3$ and $\Sigma = \R_t \times \Gamma$ we find $\Lambda^{2}_{+} (M_4) = \R_t \times T^* S^3$
and $\Sigma \; \times \!\!\!\! \tilde{~}~ \; \R_+ \cong \R_t \times \Gamma \times \R_+$.

Of particular interest is the cobordism representing the connected sum with the unknot \eqref{HHHunknot}
which can fully characterize the 2d ``effective theory'' that lives on patches of a general foam $\Sigma$.


\subsection{LG theory on ``time $\times$ knot''}
\label{sec:LGlocal}

Our next goal is to describe the effective 2d theory that ``lives'' on the world-sheet $\Sigma$
of surface operators which appear in categorification of quantum group invariants.

\FIGURE[thb]{\includegraphics[width=5.3cm]{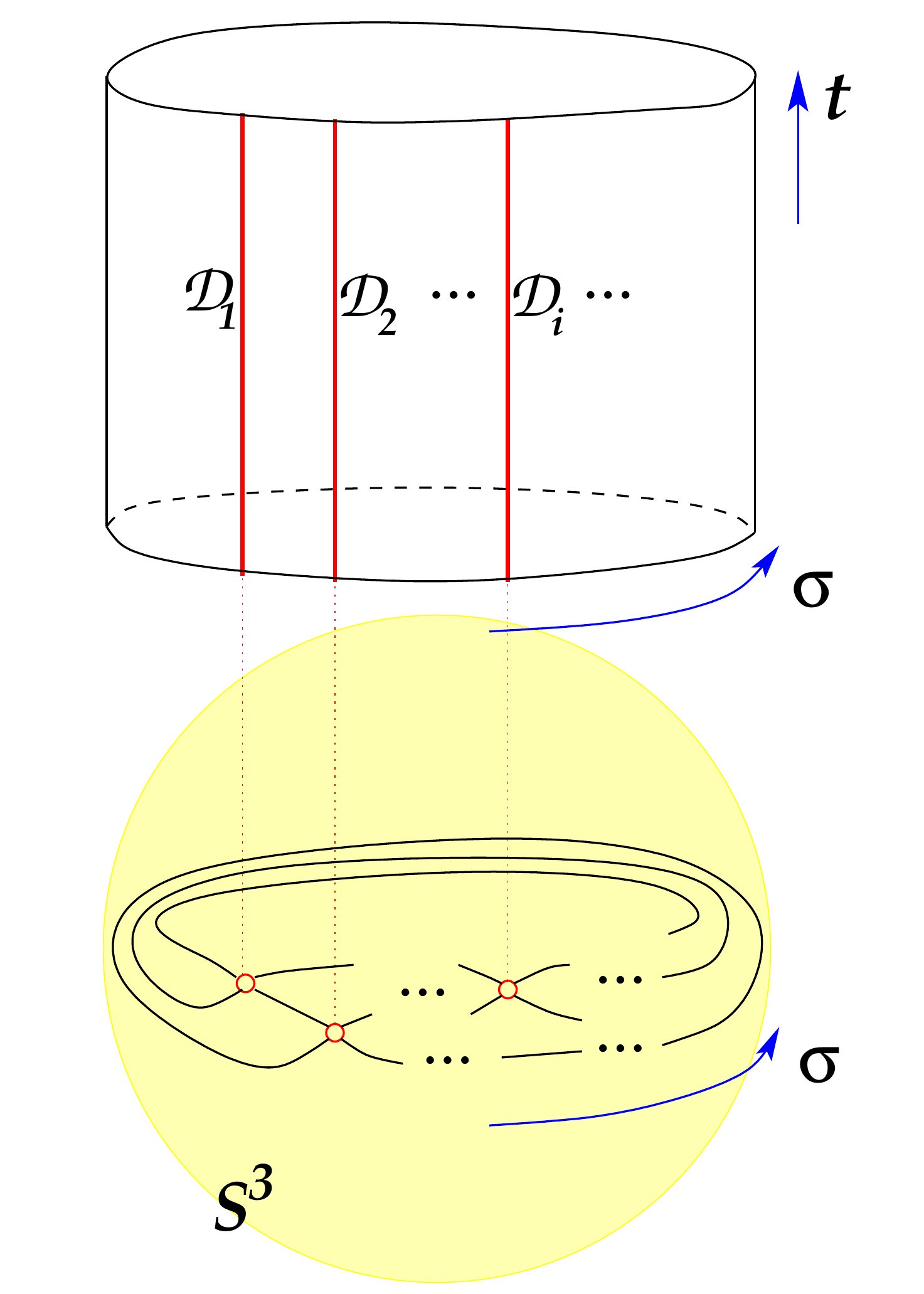}
\caption{\label{fig:LGdefects1}
A knot or a link $K$ can be represented as a closure of a braid
that runs along one of the directions in $M_3$ ($=S^3$ here).
Then, crossings or braidings (denoted by ``$\circ$'') of a surface operator in extra dimensions of $M_3$
``project'' to interfaces in 2d Landau-Ginzburg model on $\R_t \times S^1_{\sigma}$.}}
The first clue comes from the early construction of knot homology groups \cite{KR1,KR2}
based on matrix factorizations.
Since matrix factorizations are known to describe boundary conditions and interfaces
in 2d Landau-Ginzburg models \cite{Brunner:2003dc,Kapustin:2003ga,Hori:2004ja,Brunner:2007qu,Brunner:2007ur,Brunner:2008fa}
it was proposed in \cite{Gukov:2005qp,Carqueville:2011zea,Gukov:2011ry} to interpret the matrix factorizations
used by Khovanov and Rozansky as interfaces in 2d Landau-Ginzburg model on what we call $\Sigma$,
{\it i.e.} on the product of the time direction $\R_t$ and the strands of a knot, or link, or tangle, as illustrated in Figure \ref{fig:LGdefects1}.

There are several ways to show this, including dualities that map intersecting branes in M-theory to intersecting D-branes
which allow to address the question of what degrees of freedom live on $\Sigma$ via the standard
tools of perturbative string theory.
The upshot of this exercise is that for each strand in the braid or link $K$ colored by $R$
one finds a 2d theory whose chiral ring agrees with the homology of the unknot \eqref{Aunknot}
colored by a representation $R$ of Lie algebra $\frak g$.

Using this physical interpretation, not only can one reproduce the superpotential $W_{{\frak g}, R} (x)$
for $\frak g = \mathfrak{sl}_N$ and $R = \Box$ used by Khovanov and Rozansky but also produce new Landau-Ginzburg
potentials for more general representations.
This includes all symmetric and anti-symmetric representations of $\frak g = \mathfrak{sl}_N$,
certain representations of exceptional Lie algebras and $\frak g = \mathfrak{so}_N$ \cite{Gukov:2005qp},
which later were used in mathematical constructions of the corresponding knot homologies,
see {\it e.g.} \cite{MR2386537,Yonezawa,Wu,Elliot:2015pra}.

In all of these constructions, the category \eqref{unknotcategory} is a category of matrix factorizations $\text{MF} (W)$
of a polynomial
\be
W = W_{R_1} + \ldots + W_{R_n}
\ee
that is a sum of potentials $W_{R_i}$, one for each marked point (or strand).
Correspondingly, the category
\be
\frak{C}_{R_1, \ldots, R_n} \; = \; \text{MF} (W) \; = \; \text{MF} (W_{R_1}) \otimes \ldots \otimes \text{MF} (W_{R_n})
\ee
is a product of categories $\text{MF} (W_{R_i})$ associated to individual strands.
As discussed in section \ref{sec:whysurf}, the Hochschild homology of the $\text{MF} (W_{R})$ is supposed to agree with the homology of the $R$-colored unknot,
\be
\CA \; = \; HH^* \big( \text{MF} (W_{R}) \big) \; = \; \CH^{R} \big( \unknot \big)
\label{AMFunknot}
\ee
if the knot homology at hand extends to a functor \eqref{categorlore}, \cf \eq{Hochschunknot}. 
The Hochschild homology can be computed as the 
the homology of the Koszul complex associated with the sequence of partial
derivatives of $W_R$. It contains the Jacobi ring $\CJ (W_R)$ and is in fact equal to it if and only if $W_R$ has only isolated singularities. Indeed, it is this case, which will be relevant for us. Hence, we get a non-trivial constraint on the superpotential $W_R$, namely that its Jacobi ring, which is also the chiral ring of the Landau-Ginzburg model associated to an $R$-colored point, has to agree with the respective knot homology. This identification has to hold on the level of algebras. (Recall that the knot homology carries an algebra structure containing information about knot cobordisms, \cf the discussion around \eqref{unknotAAA}.)

Beyond this, also deformations of Khovanov-Rozansky homology groups categorifying quantum $\mathfrak{sl}_N$ invariants \cite{MR2173845,MR2174270,Gornik,Dunfield:2005si,MR2336253,Rose:2015pla} must be correctly incorporated in the structure of the category \eqref{unknotcategory} associated to marked points. Indeed, this is not entirely unrelated from the algebra structure: differentials in spectral sequences that relate different variants of knot homology are often represented by
generators of the algebra \eqref{Aunknot}.

In the Landau-Ginzburg setup, such deformations should be realized by relevant deformations of the Landau-Ginzburg potential $W_R$, providing further consistency checks for the proposed Landau-Ginzburg description. Moreover, also the physically motivated generalizations of the original Khovanov-Rozansky construction to other representations mentioned above 
give further credence to the Landau-Ginzburg approach.

In addition, parallel developments in the study of surface operators led to a number of alternative
descriptions of the chiral ring \eqref{AMFunknot} which, in all cases of our interest, agrees with
the chiral ring of the Landau-Ginzburg model with the superpotential $W_R$.
Indeed, as we reviewed in section \ref{sec:surface}, the same brane configurations \eqref{thetwocobord}--\eqref{2branesnonstatic}
that we use for categorification of quantum group invariants describe codimension-$2$ and codimension-$4$ defects in 6d theory on M5-branes. Thus, one can use the description of surface operators as coupled 2d-4d systems to describe the effective 2d theory on $\Sigma$.

Because the question about the chiral ring $\CA$ of the 2d theory on $\Sigma$ is completely local
(in a sense that it does not depend on the geometry away from $\Sigma$ as well as position along $\Sigma$),
we can take $M_4 = \R^4$ and $\Sigma = \R^2$.
Then, the question basically reduces to the study of the chiral ring in 2d theory on the surface operator labeled by $R$
in 4d $\CN=4$ gauge theory on $M_4$ that we already analyzed in section \ref{sec:surface}.
In this paper, we are mainly interested in the case of the gauge group $G = SU(N)$ and
its $k$-th anti-symmetric representation $R = \Lambda^k \C^N$.
The corresponding surface operators have Levy type $\mathbb{L} = SU(U(k) \times U(N-k))$
and in the description as 2d-4d coupled systems {\it \`a la} \eqref{S2d4d},
the 2d theory on $\Sigma$ is a $\CN=(4,4)$ sigma-model with hyper-K\"ahler target space $T^* (G/\mathbb{L}) = T^* Gr(k,N)$.
Topological twist of the ambient theory on $M_4$ induces a topological twist of the 2d theory on $\Sigma$,
which then becomes a 2d TQFT with the chiral ring
\be
\CA \; = \; H^* (Gr(k,N))
\ee
given by the {\it classical} cohomology of the Grassmannian, in agreement with \eqref{unknotchiralA} and \eqref{AHGr}.
A 2d TQFT with the same chiral ring can be obtained by a topological B-twist of
the ${\mathcal N}=(2,2)$ Landau-Ginzburg model with $k$ chiral superfields $x_1,\ldots,x_k$
of $U(1)_R$-charge $q={2\over N+1}$ and the superpotential
\be\label{superpotential}
W_0(x_1,\ldots,x_k)=x_1^{N+1}+\ldots+x_k^{N+1}\,.
\ee
More precisely, a change of variables from the $x_i$ to the elementary symmetric
polynomials\footnote{$\sigma_j(x_1,\ldots,x_k)=\sum_{1\leq i_1<\ldots< i_k\leq k}x_{i_1}\ldots x_{i_j}$}
$X_i=\sigma_i(x_1,\ldots,x_k)$ (see e.g.~section 8.3 of \cite{Cecotti:1992rm}):
\be\label{symmetrization}
(x_1,\ldots,x_k)\longmapsto \left(X_1=\sigma_1(x_1,\ldots,x_k),\ldots,X_k=\sigma_k(x_1,\ldots,x_k)\right)
\ee
gives rise to a new Landau-Ginzburg model denoted by $LG_k$. Its chiral superfields $X_i$ have
$U(1)_R$-charge $q_i={2i\over N+1}$, and its superpotential $W=W(X_1,\ldots,X_k)$ is just $W_0$ expressed in terms of the $X_i$, \ie $W(\sigma_1(x_1,\ldots,x_k),\ldots,\sigma_k(x_1,\ldots,x_k))=W_0(x_1,\ldots,x_k)$.
It is still quasi-homogeneous. In the IR, $LG_k$ flows to a superconformal field theory of central charge\footnote{The central charge associated to LG-models can be obtained as $c=3\sum_i(1-q_i)$.}
\be
c={3k(N-k)\over N+1}\,,
\ee
which is believed to be the level-$1$ Kazama-Suzuki model associated to the Grassmannian $Gr(k,N)$.
The chiral ring of $LG_{k}$ is the Jacobi ring of $W(X_1,\ldots,X_k)$ and indeed agrees with 
the classical cohomology ring  $H^*(Gr(k,N))$ of the Grassmannian\cite{Witten:1993xi}. The chiral superfields $X_i$
correspond to the Chern classes $c_i$ of the tautological bundle over $Gr(k,N)$.

Note that the M-theory setup features two $U(1)$-symmetries induced by rotations in $TN_4$.
One of them, namely the $q$-grading, descends to the $U(1)_R$-symmetry of the Landau-Ginzburg model.
The other one will not play a role in our discussion.

\subsection{Junctions and LG interfaces}
\label{sec:junctions}

Let us now turn to the Landau-Ginzburg description of junctions of surface operators.
We are interested in junctions  of surface operators which are created when the stack of $k$ M5$'$-branes
is split up into two stacks of $k_1$ and $k_2=k-k_1$ M5$'$-branes, \cf Figure~\ref{fig:junction4} and Figure~\ref{fig:Junction1}.
\FIGURE[h!]{\includegraphics[width=4.5cm]{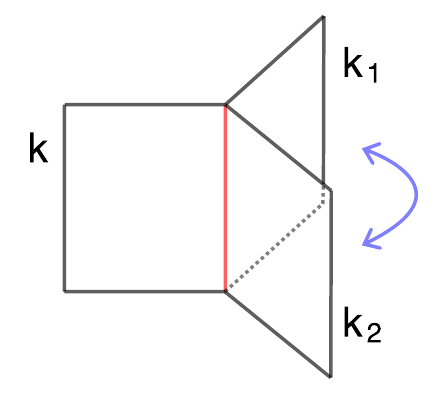}
\caption{\label{fig:Junction1}
Junction between $LG_k$, $LG_{k_1}$ and $LG_{k_2=k-k_1}$. Upon folding it can be described by an interface $\I_k^{k_1,k_2}$ between $LG_k$ and $LG_{k_1}\otimes LG_{k_2}$.}
}
As we already discussed below \eqref{euMP}, from the point of view of the Grassmannian sigma-model
this splitting corresponds to the following boundary condition at the junction:
the $k_1$- and $k_2$-dimensional subspaces defining the theory after the splitting are orthogonal
at the point of splitting and span the $k=(k_1+k_2)$-dimensional subspace defining the theory before the splitting.
This condition can be described by means of the correspondence \eqref{correspondence} between (products of) Grassmannians.
It identifies the Chern classes $c^{(k)}_1,\ldots,c^{(k)}_k$ of the tautological bundle over $Gr(k,N)$ with the symmetrizations
\be\label{relchernclasses}
c^{(k)}_i=\sum_{j=0}^ic^{(k_1)}_jc^{(k_2)}_{i-j}
\ee
of the Chern classes $c^{(k_i)}_j$ of the tautological bundles over $Gr(k_i,N)$, see Appendix \ref{appcorrespondence}.

This translates to a rather simple identification of chiral fields between the respective Landau-Ginzburg models:
We regard the junction via the folding trick as an interface $\I_k^{k_1,k_2}$ between
Landau-Ginzburg models $LG_k$ on one side and $LG_{k_1}\otimes LG_{k_2}$ on the other.
Both these theories can be realized
by means of a change of variables from one and the same model, $LG_{1}^{\otimes k}$, the Landau-Ginzburg model with
$k$ chiral superfields $x_1,\ldots, x_k$ and superpotential \eq{superpotential}. On the left side of the interface, in the model $LG_k$, the chiral superfields $X_i$ are obtained as the symmetrization of all the variables $x_i$, \cf \eq{symmetrization}, and correspond to the Chern classes $c^{(k)}_i$ of $Gr(k,N)$. On the right side of the interface, the superfields of the model $LG_{k_1}\otimes LG_{k_2}$ are obtained by symmetrizing the first  $k_1$ variables $(x_1,\ldots,x_{k_1})$ and  the last $k_2=k-k_1$ variables $(x_{k_1+1},\ldots,x_{k})$ separately:
\bea\label{partialsymmetrization}
&&(x_1,\ldots,x_k)\longmapsto(Y_1,\ldots,Y_k)=\\
&&\quad(\sigma_1(x_1,\ldots,x_{k_1}),\ldots,\sigma_{k_1}(x_1,\ldots,x_{k_1}),\sigma_1(x_{k_1+1},\ldots,x_k),\ldots,\sigma_{k_2}(x_{k_1+1},\ldots,x_k))\,.\nonumber
\eea
The $Z_i=Y_i$, $i=1,\ldots,k_1$ and $Z\p_j=Y_{k_1+j}$, $j=1,\ldots,k_2$ are the superfields of the factor models $LG_{k_1}$ and $LG_{k_2}$, respectively, and can be identified with the Chern classes $c_i^{(k_1)}$ and $c_j^{(k_2)}$.
In terms of these fields, the
junction condition \eq{relchernclasses} reads\footnote{Here one sets $Z_0=1=Z\p_0$ and $Z_i=0$ if $i$ is out of bounds ($i<0$ or $i>k_1$),
$Z\p_j=0$ if $j$ is out of bounds ($j<0$ or $j>k_2$).}
\be
X_i=\sigma_i(x_1,\ldots,x_k)=f_i(Y_1,\ldots,Y_k)=\sum_{j=0}^i Z_j Z\p_{i-j}\,.
\ee
It trivially identifies the underlying models $LG_1^{\otimes k}$ on both sides of the interface, and identifies the more symmetric variables $X_i$ on its left with the respective symmetrization of the less symmetric variables $Y_i$ on its right:
\bea\label{varrel}
X_1&=&\sigma_1(x_1,\ldots,x_k)=f_1(Y_1,\ldots,Y_k)=Y_1+Y_{k_1+1}\,,\nonumber\\
X_2&=&\sigma_2(x_1,\ldots,x_k)=f_2(Y_1,\ldots,Y_k)=Y_2+Y_{k_1+2}+Y_1Y_{k_1+1}\,,\\
&\ldots&\nonumber
\eea

This discussion immediately generalizes to more complicated configurations. The junction created by splitting up
the stack of M5$'$-branes into $r>2$ stacks of multiplicities $k_1,\ldots,k_r$, respectively,
on the level of the Grassmannian sigma-models is described by the correspondence
via the partial flag manifold $Fl(k_1,k_1+k_2,\ldots,k_1+\ldots+k_r=k,N) = G/\mathbb{L}$
with the Levi subgroup $\mathbb{L} = S \big( \prod_i U(k_i) \big)$,
{\it i.e.} the space of flags\footnote{See {\it e.g.} \cite{Gadde:2013dda} for further details
on topology of flag varieties and their coupling to 4d $\CN=4$ and $\CN=2$ gauge theories.}
\be
\C^{k_1} \subset \C^{k_1 + k_2} \subset \cdots \subset \C^{N}
\ee
\FIGURE[h!]{\includegraphics[width=4.5cm]{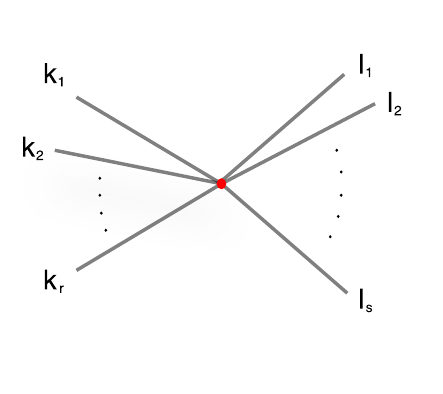}
\caption{\label{fig:Genjunction}
Interface $\I_{k_1,\ldots,k_r}^{l_1,\ldots,l_s}$ between $LG_{k_1}\otimes\ldots\otimes LG_{k_r}$ and $LG_{l_1}\otimes\ldots\otimes LG_{l_s}$. 
}}
As before, this translates to a simple condition at the interface $\I_k^{k_1,\ldots,k_r}$ between the respective Landau-Ginzburg models, $LG_k$ on one side and $LG_{k_1}\otimes\ldots\otimes LG_{k_r}$ on the other. Again, both models can be obtained by different changes of variables from one and the same underlying model, $LG_1^{\otimes k}$, which is trivially identified by the interface. The superfields $X_1,\ldots,X_k$ of $LG_k$ are the symmetrization of all the fields $x_1,\ldots,x_k$, whereas the fields $Y_1,\ldots,Y_k$ of $LG_{k_1}\otimes\ldots\otimes LG_{k_r}$ are obtained by separately symmetrizing the sets of
variables $(x_1,\ldots,x_{k_1})$, $(x_{k_1+1},\ldots,x_{k_1+k_2})$, $\ldots$, $(x_{k_1+\ldots+k_{r-1}+1},\ldots,x_k)$.
The interface then expresses the more symmetric polynomials $X_i$ in terms of the less symmetric ones $Y_j$:
\be
X_i=\sigma_i(x_1,\ldots,x_k)=f_i\p(Y_1,\ldots,Y_k)\,.
\ee

From this it is also evident how to describe a junction at which $r$ stacks of M5$'$-branes of multiplicities $k_1,\ldots,k_r$ join and immediately split up into $s$ stacks or multiplicities $l_1,\ldots,l_s$, \cf Figure \ref{fig:Genjunction}. (In this and the following figures we will suppress the time direction. The surface operators will be represented by lines, which will be labeled by the multiplicity of the respective stack of M5$'$-branes. Surface operators also have an orientation, which, in case of ambiguity will be indicated by arrows. The junctions discussed here, which arise by splitting and joining of stacks of M5$'$-branes have the property that the sum of labels of incoming lines equals the sum of labels of outgoing lines: M5$'$-branes cannot end in junctions.)
The junction of Figure \ref{fig:Genjunction} can be realized by an interface $\I_{k_1,\ldots,k_r}^{l_1,\ldots,l_s}$ between
$LG_{k_1}\otimes\ldots\otimes LG_{k_r}$ on the left side and $LG_{l_1}\otimes\ldots\otimes LG_{l_s}$
on the right. Since $\sum_i k_i=k=\sum_j l_j$, the models on either side of the interface arise from one and the same model $LG_1^{\otimes k}$ by means of different symmetrizations of variables.
On the left, the sets
$(x_1,\ldots,x_{k_1})$, $(x_{k_1+1},\ldots,x_{k_1+k_2})$, $\ldots$, $(x_{k_1+\ldots+k_{r-1}+1},\ldots,x_k)$
are symmetrized separately giving rise to the chiral superfields $(X_1,\ldots,X_k)$, whereas on the right the superfields $(Y_1,\ldots,Y_k)$ are obtained by symmetrizing the
$(x_1,\ldots,x_{l_1})$, $(x_{l_1+1},\ldots,x_{l_1+l_2})$, $\ldots$, $(x_{l_1+\ldots+l_{s-1}+1},\ldots,x_k)$. The interface now relates the fields on the two sides of the defect by expressing the symmetric polynomials
$\sigma_i(x_1,\ldots,x_k)$ in terms of the partially symmetrized ones on either side:
\bea
f\p_1(X_1,\ldots,X_k) =&\sigma_1(x_1,\ldots,x_k)&=f^{\prime\prime}_1(Y_1,\ldots,Y_k)\nonumber\\
f\p_2(X_1,\ldots,X_k) =&\sigma_2(x_1,\ldots,x_k)&=f^{\prime\prime}_2(Y_1,\ldots,Y_k)\label{genid}\\
&\ldots&\,,\nonumber
\eea
\FIGURE[t]{\includegraphics[width=11cm]{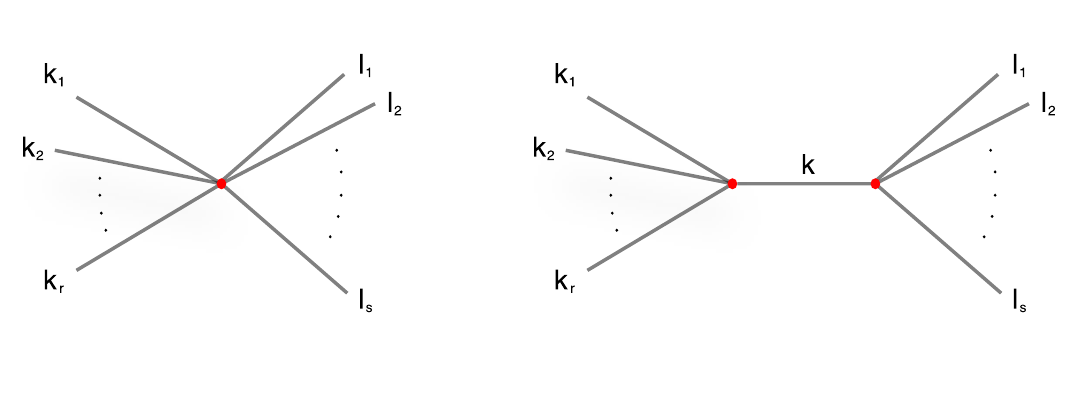}
\caption{\label{fig:Factorization}
Junction described by $\I_{k_1,\ldots,k_r}^{l_1,\ldots,l_s}$ (left) factorizes as $\I_{k_1,\ldots,r_r}^k*\I_k^{l_1,\ldots,l_s}$ over $LG_k$ (right).}}
In particular, this condition is the composition of the identifications imposed by the interfaces $\I_{l_1,\ldots,l_s}^k$
describing the splitting of one stack into $s$ stacks and the one imposed by the interface $\I_k^{k_1,\ldots,k_r}$ describing the joining of the $r$ stacks into one stack. Thus, the respective interface can be obtained by fusion
\be
\I_{k_1,\ldots,k_r}^{l_1,\ldots,l_s}=\I_{k_1,\ldots,r_r}^k*\I_k^{l_1,\ldots,l_s}\,
\ee
of these two interfaces over $LG_k$, \cf Figure \ref{fig:Factorization}.

In fact, more generally, the interfaces $\I_{k_1,\ldots,k_r}^{l_1,\ldots,l_s}$ factorize into those interfaces
describing trivalent junctions (where either two stacks of M5$'$-branes join into one or one stack splits up into two stacks),
see Figure \ref{fig:Trivfact}.
\FIGURE[h]{\includegraphics[width=11cm]{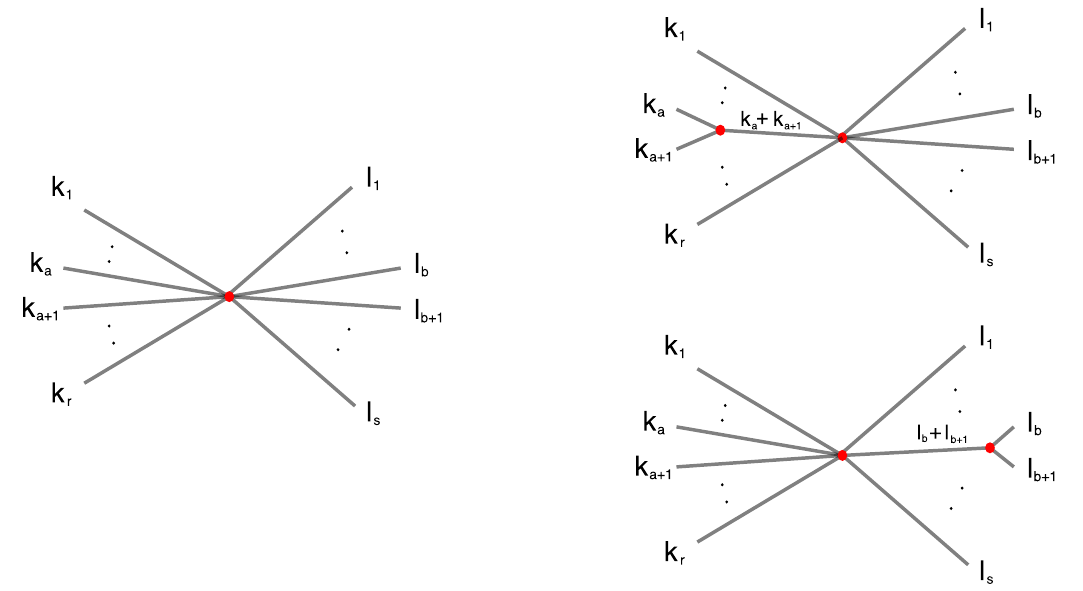}
\caption{\label{fig:Trivfact}
Junction (left) can be factorized into trivalent junctions (right).}}

\subsection{Junctions and matrix factorizations}

In the following, we will study the junctions of surface operators introduced in the last section in more detail
using an elegant representation of interfaces between Landau-Ginzburg models in terms of matrix factorizations \cite{Brunner:2007qu}. This representation in particular lends itself easily to the description of fusion. (Of course one could equivalently use the language of correspondences in Grassmannian sigma model to do the analysis.)

A matrix factorization over a polynomial ring $\mathcal{R}$ of a polynomial $W\in\mathcal{R}$ is a pair of square matrices $p_0$ and $p_1$ with entries in $\mathcal{R}$, which multiply to $W$ times the identity matrix:
\be
{\mathcal P}:\quad P_1\cong{\mathcal R}^r\overset{p_1}{\underset{p_0}{\longrightleftarrows}} {\mathcal R}^r\cong P_0\,,\qquad
p_1p_0=W\id_{P_0}\,,\quad
p_0p_1=W\id_{P_1}\,.
\ee
A B-type supersymmetric interface connecting a Landau-Ginzburg model with chiral superfields $X_1,\ldots,X_k$ and superpotential $W(X_1,\ldots,X_k)$ to one with superfields $Y_1,\ldots,Y_{l}$
and superpotential $W\p(Y_1,\ldots,Y_{l})$ is determined by a matrix factorization of the difference
$W(X_1,\ldots,X_k)-W\p(Y_1,\ldots,Y_{l})$ of superpotentials over the ring $\mathcal{R}=\C[X_1,\ldots,X_k,Y_1,\ldots,Y_{l}]$ of polynomials   in the chiral superfields on both sides.

From this description it is easy to read off certain properties of the interface, such as the chiral ring of operators on the interface, which is given by the BRST-cohomology on ${\rm End}_{\mathcal{R}}(P_0\oplus P_1)$, where the BRST-operator $Q$ acts on  $\Phi\in{\rm End}_{\mathcal{R}}(P_0\oplus P_1)$ by graded commutator
\be
Q\Phi=p\,\Phi-\sigma\Phi\sigma\, p
\ee
with the operator
\be
p=\left(\begin{array}{cc} 0 & p_1 \\p_0 & 0\end{array}\right)\,.
\ee
The $\Z_2$-grading (fermion number) is given by
\be
\sigma=\left(\begin{array}{cc} \id_{P_0} & 0 \\0 & -\id_{P_1}\end{array}\right)\,.
\ee
Correlation functions of chiral primary fields in the presence of such interfaces can be calculated by means of a residue formula \cite{Kapustin:2003ga}.

\FIGURE[t]{\includegraphics[width=10cm]{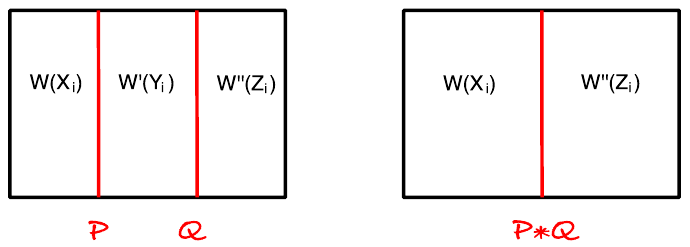}
\caption{\label{fig:mf-fusion}
Fusion of interface $\mathcal{P}$ between LG models with superpotentials $W(X_i)$ and $W\p(Y_i)$, and
$\mathcal{Q}$ between LG models with superpotentials $W\p(Y_i)$ and $W^{\prime\prime}(Z_i)$ (left) gives rise to
interface $\mathcal{P}*\mathcal{Q}$ between LG models with superpotentials $W(X_i)$ and $W^{\prime\prime}(Z_i)$ (right).}}
Moreover, fusion of interfaces ${\mathcal P}$ and ${\mathcal Q}$ separating respectively a Landau-Ginzburg model with superpotential $W(X_1,\ldots,X_r)$ from one with superpotential $W\p(Y_1,\ldots,Y_s)$, and the model with superpotential $W\p(Y_1,\ldots,Y_s)$ from one with superpotential $W^{\prime\prime}(Z_1,\ldots,Z_t)$,
\cf Figure \ref{fig:mf-fusion}, has a very simple description in terms of matrix factorizations. The interfaces are described by two matrix factorizations
\be
{\mathcal P}:\; P_1\overset{p_1}{\underset{p_0}{\longrightleftarrows}} P_0\quad{\rm and}\quad
{\mathcal Q}:\; Q_1\overset{q_1}{\underset{q_0}{\longrightleftarrows}} Q_0
\ee
of $W(X_1,\ldots,X_r)-W\p(Y_1,\ldots,Y_s)$ and $W\p(Y_1,\ldots,Y_s)-W^{\prime\prime}(Z_1,\ldots,Z_t)$, respectively.
The fused interface ${\mathcal P}*{\mathcal Q}$ separates the Landau-Ginzburg models with superpotential $W(X_1,\ldots,X_r)$ from the one with superpotential $W^{\prime\prime}(Z_1,\ldots,Z_t)$. It is described by the tensor product matrix factorization
\be
\left({\mathcal P}\otimes {\mathcal Q}\right)_1:=\left(\begin{array}{c}P_1 \otimes Q_0\\ \oplus \\P_0\otimes Q_1\end{array}\right)
\overset{\left(\begin{array}{cc}p_1\otimes \id_{Q_0} & -\id_{P_0}\otimes q_1\\
\id_{P_1}\otimes q_0 & p_0\otimes \id_{Q_1}\end{array}\right)}
{\underset{\left(\begin{array}{cc}p_0\otimes \id_{Q_0} & \id_{P_1}\otimes q_1\\
-\id_{P_0}\otimes q_0 & p_1\otimes \id_{Q_1}\end{array}\right)
}{\llongrightleftarrows}}
\left(\begin{array}{c} P_0\otimes Q_0\\ \oplus \\ P_1\otimes Q_1 \end{array}\right)=:\left({\mathcal P}\otimes{\mathcal Q}\right)_0\,.
\ee
This is a matrix factorization of the sum
\bea
&&\left(W(X_1,\ldots,X_r)-W\p(Y_1,\ldots,Y_s)\right) +
\left(W\p(Y_1,\ldots,Y_s)-W^{\prime\prime}(Z_1,\ldots,Z_t)\right)\\
&&\qquad\qquad\qquad\qquad\qquad\qquad\qquad\qquad\qquad\qquad=W(X_1,\ldots,X_r)-W^{\prime\prime}(Z_1,\ldots,Z_t)\,,\nonumber
\eea
as it should be. It still involves the
fields $Y_i$ of the Landau-Ginzburg model squeezed in between the interfaces, which are
promoted to new degrees of freedom on the interface. Therefore, a priori it is a matrix factorization of infinite rank over
$\C[X_1,\ldots,X_r,Z_1,\ldots,Z_t]$.
However, it can be shown that by splitting off ``trivial'' matrix factorizations\footnote{matrix factorizations of the form $1 \cdot W$, are physically trivial} these matrix factorizations always reduce to finite rank.
See \cite{Brunner:2007qu,DM11,Carqueville:2011zea} for a more detailed discussion of fusion in this context.

The Landau-Ginzburg models we are interested in here exhibit a $U(1)_R$-symmetry, and the interfaces also preserve this symmetry. This means that the modules $P_0$ and $P_1$ carry representations $\rho_0$ and $\rho_1$ of $U(1)_R$, respectively, which are compatible with the $\mathcal{R}$-module structure, such that $p_0$ and $p_1$ are homogeneous of charge $1$\footnote{The superpotentials which are factorized by $p_0$ and $p_1$ have $U(1)_R$-charge $2$.}. The respective matrix factorization is called ``graded''. For ease of notation we
will rescale the $U(1)_R$-charges by the degree $N+1$ of the potential. In these units the superpotential has charge $2N+2$, the maps $p_i$ have charge $N+1$, and the fields $X_i$ defined in \eq{symmetrization} have charge $2i$. All charges appearing will be integral.

Now, the interface $\I_k^{k_1,k_2}$ associated to the trivalent junction in Figure \ref{fig:Junction1} separates the Landau-Ginzburg model $LG_k$ with superpotential $W(X_1,\ldots,X_k)$ (the superpotential of \eq{superpotential} written in the symmetrized fields  \eq{symmetrization}), from the Landau-Ginzburg model $LG_{k_1}\otimes LG_{k_2}$, which has superpotential $W(f_1,\ldots,f_k)$ (the same superpotential, but written in the only partially symmetrized fields \eq{partialsymmetrization}). Here $f_i=f_i(Y_1,\ldots,Y_k)$ is the map from \eq{varrel} symmetrizing the partially symmetrized $Y_i$. The matrix factorization of $W(X_1,\ldots,X_k)-W(f_1,\ldots,f_k)$ which imposes the identification \eq{varrel} is now easy to construct. It is given by a Koszul type matrix factorization, \cf \cite{Brunner:2007qu}.

One can find homogeneous polynomials
$U_i(X_1,\ldots,X_k,Y_1,\ldots,Y_k)$, $i=1,\ldots,k$, such that
\be\label{polydiff}
W(X_1,\ldots,X_k)-W(f_1,\ldots,f_k)
=\sum_{i=1}^k(X_i-f_i(Y_1,\ldots,Y_k))\,U_i(X_1,\ldots,X_k,Y_1,\ldots,Y_k)\,.
\ee
One particular choice\footnote{Other choices lead to equivalent matrix factorizations.}
of such polynomials is given by
\be
U_i={1\over X_i-f_i}\left(W(f_1,\ldots,f_{i-1},X_i,\ldots,X_k)-
 W(f_1,\ldots,f_{i},X_{i+1},\ldots,X_k)\right)\,.
\ee
The matrix factorization describing the interface is then given by the tensor product
\be\label{koszulfact}
\I_{k}^{k_1,k_2}=\left(\bigotimes_{i=1}^k {\mathcal P}^i\right)\{-k_1k_2\}
\ee
of the rank-$1$ matrix factorizations
\be
{\mathcal P}^{i}:\quad P^i_1\cong{\mathcal R}\{2i-N-1\}\overset{p^i_1=(X_i-f_i)}{\underset{p^i_0=U_i}{\longrightleftarrows}} {\mathcal R}\{0\}\cong P^i_0
\ee
over ${\mathcal R}=\C[X_1,\ldots,X_k,Y_1,\ldots,Y_k]$. Since, under the tensor product of matrix factorizations, the factorized polynomials behave additively, \eq{koszulfact} is indeed a matrix factorization of \eq{polydiff} and has rank $r=2^{k-1}$. Here, with the notation ${\mathcal R}\{a\}$ we specify the $U(1)_R$-representations on ${\mathcal R}$ by indicating the charge $a$ of $1\in{\mathcal R}$; and a matrix factorization ${\mathcal Q}\{a\}$, as in \eq{koszulfact}, denotes the matrix factorization ${\mathcal Q}$ where all
$U(1)_R$-charges are shifted by $a$\footnote{Such a shift only changes the $U(1)_R$-charges of interface changing fields.}.

This construction immediately generalizes to the more general junctions $\I_{k_1,\ldots,k_r}^{l_1,\ldots,l_s}$
of Figure \ref{fig:Genjunction}. The respective interface separates the Landau-Ginzburg model
$LG_{k_1}\otimes\ldots\otimes LG_{k_r}$ on the left from $LG_{l_1}\otimes\ldots\otimes LG_{l_s}$ on the right.
Both models are obtained from the Landau-Ginzburg model of \eq{superpotential} by means of different symmetrizations of the variables $x_1,\ldots,x_k$. On the left, the sets of variables $(x_1,\ldots,x_{k_1})$, $(x_{k_1+1},\ldots,x_{k_1+k_2})$, $\ldots$, $(x_{k_1+\ldots+k_{r-1}+1},\ldots,x_k)$ are symmetrized separately, giving rise to the superfields $(X_1,\ldots,X_k)$, whereas on the right the superfields $(Y_1,\ldots,Y_k)$ are obtained by symmetrizing
the sets of variables
$(x_1,\ldots,x_{l_1})$, $(x_{l_1+1},\ldots,x_{l_1+l_2})$, $\ldots$, $(x_{l_1+\ldots+l_{s-1}+1},\ldots,x_k)$.
The superpotentials on the two sides are just given by \eq{superpotential}, expressed in terms of the respective superfields on the two sides. On the left it is $W(f\p_1,\ldots,f\p_k)$, and on the right  $W(f^{\prime\prime}_1,\ldots,f^{\prime\prime}_k)$, where $f\p_i$ and $f^{\prime\prime}_i$ are the maps \eq{genid} symmetrizing the partially symmetric $(X_1,\ldots,X_k)$ and $(Y_1,\ldots,Y_k)$, respectively.

As for the trivalent case, the junction condition \eq{genid} can now be implemented by a Koszul matrix factorization.
The
polynomials
\be
U_i={1\over f\p_i-f^{\prime\prime}_i}\left(W(f^{\prime\prime}_1,\ldots,f^{\prime\prime}_{i-1},f\p_i,\ldots,f\p_k)-W(f^{\prime\prime}_1,\ldots,f^{\prime\prime}_i,f\p_{i+1},\ldots,f\p_k)\right)\,,
\ee
satisfy
\be
W(f\p_1,\ldots,f\p_k)-W(f^{\prime\prime}_1,\ldots,f^{\prime\prime}_k)
=\sum_{i=1}^k(f\p_i-f^{\prime\prime}_i)\,U_i\,,
\ee
and the tensor product
\be\label{genlgdef}
\I_{k_1,\ldots,k_r}^{l_1,\ldots,l_s}=\left(\bigotimes_{i=1}^k {\mathcal P}^i\right){\scriptstyle \left\{{ -\sum_{1\leq a<b\leq k}l_al_b}\right\}}
\ee
of the rank-one matrix factorizations
\be
{\mathcal P}^{i}:\quad P^i_1\cong{\mathcal R}\{2i-N-1\}\overset{p^i_1=(f\p_i-f^{\prime\prime}_i)}{\underset{p^i_0=U_i}{\longrightleftarrows}} {\mathcal R}\{0\}\cong P^i_0
\ee
implements the identification \eq{genid} across the interface.\footnote{Indeed, the 
Landau-Ginzburg defect $\I_{k_1,\ldots,k_r}^{l_1,\ldots,l_s}$ for the case $s=1$, $k_i=1$ was already mentioned in \cite{Behr:2012xg}.}

\subsection{Junctions and categorification of quantum groups}

\FIGURE[htb]{\includegraphics[width=4cm]{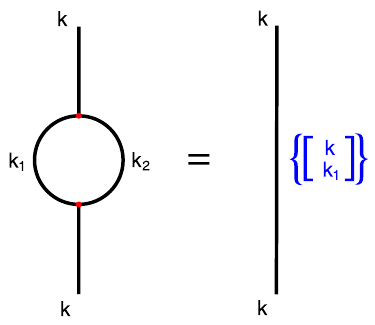}
\caption{\label{fig:bubblerel}
Fusion of 
$\I^k_{k_1,k_2}$ and $\I^{k_1,k_2}_k$ in $LG_{k_1}\otimes LG_{k_2}$ produces copies of the identity defect 
in $LG_k$.}}
The interfaces described in the last section satisfy many interesting properties.
For instance,
from the construction it is evident, that they factorize through a surface operator $LG_k$, \ie
\be
\I_{k_1,\ldots,k_r}^{l_1,\ldots,l_s}=
\I^{l_1,\ldots,l_s}_k*\I_{k_1,\ldots,k_r}^k\,,
\ee
\cf Figure \ref{fig:Factorization}, and that more generally, they factorize into interfaces corresponding to trivalent junctions, as displayed in Figure \ref{fig:Trivfact}.

Moreover, fusing the interfaces
$\I^k_{k_1,k_2}$ and $\I^{k_1,k_2}_k$ ($k=k_1+k_2$) in $LG_{k_1}\otimes LG_{k_2}$
produces a number of copies of the identity defect $\Id_k=\I_k^k$ in $LG_k$:
\be\label{bubblerel}
\I^k_{k_1,k_2}*\I^{k_1,k_2}_k=
\Id_k\left\{k \brack k_1\right\}\,,
\ee
see Figure \ref{fig:bubblerel}. Here, we use the notation
$
\P\{q^{a_1}+\ldots+q^{a_r}\}:=\P\{a_1\}\oplus\ldots\oplus\P\{a_r\}
$
for a matrix factorization $\P$, and, as before,
${k\brack k_1}={[k]!\over[k_1]![k-k_1]!}$ denotes the quantum binomial coefficient with
$[k]!=[k][k-1]\cdots[1]$, $[k]={q^k-q^{-k}\over q-q^{-1}}=q^{k-1}+q^{k-3}+\ldots+q^{-k+1}$.
A simple proof of this relation can be found in Appendix \ref{appbubble}.

Let
\bea
\qE_{k_1,k_2}&:=&\left(\Id_{k_1-1}\otimes \I_{1,k_2}^{k_2+1}\right)
*\left(\I_{k_1}^{k_1-1,1}\otimes\Id_{k_2}\right)
\\
\qF_{k_1,k_2}&:=&
\left(\I_{k_1,1}^{k_1+1}\otimes \Id_{k_2-1}\right)
*\left(\Id_{k_1}\otimes\I_{k_2}^{1,k_2-1}\right)\,,\nonumber
\eea
be the combination of interfaces represented in Figure \ref{fig:qgenerators}. Then, it is not difficult 
%
to
see (\cf Appendix \ref{app:qrel}) that
\be
\qE_{k_1+1,k_2-1}*\qF_{k_1,k_2} \cong
\qF_{k_1-1,k_2+1}*\qE_{k_1,k_2}\oplus \Id_{k_1,k_2}\{\left[k_2-k_1\right]\}
\ee
for $k_1\leq k_1$ and
\be\label{qrel1}
\qF_{k_1-1,k_2+1}*\qE_{k_1,k_2} \cong
\qE_{k_1+1,k_2-1}*\qF_{k_1,k_2}\oplus \Id_{k_1,k_2}\{\left[k_1-k_2\right]\}
\ee
for $k_1\geq k_2$. 
\FIGURE[h]{
\includegraphics[width=5cm]{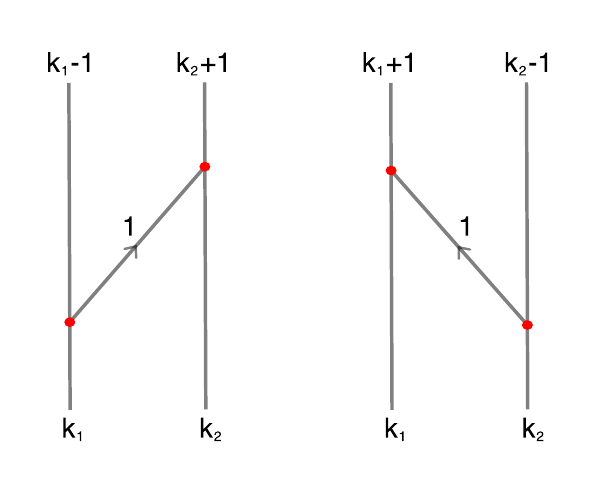}
\caption{\label{fig:qgenerators}
Configurations of surface operators categorifying the quantum group generators: $\qE_{k_1,k_2}$ (left)
and $\qF_{k_1,k_2}$ (right).}}
Here $\Id_{k_1,k_2}=\Id_{k_1}\otimes \Id_{k_2}$ denotes the identity defect
in the tensor product $LG_{k_1}\otimes LG_{k_2}$. But this is nothing but a categorification of the
relation $[E,F]1_n=[n]1_n$ of the quantum group $\dot{\mathbf{U}}_q(\mathfrak{sl}_2)$, \cf \eq{qsl2rel2}.
Here, the difference $n=k_2-k_1$ plays the role of a $\mathfrak{sl}_2$ weight.

In fact, the matrix factorizations $\I_{k_1,\ldots,k_r}^{l_1,\ldots,l_s}$ have already appeared in the construction of $\mathfrak{sl}_N$-link homology in the math literature \cite{Wu}. There it was shown that they satisfy relations categorifying the MOY relations depicted in Figure \ref{fig:rels}. This means in particular, that the junctions of surface operators defined in section \ref{sec:junctions} provide a categorification of the $\dot{\mathbf{U}}_q(\mathfrak{sl}_m)$ representation obtained by ladders of Wilson lines in Chern-Simons theory, \cf section \ref{sec:CS}. This 2-representation was analysed from a mathematical point of view in \cite{MY13}.
Building blocks are configurations of $m$ parallel stacks of M5$'$-branes, on which the quantum group generators $\qE_i$
act by  peeling off one M5$'$-brane from the $i$-th stack and migrating it to the $(i+1)$-st stack, whereas the $\qF_i$ are realized by peeling off one M5$'$-brane from the $(i+1)$-st and moving it to the $i$-th stack.

The 2-category structure is given as follows: Objects are $m$-tuples $(k_1,\ldots,k_m)$, $0\leq k_i\leq N$, corresponding to $m$ parallel stacks of M5$'$-branes of multiplicities $k_i$. From a 2d perspective they are realized by tensor products $LG_{k_1}\otimes\ldots\otimes LG_{k_m}$ of the Landau-Ginzburg theories on the respective world-volumes. 1-Morphisms correspond to configurations joining such stacks. In 2d they are interfaces  between the respective Landau-Ginzburg models. The $2$-morphisms correspond to interface changing fields.


\section{What's next?}
\label{sec:questions}

We conclude with a list of questions and possible generalizations, some of which have already been mentioned in the text:

\begin{itemize}
\item Even though the current work shows how the BPS junctions of surface operators realize the generators and elementary relations of the 2-category $\dot \CU$, it still remains unproven whether these junctions reproduce the entire set of KLR algebra relations, or equivalently, the foam relations given in \cite{QR14}. 

\item
It would be interesting to verify the existence of BPS junctions of surface operators by alternative methods
(in particular, by constructing explicit solutions to the field equations)
and extend the classification presented here to more general 4d theories and more general surface operators.

\item
For example, it would be interesting to construct holographic duals of surface operator junctions proposed here.

\item 
Also it would be interesting to investigate, whether the categories of junctions of surface operators can be reconstructed out of the representation varieties appearing in the discussion of the charge conservation condition at junctions in section \ref{sec:Horn}.

\item
Mathematically, other types of junctions (for other types of surface operators and LG models) also lead to diagrammatic algebras
and may be worth exploring.

\item
As a generalization in a different direction, one may keep the same types of surface operators and LG interfaces,
but consider changing space-time geometry, in particular, the geometry of the fivebrane world-volume.
Indeed, in most of our discussion we assumed $M_3 = \R^3$, and one immediate generalization would be
to consider $M_3 = S^1 \times \R^2$ which would result in planar diagrammatics on a cylinder.
More generally, one might consider junctions of surface operators in $M_4 = \tilde \Sigma \times \R^2$
which, via the arguments of this paper, would lead to LG models and planar diagrammatics on a Riemann surface $\tilde \Sigma$.

\item
The analysis of section~\ref{sec:surface} suggests to carry out the analogue of \cite{MR3190356,MR3125899}
for the complexified gauge group $G_{\C}$ and to compare Poincar\'e polynomials of the resulting moduli spaces $\CM (G_{\C}, \Gamma)$
to MOY polynomials of the planar graphs $\Gamma$.

\item
In section \ref{sec:LG} we already mentioned two important generalizations to supergroups and symmetric representations
that should categorify the super $q$-Howe duality 
%
\cite{QS15}
 and symmetric $q$-Howe duality \cite{SymmetricHowe}, respectively.

\item
It would be interesting to work out a precise dictionary between ``fake bubbles'' of the diagrammatic approach \cite{Lauda}
and relevant deformations of the Landau-Ginzburg potential \cite{Gukov:2005qp}, both of which are related to differentials
on Khovanov-Rozansky homology.

\end{itemize}

\noindent
We plan to tackle some of these questions in the future work.


\acknowledgments{We thank A.~Lauda and D.~Rose for many patient and very helpful explanations, as well as J.~Kamnitzer, M.~Khovanov, A.~Lobb,
M.~Mackaay, L.~Rozansky, M.~Stosic, C.~Stroppel, C.~Teleman, B.~Webster, and P.~Wedrich
for helpful comments and advice. We also thank participants of the ``Knot homologies, BPS states, and SUSY gauge theories" program at the Simons Center for Geometry and Physics for useful discussions.

The work of S.G. is funded in part by the DOE Grant DE-SC0011632 and the Walter Burke Institute for Theoretical Physics.}


\appendix

\section{Wilson lines and categories $N$Web}
\label{appendix:derivations}

In this section, we derive the remaining relations of Figure \ref{fig:rels}. Before proceeding to the derivations, however, it is necessary to resolve an ambiguity in our normalization. Let us denote the invariant tensors at the junctions of Figure \ref{fig:normalization} as $\epsilon_{k_{1},k_{2}}$ and $\tilde{\epsilon}_{k_{1},k_{2}}$, respectively. Since the RHS of Figure \ref{fig:normalization} only determines the normalization of their product, it remains the same if one invariant tensor is multiplied by a complex number and the other by the inverse. Thus, there is an ambiguity in the ``relative'' normalization among the invariant tensors from different types of junctions. Such ambiguity would not be observed if $\epsilon$'s and $\tilde{\epsilon}$'s always appear in pairs, but they do not in Figure \ref{fig:rels}(a), \ref{fig:rels}(d) and \ref{fig:rels}(f).

\begin{figure} [b] \centering
\includegraphics {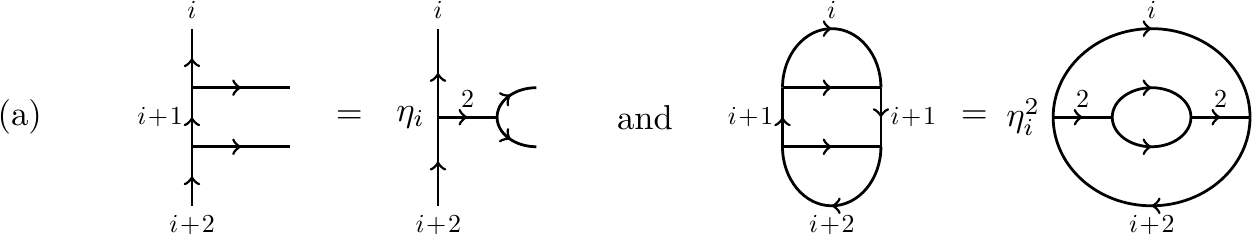} \\[1.5ex]
\includegraphics {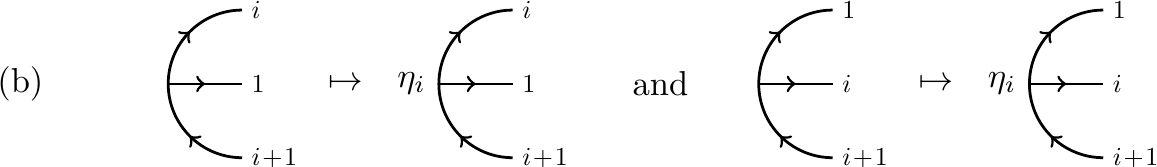}
\caption{(a) Definition of $\eta_{i}$ and the closed Wilson lines which determine the value of $\eta_{i}^{2}$. (b) Renormalization of junctions which involve the Wilson lines in $\square$.}
\label{fig:normalization2}
\end{figure}

We claim that the junctions can be normalized so as to satisfy the relations of Figure \ref{fig:rels}. Let us first normalize those which involve Wilson lines in the defining representations. The open Wilson lines of Figure \ref{fig:normalization2}(a) are proportional to each other, since the associated Hilbert space is 1-dimensional. Renormalize the invariant tensors $\epsilon_{i,1}$ and $\epsilon_{1,i}$ via multiplication by $\eta_{i}$ (\cf Figure \ref{fig:normalization2}(b)), and renormalize $\tilde{\epsilon}_{i,1}$ and $\tilde{\epsilon}_{1,i}$ accordingly via multiplication by $1/\eta_{i}$. This proves Figure \ref{fig:rels}(a) for $j=k=1$.

\begin{figure} [t] \centering
\includegraphics {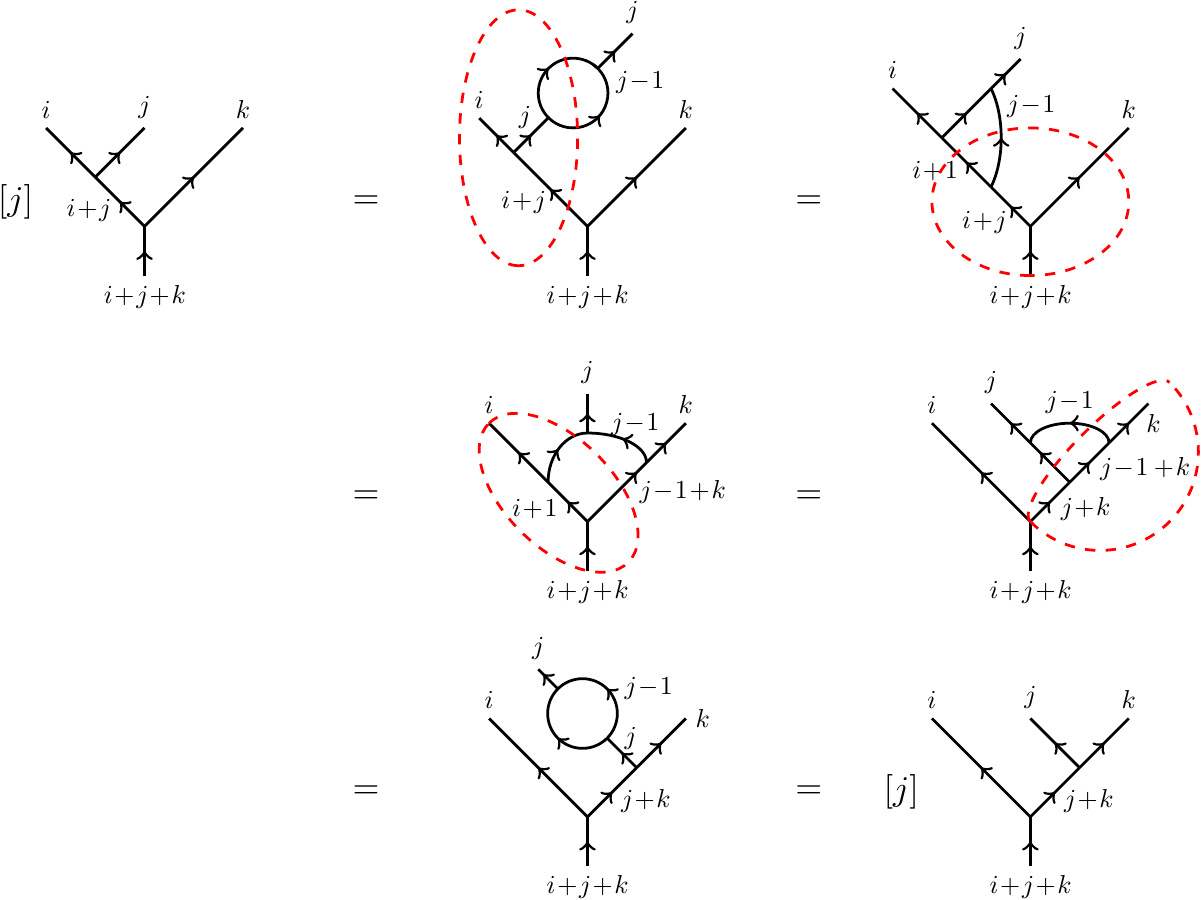}
\caption{Induction on $j$. Apply the base case $j=1$ and the induction hypothesis for $j-1$ in the red dashed cirlces.}
\label{fig:induction}
\end{figure}

Inductively renormalize the junctions of higher rank representations. Consider Wilson lines of Figure \ref{fig:rels}(a) with $j=1$, and assume WLOG that $i \geq k$. Since the associated Hilbert space is 1-dimensional, the Wilson lines on each side are proportional to each other. Renormalizing $\epsilon_{i,k+1}$ by absorbing the proportionality constant, we have turned the relation into an identity and determined the relative normalization between $\epsilon_{i,k+1}$ and $\epsilon_{i+1,k}$. Proceed recursively until we reach $\epsilon_{i+k,1}$, whose normalization is fixed by Figure \ref{fig:normalization2}(b). Applying this procedure for all $i+k$, the relative normalizations among the invariant tensors can be determined such that Figure \ref{fig:rels}(a) holds for $j=1$. Finally, induction on $j$ shows that Figure \ref{fig:rels}(a) holds for any $i,j$ and $k$ (\cf Figure \ref{fig:induction}.)

Now that we have fixed the normalization of junctions, let us proceed to derive the remaining relations. Figure \ref{fig:rels}(b) is rather straightforward, once we close the open ends by connecting them.

The three vectors of Figure \ref{fig:rels}(c) satisfy a linear relation, as the associated Hilbert space is 2-dimensional. Close the open ends in two inequivalent ways (\cf Figure \ref{fig:almostEF}.) The resulting closed Wilson lines have computable expectation values via the direct sum formula, Figure \ref{fig:rels}(a) and Figure \ref{fig:rels}(b).

Figure \ref{fig:rels}(d) follows from Figures \ref{fig:rels}(a), (b) and (e). First, apply Figure \ref{fig:rels}(a) and \ref{fig:rels}(b) to replace the right-moving Wilson line labeled by $\wedge^{n}\square$ (the left-moving Wilson line labeled by $\wedge^{l+n-1}\square$) by $n$ right-moving Wilson lines labeled by $\square$ ($l\!+\!n\!-\!1$ left-moving Wilson lines labeled by $\square$). Recursively apply Figure \ref{fig:rels}(e) until the resulting Wilson lines are those of the RHS, and the coefficients follow.

Figure \ref{fig:rels}(f) also follows from Figures \ref{fig:rels}(a), (b) and (e). First, apply Figure \ref{fig:rels}(a) on LHS with $(i,j,k) = (m\!-\!2,1,1)$. Use Figure \ref{fig:rels}(a) and (b) to replace the upper Wilson line labeled by $\wedge^{j}\square$ by $j$ Wilson lines labeled by $\square$. Then, recursively apply Figure \ref{fig:rels}(e) until the resulting Wilson lines differ from those of the RHS only by a single right-moving Wilson line labeled by $\wedge^{2}\square$. Use Figure \ref{fig:rels}(a) and (b) to replace the latter 
by two Wilson lines labeled by $\square$, and the coefficients follow. This concludes our proof of Figure \ref{fig:rels}.

\section{Domain walls, junctions and Grassmannians}
\label{app:domainwalls}

For $N>r$,
and for generic values of mass parameters $m_i$,
4d $\CN=2$ SQCD theory has
massive isolated vacua (with $i_n < i_{n+1}$):
\be
\phi \; = \; \text{diag} (m_{i_1}, m_{i_2}, \ldots, m_{i_{r}})
\ee
in which the gauge group $U(r)$ is completely broken and $SU(N_f) \to U(1)^{N_f - 1}$.
When all masses are real --- as \eg in 5d lift of this theory --- one can also assume $m_i < m_{i+1}$.
Clearly, the set of such vacua is in one-to-one correspondence with Schubert cells $\CS_I$ in $Gr (r,N)$
or, equivalently, vertices of the moment graph associated to the $T = U(1)^N$ action on $Gr(r,N)$
discussed around \eq{Tequiv}. 

Note, that the mass parameters $m_i$ can be identified with the equivariant parameters for the torus action.
For instance, $U(2)$ theory with $N_f = 4$ has six vacua \eqref{Gr24Hasse}, also shown in Figure~\ref{fig:junction2}.

While vacua of 4d $\CN=2$ SQCD correspond to vertices of the moment graph,
elementary (not composite) domain walls that interpolate between vacua labeled by $I$ and $J$ correspond
to edges of the moment graph connecting vertices $I$ and $J$. Two vertices $I \ne J$ are joined by an edge if and only if
\be
| I \cup J | \; = \; r-1
\ee
in which case this edge is labeled by $m_i - m_j$, where $i$ and $j$ are the two elements in the symmetric difference of $I$ and $J$.
For instance, $U(2)$ theory with $N_f = 4$ has 12 elementary walls that correspond to edges of the octahedron in Figure~\ref{fig:junction2}.
In general, each elementary wall carries $U(1) \subset U(r)$ symmetry left unbroken by
$r-1$ components of the Higgs field which remain non-zero in the core of the wall in the weak coupling limit.
For this reason, such elementary walls are called {\it abelian} \cite{Eto:2005fm}.
For special values of mass parameters, one finds walls that support non-abelian symmetries \cite{Shifman:2003uh,Eto:2008dm}.

The BPS condition relates the domain wall tension $T$ and its orientation, say, in the $(x^1,x^2)$ plane
to the absolute value and phase of $m_i - m_j$, respectively:
\be
m_i - m_j = T e^{i \theta}
\label{mmangle}
\ee
where $\theta$ is the angle in the $(x^1,x^2)$ plane.
In particular, if all mass parameters $m_i$ are real, the BPS condition means that walls must be parallel in four-dimensional space-time
and their ordering in $x^2$ direction
precisely follows the closure ordering of Schubert cells \eq{partialorder}.
When mass parameters are complex, on the other hand, in general one finds a web (or network) of walls at angles
determined by \eqref{mmangle}, see \eg \cite{Eto:2005cp} for a nice ``adiabatic'' argument (for small values of $\text{Im} (m_i)$).

Webs or networks of domain walls involve junctions, which can be either {\it abelian} or {\it non-abelian},
depending on the unbroken gauge symmetry in the core of the junction at weak coupling \cite{Eto:2005fm}.
Specifically, a trivalent junction of walls separating vacua labeled by Schubert symbols
$I_1 = \langle a \, \underline{\cdots} \rangle$, $I_2 = \langle b \, \underline{\cdots} \rangle$, and $I_3 = \langle c \, \underline{\cdots} \rangle$
has vanishing components of the Higgs field $\phi_a$, $\phi_b$, and $\phi_c$ in the core of the junction and,
therefore, carries $U(1)$ gauge symmetry left unbroken by the remaining $r-1$ non-zero components of the Higgs field.
On the other hand, a junction of walls between vacua labeled by
$I_1 = \langle ab \, \underline{\cdots} \rangle$, $I_2 = \langle bc \, \underline{\cdots} \rangle$,
and $I_3 = \langle ac \, \underline{\cdots} \rangle$
has only $r-2$ non-zero components of the Higgs field at the core and, therefore, carries $U(2)$ unbroken symmetry.
This discussion easily generalizes to other types 
of walls and junctions.

\section{LG Interfaces and the cohomology of Grassmannians}
\label{appcorrespondence}
The cohomology of the Grassmannian $Gr(k,N)$ can be described as the polynomial ring in the Chern classes $c_1^{(k)},\ldots,c_k^{(k)}$ of the tautological bundle and the Chern classes $\bar{c}^{(k)}_1,\ldots,\bar{c}^{(k)}_{N-k}$ of the complementary bundle modulo the ideal $I_{k,N}$ generated by the relations
\be
(1+tc^{(k)}_1+\ldots+t^kc^{(k)}_k)(1+t\bar{c}^{(k)}_1+\ldots+t^{N-k}\bar{c}^{(k)}_{N-k})=1\,.
\ee
Similarly, the cohomology of the partial flag variety $Fl(k_1,k=k_1+k_2,N)$ can be realized as polynomial ring in $c_1,\ldots,c_{k_1},c\p_1,\ldots,c\p_{k_2},\bar{c}_1,\ldots,\bar{c}_{N-k}$ modulo the ideal $I_{k_1,k,N}$ generated by the relations
\be
(1+tc_1+\ldots+t^{k_1}c_{k_1})(1+tc\p_1+\ldots+t^{k_2}c\p_{k_2})(1+t\bar{c}+\ldots+t^{N-k}\bar{c}_{N-k})=1\,.
\ee
Now \eq{correspondence} maps
\bea
H^*(Gr(k_1,N)\times Gr(k_2,N))&\longrightarrow&H^*(Fl(k_1,k,N))\\
c^{(k_1)}_i&\longmapsto & c_i\nonumber\\
\bar{c}^{(k_1)}_i&\longmapsto & \sum_{j=0}^i c\p_j\bar{c}_{i-j}\nonumber\\
c^{(k_2)}_i&\longmapsto & c\p_i\nonumber\\
\bar{c}^{(k_2)}_i&\longmapsto & \sum_{j=0}^i c_j\bar{c}_{i-j}\nonumber
\eea
and
\bea
H^*(Gr(k,N))&\longrightarrow&H^*(Fl(k_1,k,N))\\
c^{(k)}_i&\longmapsto & \sum_{j=0}^ic_jc\p_{i-j}\nonumber\\
\bar{c}^{(k)}_i&\longmapsto & \bar{c}_i\nonumber\,.
\eea
Hence, the correspondence \eq{correspondence} identifies $c^{(k)}_i$ with $\sum_{j=0}^i c^{(k_1)}_j c^{(k_2)}_{i-j}$.

\section{LG Interfaces and 2-categories $N$Foam}
The category $N$Foam can be described in terms of generators and relations \cite{QR14}. In this appendix we exemplary check that the realization we propose in terms of surface operators satisfies two such relations.

\subsection{Derivation of the bubble relation}
\label{appbubble}

To derive the bubble formula \eq{bubblerel}, we make use of a correspondence between
matrix factorizations of $W$ over a polynomial ring ${\mathcal R}$ and finitely generated modules over the respective hypersurface ring $\hat{\mathcal R}:={\mathcal R}/(W)$. Namely, free resolutions of the latter always turn $2$-periodic after finitely many steps, with the $2$-periodic part given by a matrix factorization of $W$, \cite{eisenbud}. Thus, questions about matrix factorizations can be turned into questions about $\hat{\mathcal R}$-modules.
This comes in handy in particular in the calculation of fusion of interfaces in Landau-Ginzburg models \cite{Brunner:2007qu}.

The matrix factorization $\I_{k}^{k_1,k_2}$ for instance is related in this way to the module
\be
M=\hat{\mathcal R}/{\mathcal J}\hat{\mathcal R}\{-k_1k_2\}\,,
\ee
over the ring $\hat{\mathcal R}=\C[X_1,\ldots,X_k,Y_1,\ldots,Y_k]/(W(X_1,\ldots,X_k)-W(f_1,\ldots,f_k))$, where $\mathcal J$ is the ideal generated by $(X_1-f_1(Y_1,\ldots,Y_k)), \ldots, (X_k-f_k(Y_1,\ldots,Y_k)$. 
Similarly $\I^{k}_{k_1,k_2}$ is related to the module
\be
M^\prime=\hat{\mathcal R}^\prime/{\mathcal J}^\prime\hat{\mathcal R}^\prime\,,
\ee
over
$\hat{\mathcal R}=\C[X^\prime_1,\ldots,X^\prime_k,Y_1,\ldots,Y_k]/(W(X^\prime_1,\ldots,X^\prime_k)-W(f_1,\ldots,f_k))$, where the ideal ${\mathcal J}^\prime$ is generated by 
$(X^\prime_1-f_1(Y_1,\ldots,Y_k)), \ldots, (X^\prime_k-f_k(Y_1,\ldots,Y_k)$.
Here, the $X_i$ and $X^\prime_i$ correspond to the fields of the incoming, respectively outgoing models $LG_k$, whereas the $Y_i$ are the fields of the intermediate model $LG_{k_1}\otimes LG_{k_2}$.

The matrix factorization of the fusion product $\I_{k_1,k_2}^k*\I^{k_1,k_2}_k$ is now given by the $2$-periodic part of the free resolution of the module $M^{\prime\prime}:=M\otimes M^\prime$ considered as a module over $\hat{\mathcal R}^{\prime\prime}:=\C[X_1,\ldots,X_k,X^\prime_1,\ldots,X^\prime_k]/(W(X_1,\ldots,X_k)-W(X^\prime_1,\ldots,X^\prime_k))$. But
\bea
M^{\prime\prime}&\cong &
\hat{\mathcal R}^{\prime\prime}\otimes\left(\C[Y_1,\ldots,Y_k]/((X_i-f_i(Y_1,\ldots,Y_k)),(X_i^\prime-f_i(Y_1,\ldots,Y_k))\right)\{-k_1k_2\}\\
&\cong&
\left(\hat{\mathcal R}^{\prime\prime}/(X_i-X_i^\prime)\right)\otimes \left(\C[Y_1,\ldots,Y_k]/(f_i(Y_1,\ldots,Y_k))\right)\{-k_1k_2\}\,.\nonumber
\eea
Now, the $Y_i$ are obtained from $(x_1,\ldots,x_k)$ by partial symmetrization with respect to permutations of respectively the first $k_1$ and the remaining $k_2$ variables, and the $f_i=\sigma_i(x_1,\ldots,x_k)$ are a basis of the completely symmetrized $(x_1,\ldots,x_k)$.
Hence,
\bea
\left(\C[Y_1,\ldots,Y_k]/(f_i(Y_1,\ldots,Y_k))\right)\{-k_1k_2\}
&\cong&\left(\C[x_1,\ldots,x_k]^{S_{k_1}\times S_{k_2}}/\C[x_1,\ldots,x_k]^{S_k}\right)\{-k_1k_2\}\,.\nonumber\\
&\cong& H^*(Gr(k_1,k))\{-k_1k_2\}\\
&\cong& \C \left\{{k\brack k_1}\right\}\,.\nonumber
\eea
The relation \eq{bubblerel} follows from the fact that
the $2$-periodic part of the Koszul resolution of
\be
\left(\hat{\mathcal R}^{\prime\prime}/(X_i-X_i^\prime)\right)
\ee
is the matrix factorization corresponding to the identity defect $\I_k^k$ of $LG_k$.

\subsection{Quantum group relations}
\label{app:qrel}

\FIGURE[htb]{\includegraphics[width=6cm]{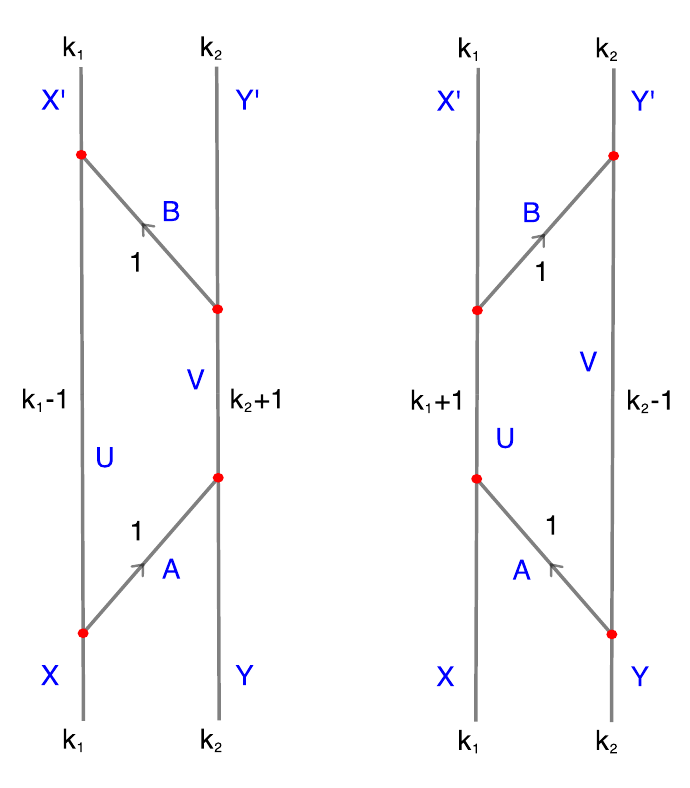}
\caption{\label{fig:qrel1}
Interface configurations associated to $\qF\qE$ (left) and $\qE\qF$ (right).}}

Here, we briefly explain how to obtain \eq{qrel1}. Pictorially, the left hand side of this equation is given by the configuration of interfaces depicted on the left of Figure \ref{fig:qrel1}. The blue symbols specify the Landau-Ginzburg fields of the respective segments of surface operators.
Let ${\mathcal R}$ be the polynomial ring in the external variables $(X_i,X_i^\prime,Y_i,Y_i^\prime)$ and
\bea
&&\widehat{\mathcal R}={\mathcal R}/(W(X^\prime_1,\ldots,X^\prime_{k_1})+W(Y^\prime_1,\ldots,Y^\prime_{k_2})\\
&&\qquad\qquad\qquad-W(X_1,\ldots,X_{k_1})-W(Y_1,\ldots,Y_{k_2}))\nonumber
\eea
the respective hypersurface ring. The $\widehat{\mathcal R}$-module corresponding to the matrix factorization on the left hand side of \eq{qrel1} is then given by
\be\label{mmodule}
M=\widehat{\mathcal R}_1/{\mathcal J}\widehat{\mathcal R}_1\left\{-k_1-k_2+1\right\}\,,
\ee
where
\be
\widehat{\mathcal R}_1=\widehat{\mathcal R}\left[A,B,U_1,\ldots,U_{k_1-1},V_1,\ldots,V_{k_2+1}\right]
\ee
is obtained by associating the internal variables $(A,B,U_i,V_i)$ to $\widehat{\mathcal R}$, and ${\mathcal J}$ is the ideal in $\widehat{\mathcal R}_1$ generated by the relations
\bea
X_i&=&\sigma_i(A,U_1,\ldots,U_{k_1-1})\label{xeq}\\
X_i^\prime&=&\sigma_i(B,U_1,\ldots,U_{k_1-1})\label{xpeq}\\
V_i&=&\sigma_i(A,Y_1,\ldots,Y_{k_2})=\sigma_i(B,Y_1^\prime,\ldots,Y_{k_2}^\prime)\label{veq}\,.
\eea
The module \eq{mmodule} can be recast in the following way.
The variables $V_i$ can be eliminated by \eq{veq}. Introducing the new variables
\be
\bar{V}_i:=Y_i-B\bar{V}_{i-1}\,,\quad \bar{V_0}=1\,,
\ee
the second part of the relations \eq{veq} can be rephrased as
\bea
Y_i&=&\sigma_i(B,\bar{V}_1,\ldots,\bar{V}_{k_2})\,,\label{yeq}\\
Y_i^\prime&=&\sigma_i(A,\bar{V}_1,\ldots,\bar{V}_{k_2})\,,\label{ypeq}\\
(A-B)\bar{V}_{k_2}&=&0\,.\label{vk2eq}
\eea
Thus, we have traded the variables $V_1,\ldots,V_{k_2+1}$ for variables $\bar{V}_1,\ldots,\bar{V}_{k_2}$ and relations \eq{veq} for \eq{yeq}, \eq{ypeq} and \eq{vk2eq}.
Defining
\be
\widehat{\mathcal R}_2=\widehat{\mathcal R}\left[A,B,U_1,\ldots,U_{k_1-1},\bar{V}_1,\ldots,\bar{V}_{k_2-1}\right]
\ee
one obtains
\be
M\cong {\widehat{\mathcal R}}_2
[V_{k_2}]
/{\mathcal J}^\prime
\widehat{\mathcal R}_2
[V_{k_2}]
\left\{-k_1-k_2+1\right\}\,,
\ee
where ${\mathcal J}^\prime$ is defined by the relations \eq{xeq}, \eq{xpeq}, \eq{yeq}, \eq{ypeq} and \eq{vk2eq}.
Next, one can use \eq{vk2eq} to eleminate $\bar{V}_{k_2}$:
\be
M\cong \underbrace{\widehat{\mathcal R}_2/
{\mathcal J}^{\prime\prime}
\widehat{\mathcal R}_2\left\{-k_1-k_2+1\right\}}_{=:M_1}\oplus
\underbrace{
\widehat{\mathcal R}_2/((A-B),{\mathcal J}^{\prime\prime})\widehat{\mathcal R}_2\left\{-k_1+k_2+1\right\}}_{=:M_2}\,.
\ee
Here ${\mathcal J}^{\prime\prime}$ is the ideal in $\widehat{\mathcal R}_2$ generated by \eq{xeq}, \eq{xpeq}, \eq{yeq} and \eq{ypeq}, where $\bar{V}_{k_2}$ is set to zero. With $A-B=0$, these equations can be solved on $M_2$.
One obtains $X_i=X_i^\prime$ and $Y_i=Y_i^\prime$, and \eq{xeq} can be used to eliminate the $U_i$.
Only the identity with $i=k_1$ in \eq{xeq} remains and gives rise to the degree $k_1$ relation $\sum_{i=0}^{k_1} X_{k_1-i}(-A)^i=0$, which truncates the possible exponents of $A$ ($X_0:=1$). Thus,
\be
M_2\cong\hat{\mathcal R}/{\mathcal J}_2\hat{\mathcal R}\left\{q^{-k_1-k_2+1}+q^{-k_1-k_2+3}+\ldots+q^{k_1-k_2-1}\right\}\,,
\ee
where ${\mathcal J}_2$ is the ideal in $\widehat{\mathcal R}$ generated by the relations
\be
0= (X_i-X_i^\prime)=(Y_i-Y_i^\prime)\,.
\ee
Now, $M_1$ is symmetric in the roles the variables $(X_i, X_i^\prime)$ and $(Y_i, Y_i^\prime)$ play. In particular, it also appears in the module $M^\prime$ corresponding to the matrix factorization on the right hand side of equation \eq{qrel1}, represented on the right of Figure \ref{fig:qrel1}. Indeed,
\be
M^\prime\cong M_1\oplus M_2^\prime\,,
\ee
with
\be
M_2^\prime\cong\hat{\mathcal R}/{\mathcal J}_2\hat{\mathcal R}\left\{q^{-k_1-k_2+1}+q^{-k_1-k_2+3}+\ldots+q^{-k_1+k_2-1}\right\}\,.
\ee
Therefore for $k_1\geq k_2$
\be
M\cong M^\prime \oplus \hat{\mathcal R}/{\mathcal J}_2\hat{\mathcal R}\Big\{\underbrace{q^{-k_1-+k_2+1}+q^{-k_1-k_2+3}+\ldots+q^{k_1-k_2-1}}_{[k_1-k_2]}\Big\}\,.
\ee
The module $\hat{\mathcal R}/{\mathcal J}_2\hat{\mathcal R}$ now corresponds to the identity defect in $LG_{k_1}\otimes LG_{k_2}$, which
proves that upon fusion the configurations of interfaces $\qE$ and $\qF$ satisfy the quantum $\mathfrak{sl}_2$-relations \eq{qrel1}.


\newpage

\bibliographystyle{JHEP_TD}
\bibliography{draft}

\providecommand{\href}[2]{#2}\begingroup\raggedright\begin{thebibliography}{100}

\bibitem{MR1030450}
G.~Moore and N.~Reshetikhin, {\it A comment on quantum group symmetry in
  conformal field theory},  {\em Nuclear Phys. B} {\bf 328} (1989), no.~3
  557--574.

\bibitem{MR2593278}
S.~Cautis, J.~Kamnitzer, and A.~Licata, {\it Categorical geometric skew {H}owe
  duality},  {\em Invent. Math.} {\bf 180} (2010), no.~1 111--159.

\bibitem{MR3263166}
S.~Cautis, J.~Kamnitzer, and S.~Morrison, {\it Webs and quantum skew {H}owe
  duality},  {\em Math. Ann.} {\bf 360} (2014), no.~1-2 351--390.

\bibitem{MR1074310}
A.~A. Beilinson, G.~Lusztig, and R.~MacPherson, {\it A geometric setting for
  the quantum deformation of GL(n)},  {\em Duke Math. J.} {\bf 61} (1990),
  no.~2 655--677.

\bibitem{MR1197834}
I.~Grojnowski and G.~Lusztig, {\it On bases of irreducible representations of
  quantum GL(n)},  in {\em Kazhdan-{L}usztig theory and related topics
  ({C}hicago, {IL}, 1989)}, vol.~139 of {\em Contemp. Math.}, pp.~167--174.
\newblock Amer. Math. Soc., Providence, RI, 1992.

\bibitem{MR1714141}
J.~Bernstein, I.~Frenkel, and M.~Khovanov, {\it A categorification of the
  {T}emperley-{L}ieb algebra and {S}chur quotients of U(sl(2)) via projective
  and {Z}uckerman functors},  {\em Selecta Math. (N.S.)} {\bf 5} (1999), no.~2
  199--241.

\bibitem{MR2305608}
I.~Frenkel, M.~Khovanov, and C.~Stroppel, {\it A categorification of
  finite-dimensional irreducible representations of quantum sl(2) and their
  tensor products},  {\em Selecta Math. (N.S.)} {\bf 12} (2006), no.~3-4
  379--431.

\bibitem{MR2503405}
H.~Zheng, {\it A geometric categorification of representations of Uq(sl(2))},
  in {\em Topology and physics}, vol.~12 of {\em Nankai Tracts Math.},
  pp.~348--356.
\newblock World Sci. Publ., Hackensack, NJ, 2008.

\bibitem{MR2729010}
A.~D. Lauda, {\it A categorification of quantum sl(2)},  {\em Adv. Math.} {\bf
  225} (2010), no.~6 3327--3424.

\bibitem{MR2525917}
M.~Khovanov and A.~D. Lauda, {\it A diagrammatic approach to categorification
  of quantum groups. {I}},  {\em Represent. Theory} {\bf 13} (2009) 309--347.

\bibitem{MR2763732}
M.~Khovanov and A.~D. Lauda, {\it A diagrammatic approach to categorification
  of quantum groups {II}},  {\em Trans. Amer. Math. Soc.} {\bf 363} (2011),
  no.~5 2685--2700.

\bibitem{KLIII}
M.~Khovanov and A.~D. Lauda, {\it A diagrammatic approach to categorification
  of quantum groups. {III}},  {\em Quantum Topol.} {\bf 1} (2010), no.~1 1--92.

\bibitem{MR2747932}
M.~Khovanov, {\it Categorifications from planar diagrammatics},  {\em Jpn. J.
  Math.} {\bf 5} (2010), no.~2 153--181.

\bibitem{Lauda}
A.~D. Lauda, {\it An introduction to diagrammatic algebra and categorified
  quantum sl(2)},  {\em Bull. Inst. Math. Acad. Sin. (N.S.)} {\bf 7} (2012),
  no.~2 165--270.

\bibitem{KamnitzerRev}
J.~Kamnitzer, {\it Categorification of Lie algebras [d'apres Rouquier,
  Khovanov-Lauda]},  {\em Ast\'erisque} (2013) Exp. No. 1072. S{\'e}minaire
  Bourbaki. Volume 2012/2013. Expos{\'e}s.

\bibitem{Rouquier}
R.~Rouquier, {\it {2-Kac-Moody algebras}},
  \href{http://xxx.lanl.gov/abs/0812.5023}{{\tt arXiv:0812.5023}}.

\bibitem{MR3426687}
A.~D. Lauda, H.~Queffelec, and D.~E.~V. Rose, {\it Khovanov homology is a skew
  {H}owe 2-representation of categorified quantum {${sl}_m$}},  {\em Algebr.
  Geom. Topol.} {\bf 15} (2015), no.~5 2517--2608.

\bibitem{QR14}
H.~Queffelec and D.~E.~V. Rose, {\it The $sl_n$ foam 2-category: a
  combinatorial formulation of Khovanov-Rozansky homology via categorical skew
  Howe duality},  \href{http://xxx.lanl.gov/abs/1405.5920}{{\tt
  arXiv:1405.5920}}.

\bibitem{Gukov:2006jk}
S.~Gukov and E.~Witten, {\it {Gauge Theory, Ramification, And The Geometric
  Langlands Program}},  \href{http://xxx.lanl.gov/abs/hep-th/0612073}{{\tt
  hep-th/0612073}}.

\bibitem{Gukov:2007ck}
S.~Gukov, {\it {Gauge theory and knot homologies}},  {\em Fortsch.Phys.} {\bf
  55} (2007) 473--490, [\href{http://xxx.lanl.gov/abs/0706.2369}{{\tt
  arXiv:0706.2369}}].

\bibitem{Gukov:2008sn}
S.~Gukov and E.~Witten, {\it {Rigid Surface Operators}},  {\em Adv. Theor.
  Math. Phys.} {\bf 14} (2010) 87--178,
  [\href{http://xxx.lanl.gov/abs/0804.1561}{{\tt arXiv:0804.1561}}].

\bibitem{Gukov:2008ve}
S.~Gukov and E.~Witten, {\it {Branes and Quantization}},  {\em Adv. Theor.
  Math. Phys.} {\bf 13} (2009) 1,
  [\href{http://xxx.lanl.gov/abs/0809.0305}{{\tt arXiv:0809.0305}}].

\bibitem{Brunner:2007qu}
I.~Brunner and D.~Roggenkamp, {\it {B-type defects in Landau-Ginzburg models}},
   {\em JHEP} {\bf 0708} (2007) 093,
  [\href{http://xxx.lanl.gov/abs/0707.0922}{{\tt arXiv:0707.0922}}].

\bibitem{Lusztig93}
G.~Lusztig, {\em Introduction to quantum groups}.
\newblock Birkh\"{a}user, Boston, 1993.

\bibitem{Witten:1988hf}
E.~Witten, {\it {Quantum Field Theory and the Jones Polynomial}},  {\em
  Comm.Math.Phys.} {\bf 121} (1989) 351--399.

\bibitem{Witten:1989wf}
E.~Witten, {\it {Gauge Theories and Integrable Lattice Models}},  {\em
  Nucl.Phys.} {\bf B322} (1989) 629.

\bibitem{Witten:1989rw}
E.~Witten, {\it {Gauge Theories, Vertex Models, and Quantum Groups}},  {\em
  Nucl.Phys.} {\bf B330} (1990) 285.

\bibitem{Wu}
H.~Wu, {\it A colored $\mathfrak{sl}(N)$ homology for links in ${S}^3$.},  {\em
  Dissertationes Math.} {\bf 499} (2014)
  [\href{http://xxx.lanl.gov/abs/0907.0695}{{\tt 0907.0695}}].

\bibitem{Crane:1994ty}
L.~Crane and I.~Frenkel, {\it {Four-dimensional topological field theory, Hopf
  categories, and the canonical bases}},  {\em J. Math. Phys.} {\bf 35} (1994)
  5136--5154, [\href{http://xxx.lanl.gov/abs/hep-th/9405183}{{\tt
  hep-th/9405183}}].

\bibitem{MR2174270}
D.~Bar-Natan, {\it Khovanov's homology for tangles and cobordisms},  {\em Geom.
  Topol.} {\bf 9} (2005) 1443--1499.

\bibitem{MR2253455}
D.~Bar-Natan and S.~Morrison, {\it The {K}aroubi envelope and {L}ee's
  degeneration of {K}hovanov homology},  {\em Algebr. Geom. Topol.} {\bf 6}
  (2006) 1459--1469.

\bibitem{MR2336253}
M.~Mackaay and P.~Vaz, {\it The universal sl(3)-link homology},  {\em Algebr.
  Geom. Topol.} {\bf 7} (2007) 1135--1169.

\bibitem{MR2173845}
E.~S. Lee, {\it An endomorphism of the {K}hovanov invariant},  {\em Adv. Math.}
  {\bf 197} (2005), no.~2 554--586.

\bibitem{MR2232858}
M.~Khovanov, {\it Link homology and {F}robenius extensions},  {\em Fund. Math.}
  {\bf 190} (2006) 179--190.

\bibitem{Gornik}
B.~Gornik, {\it {Note on Khovanov link cohomology}},
  \href{http://xxx.lanl.gov/abs/math/0402266}{{\tt math/0402266}}.

\bibitem{MR2826932}
H.~Wu, {\it Generic deformations of the colored sl(N)-homology for links},
  {\em Algebr. Geom. Topol.} {\bf 11} (2011), no.~4 2037--2106.

\bibitem{MR2916277}
A.~Lobb, {\it A note on {G}ornik's perturbation of {K}hovanov-{R}ozansky
  homology},  {\em Algebr. Geom. Topol.} {\bf 12} (2012), no.~1 293--305.

\bibitem{Rose:2015pla}
D.~E.~V. Rose and P.~Wedrich, {\it {Deformations of colored sl(N) link
  homologies via foams}},  \href{http://xxx.lanl.gov/abs/1501.0256}{{\tt
  arXiv:1501.0256}}.

\bibitem{Gorsky:2013jxa}
E.~Gorsky, S.~Gukov, and M.~Stosic, {\it {Quadruply-graded colored homology of
  knots}},  \href{http://xxx.lanl.gov/abs/1304.3481}{{\tt arXiv:1304.3481}}.

\bibitem{Khovanov}
M.~Khovanov, {\it {A categorification of the Jones polynomial}},  {\em Duke
  Math. J.} {\bf 101} (2000) 359–426.

\bibitem{MR2729272}
J.~Rasmussen, {\it Khovanov homology and the slice genus},  {\em Invent. Math.}
  {\bf 182} (2010), no.~2 419--447.

\bibitem{Gukov:2014gja}
S.~Gukov, {\it {Surface Operators}},
  \href{http://xxx.lanl.gov/abs/1412.7127}{{\tt arXiv:1412.7127}}.

\bibitem{Alford:1992yx}
M.~G. Alford, K.-M. Lee, J.~March-Russell, and J.~Preskill, {\it {Quantum field
  theory of nonAbelian strings and vortices}},  {\em Nucl.Phys.} {\bf B384}
  (1992) 251--317, [\href{http://xxx.lanl.gov/abs/hep-th/9112038}{{\tt
  hep-th/9112038}}].

\bibitem{Rozansky:2003hz}
L.~Rozansky, {\it {Topological A models on seamed Riemann surfaces}},  {\em
  Adv.Theor.Math.Phys.} {\bf 11} (2007)
  [\href{http://xxx.lanl.gov/abs/hep-th/0305205}{{\tt hep-th/0305205}}].

\bibitem{MR2322554}
M.~Khovanov and L.~Rozansky, {\it Topological {L}andau-{G}inzburg models on the
  world-sheet foam},  {\em Adv. Theor. Math. Phys.} {\bf 11} (2007), no.~2
  233--259.

\bibitem{Lerche:1989uy}
W.~Lerche, C.~Vafa, and N.~P. Warner, {\it {Chiral Rings in N=2 Superconformal
  Theories}},  {\em Nucl.Phys.} {\bf B324} (1989) 427.

\bibitem{Witten:1993yc}
E.~Witten, {\it {Phases of N=2 theories in two-dimensions}},  {\em Nucl.Phys.}
  {\bf B403} (1993) 159--222,
  [\href{http://xxx.lanl.gov/abs/hep-th/9301042}{{\tt hep-th/9301042}}].

\bibitem{Witten:1997sc}
E.~Witten, {\it {Solutions of four-dimensional field theories via M theory}},
  {\em Nucl.Phys.} {\bf B500} (1997) 3--42,
  [\href{http://xxx.lanl.gov/abs/hep-th/9703166}{{\tt hep-th/9703166}}].

\bibitem{Gadde:2013sca}
A.~Gadde, S.~Gukov, and P.~Putrov, {\it {Fivebranes and 4-manifolds}},
  \href{http://xxx.lanl.gov/abs/1306.4320}{{\tt arXiv:1306.4320}}.

\bibitem{Xie:2013lca}
D.~Xie, {\it {Higher laminations, webs and N=2 line operators}},
  \href{http://xxx.lanl.gov/abs/1304.2390}{{\tt arXiv:1304.2390}}.

\bibitem{Xie:2013vfa}
D.~Xie, {\it {Aspects of line operators of class S theories}},
  \href{http://xxx.lanl.gov/abs/1312.3371}{{\tt arXiv:1312.3371}}.

\bibitem{Bullimore:2013xsa}
M.~Bullimore, {\it {Defect Networks and Supersymmetric Loop Operators}},  {\em
  JHEP} {\bf 1502} (2015) 066, [\href{http://xxx.lanl.gov/abs/1312.5001}{{\tt
  arXiv:1312.5001}}].

\bibitem{Coman:2015lna}
I.~Coman, M.~Gabella, and J.~Teschner, {\it {Line operators in theories of
  class S, quantized moduli space of flat connections, and Toda field theory}},
   \href{http://xxx.lanl.gov/abs/1505.0589}{{\tt arXiv:1505.0589}}.

\bibitem{Tachikawa:2015iba}
Y.~Tachikawa and N.~Watanabe, {\it {On skein relations in class S theories}},
  \href{http://xxx.lanl.gov/abs/1504.0012}{{\tt arXiv:1504.0012}}.

\bibitem{Witten:1997ep}
E.~Witten, {\it {Branes and the dynamics of QCD}},  {\em Nucl.Phys.} {\bf B507}
  (1997) 658--690, [\href{http://xxx.lanl.gov/abs/hep-th/9706109}{{\tt
  hep-th/9706109}}].

\bibitem{Frenkel:2015rda}
E.~Frenkel, S.~Gukov, and J.~Teschner, {\it {Surface Operators and Separation
  of Variables}},  \href{http://xxx.lanl.gov/abs/1506.0750}{{\tt
  arXiv:1506.0750}}.

\bibitem{MR1204322}
L.~C. Jeffrey and J.~Weitsman, {\it Bohr-{S}ommerfeld orbits in the moduli
  space of flat connections and the {V}erlinde dimension formula},  {\em Comm.
  Math. Phys.} {\bf 150} (1992), no.~3 593--630.

\bibitem{MR1671192}
S.~Agnihotri and C.~Woodward, {\it Eigenvalues of products of unitary matrices
  and quantum {S}chubert calculus},  {\em Math. Res. Lett.} {\bf 5} (1998),
  no.~6 817--836.

\bibitem{MR1856023}
P.~Belkale, {\it Local systems on P(1)-S for S a finite set},  {\em Compositio
  Math.} {\bf 129} (2001), no.~1 67--86.

\bibitem{MR2008438}
C.~Teleman and C.~Woodward, {\it Parabolic bundles, products of conjugacy
  classes and {G}romov-{W}itten invariants},  {\em Ann. Inst. Fourier
  (Grenoble)} {\bf 53} (2003), no.~3 713--748.

\bibitem{Ressayre}
N.~Ressayre, {\it {On the quantum Horn problem}},
  \href{http://xxx.lanl.gov/abs/1310.7331}{{\tt arXiv:1310.7331}}.

\bibitem{BelkaleKumar}
P.~Belkale and S.~Kumar, {\it {The multiplicative eigenvalue problem and
  deformed quantum cohomology}},  \href{http://xxx.lanl.gov/abs/1310.3191}{{\tt
  arXiv:1310.3191}}.

\bibitem{MR3190356}
A.~Lobb and R.~Zentner, {\it The quantum sl(N) graph invariant and a moduli
  space},  {\em Int. Math. Res. Not. IMRN} (2014), no.~7 1956--1972.

\bibitem{MR3125899}
J.~Grant, {\it The moduli problem of {L}obb and {Z}entner and the colored sl(N)
  graph invariant},  {\em J. Knot Theory Ramifications} {\bf 22} (2013), no.~10
  1350060, 16.

\bibitem{MR1454400}
A.~Bertram, {\it Quantum {S}chubert calculus},  {\em Adv. Math.} {\bf 128}
  (1997), no.~2 289--305.

\bibitem{Witten:1993xi}
E.~Witten, {\it {The Verlinde algebra and the cohomology of the Grassmannian}},
   \href{http://xxx.lanl.gov/abs/hep-th/9312104}{{\tt hep-th/9312104}}.

\bibitem{Gukov:2015sna}
S.~Gukov and D.~Pei, {\it {Equivariant Verlinde formula from fivebranes and
  vortices}},  \href{http://xxx.lanl.gov/abs/1501.0131}{{\tt arXiv:1501.0131}}.

\bibitem{MR2231042}
L.~Mihalcea, {\it Equivariant quantum {S}chubert calculus},  {\em Adv. Math.}
  {\bf 203} (2006), no.~1 1--33.

\bibitem{Eto:2005mx}
M.~Eto, Y.~Isozumi, M.~Nitta, K.~Ohashi, K.~Ohta, {\em et.~al.}, {\it {D-brane
  configurations for domain walls and their webs}},  {\em AIP Conf.Proc.} {\bf
  805} (2006) 354--357, [\href{http://xxx.lanl.gov/abs/hep-th/0509127}{{\tt
  hep-th/0509127}}].

\bibitem{Eto:2006mz}
M.~Eto, T.~Fujimori, Y.~Isozumi, M.~Nitta, K.~Ohashi, {\em et.~al.}, {\it
  {Non-Abelian vortices on cylinder: Duality between vortices and walls}},
  {\em Phys.Rev.} {\bf D73} (2006) 085008,
  [\href{http://xxx.lanl.gov/abs/hep-th/0601181}{{\tt hep-th/0601181}}].

\bibitem{Eto:2008dm}
M.~Eto, T.~Fujimori, M.~Nitta, K.~Ohashi, and N.~Sakai, {\it {Domain walls with
  non-Abelian clouds}},  {\em Phys.Rev.} {\bf D77} (2008) 125008,
  [\href{http://xxx.lanl.gov/abs/0802.3135}{{\tt arXiv:0802.3135}}].

\bibitem{QS15}
H.~{Queffelec} and A.~{Sartori}, {\it {Mixed quantum skew Howe duality and link
  invariants of type A}},  \href{http://xxx.lanl.gov/abs/1504.0122}{{\tt
  arXiv:1504.0122}}.

\bibitem{SymmetricHowe}
D.~E.~V. Rose and D.~Tubbenhauer, {\it {Symmetric webs, Jones-Wenzl recursions
  and q-Howe duality}},  \href{http://xxx.lanl.gov/abs/1501.0091}{{\tt
  arXiv:1501.0091}}.

\bibitem{OShfk}
P.~Ozsvath and Z.~Szabo, {\it {Holomorphic disks and knot invariants}},  {\em
  Adv. Math.} {\bf 186} (2004) 58--116,
  [\href{http://xxx.lanl.gov/abs/math/0209056}{{\tt math/0209056}}].

\bibitem{RasmussenHFK}
J.~Rasmussen, {\it {Floer homology and knot complements}},
  \href{http://xxx.lanl.gov/abs/math/0306378}{{\tt math/0306378}}.

\bibitem{Gukov:2004hz}
S.~Gukov, A.~S. Schwarz, and C.~Vafa, {\it {Khovanov-Rozansky homology and
  topological strings}},  {\em Lett.Math.Phys.} {\bf 74} (2005) 53--74,
  [\href{http://xxx.lanl.gov/abs/hep-th/0412243}{{\tt hep-th/0412243}}].

\bibitem{GV-I}
R.~Gopakumar and C.~Vafa, {\it M-Theory and Topological Strings-I},
  \href{http://xxx.lanl.gov/abs/hep-th/9809187v1}{{\tt hep-th/9809187v1}}.

\bibitem{GV-II}
R.~Gopakumar and C.~Vafa, {\it M-Theory and Topological Strings-II},
  \href{http://xxx.lanl.gov/abs/hep-th/9812127v1}{{\tt hep-th/9812127v1}}.

\bibitem{AS}
M.~Aganagic and S.~Shakirov, {\it {Knot Homology from Refined Chern-Simons
  Theory}},  \href{http://xxx.lanl.gov/abs/1105.5117}{{\tt arXiv:1105.5117}}.

\bibitem{IK11}
A.~Iqbal and C.~Kozcaz, {\it {Refined Hopf Link Revisited}},
  \href{http://xxx.lanl.gov/abs/1111.0525}{{\tt arXiv:1111.0525}}.

\bibitem{Cherednik}
I.~Cherednik, {\it {Jones polynomials of torus knots via DAHA}},
  \href{http://xxx.lanl.gov/abs/1111.6195}{{\tt arXiv:1111.6195}}.

\bibitem{Nakajima:2012gx}
H.~Nakajima, {\it {Refined Chern-Simons theory and Hilbert schemes of points on
  the plane}},  \href{http://xxx.lanl.gov/abs/1211.5821}{{\tt
  arXiv:1211.5821}}.

\bibitem{Samuelson:2014}
P.~Samuelson, {\it {A topological construction of Cherednik's
  $\mathfrak{sl}\_2$ torus knot polynomials}},
  \href{http://xxx.lanl.gov/abs/1408.0483}{{\tt arXiv:1408.0483}}.

\bibitem{Witten:2011zz}
E.~Witten, {\it {Fivebranes and Knots}},
  \href{http://xxx.lanl.gov/abs/1101.3216}{{\tt arXiv:1101.3216}}.

\bibitem{Haydys:2010dv}
A.~Haydys, {\it {Fukaya-Seidel category and gauge theory}},
  \href{http://xxx.lanl.gov/abs/1010.2353}{{\tt arXiv:1010.2353}}.

\bibitem{Gukov:2005qp}
S.~Gukov and J.~Walcher, {\it {Matrix factorizations and Kauffman homology}},
  \href{http://xxx.lanl.gov/abs/hep-th/0512298}{{\tt hep-th/0512298}}.

\bibitem{Carqueville:2011zea}
N.~Carqueville and D.~Murfet, {\it {Computing Khovanov-Rozansky homology and
  defect fusion}},  {\em Topology} {\bf 14} (2014) 489--537,
  [\href{http://xxx.lanl.gov/abs/1108.1081}{{\tt arXiv:1108.1081}}].

\bibitem{Fuji:2012pi}
H.~Fuji, S.~Gukov, M.~Stosic, and P.~Sulkowski, {\it {3d analogs of
  Argyres-Douglas theories and knot homologies}},  {\em JHEP} {\bf 1301} (2013)
  175, [\href{http://xxx.lanl.gov/abs/1209.1416}{{\tt arXiv:1209.1416}}].

\bibitem{Chung:2014qpa}
H.-J. Chung, T.~Dimofte, S.~Gukov, and P.~Sulkowski, {\it {3d-3d Correspondence
  Revisited}},  \href{http://xxx.lanl.gov/abs/1405.3663}{{\tt
  arXiv:1405.3663}}.

\bibitem{engineering}
S.~Katz, A.~Klemm, and C.~Vafa, {\it Geometric Engineering of Quantum Field
  Theories},  {\em Nucl. Phys.} {\bf B497} (1997)
  [\href{http://xxx.lanl.gov/abs/hep-th/9609239}{{\tt hep-th/9609239}}].

\bibitem{DGH}
T.~Dimofte, S.~Gukov, and L.~Hollands, {\it {Vortex Counting and Lagrangian
  3-manifolds}},  {\em Lett.Math.Phys.} {\bf 98} (2011) 225--287,
  [\href{http://xxx.lanl.gov/abs/1006.0977}{{\tt arXiv:1006.0977}}].

\bibitem{KR2}
M.~Khovanov and L.~Rozansky, {\it Matrix factorizations and link homology.
  {II}},  {\em Geom. Topol.} {\bf 12} (2008), no.~3 1387--1425,
  [\href{http://xxx.lanl.gov/abs/math.QA/0}{{\tt arXiv:math.QA/0}}].

\bibitem{MR2100691}
M.~Khovanov, {\it sl(3) link homology},  {\em Algebr. Geom. Topol.} {\bf 4}
  (2004) 1045--1081.

\bibitem{Aganagic:2013jpa}
M.~Aganagic, T.~Ekholm, L.~Ng, and C.~Vafa, {\it {Topological Strings, D-Model,
  and Knot Contact Homology}},  {\em Adv.Theor.Math.Phys.} {\bf 18} (2014)
  827--956, [\href{http://xxx.lanl.gov/abs/1304.5778}{{\tt arXiv:1304.5778}}].

\bibitem{Fuji:2013rra}
H.~Fuji and P.~Sulkowski, {\it {Super-A-polynomial}},
  \href{http://xxx.lanl.gov/abs/1303.3709}{{\tt arXiv:1303.3709}}.

\bibitem{KR1}
M.~Khovanov and L.~Rozansky, {\it Matrix factorizations and link homology},
  {\em Fund. Math.} {\bf 199} (2008), no.~1 1--91,
  [\href{http://xxx.lanl.gov/abs/math/0401268}{{\tt math/0401268}}].

\bibitem{Brunner:2003dc}
I.~Brunner, M.~Herbst, W.~Lerche, and B.~Scheuner, {\it {Landau-Ginzburg
  realization of open string TFT}},  {\em JHEP} {\bf 0611} (2006) 043,
  [\href{http://xxx.lanl.gov/abs/hep-th/0305133}{{\tt hep-th/0305133}}].

\bibitem{Kapustin:2003ga}
A.~Kapustin and Y.~Li, {\it {Topological correlators in Landau-Ginzburg models
  with boundaries}},  {\em Adv.Theor.Math.Phys.} {\bf 7} (2004) 727--749,
  [\href{http://xxx.lanl.gov/abs/hep-th/0305136}{{\tt hep-th/0305136}}].

\bibitem{Hori:2004ja}
K.~Hori and J.~Walcher, {\it {F-term equations near Gepner points}},  {\em
  JHEP} {\bf 0501} (2005) 008,
  [\href{http://xxx.lanl.gov/abs/hep-th/0404196}{{\tt hep-th/0404196}}].

\bibitem{Brunner:2007ur}
I.~Brunner and D.~Roggenkamp, {\it {Defects and bulk perturbations of boundary
  Landau-Ginzburg orbifolds}},  {\em JHEP} {\bf 0804} (2008) 001,
  [\href{http://xxx.lanl.gov/abs/0712.0188}{{\tt arXiv:0712.0188}}].

\bibitem{Brunner:2008fa}
I.~Brunner, H.~Jockers, and D.~Roggenkamp, {\it {Defects and D-Brane
  Monodromies}},  {\em Adv.Theor.Math.Phys.} {\bf 13} (2009) 1077--1135,
  [\href{http://xxx.lanl.gov/abs/0806.4734}{{\tt arXiv:0806.4734}}].

\bibitem{Gukov:2011ry}
S.~Gukov and M.~Sto$\check{\text{s}}$i$\acute{\text{c}}$, {\it {Homological
  Algebra of Knots and BPS States}},  {\em Proc.Symp.Pure Math.} {\bf 85}
  (2012) 125--172, [\href{http://xxx.lanl.gov/abs/1112.0030}{{\tt
  arXiv:1112.0030}}].

\bibitem{MR2386537}
M.~Khovanov and L.~Rozansky, {\it Virtual crossings, convolutions and a
  categorification of the SO(2N) {K}auffman polynomial},  {\em J. G\"okova
  Geom. Topol. GGT} {\bf 1} (2007) 116--214.

\bibitem{Yonezawa}
Y.~Yonezawa, {\it {Quantum ($sl_n,\wedge V_n$) link invariant and matrix
  factorizations,}},  \href{http://xxx.lanl.gov/abs/0906.0220}{{\tt
  arXiv:0906.0220}}.

\bibitem{Elliot:2015pra}
R.~Elliot and S.~Gukov, {\it {Exceptional knot homology}},
  \href{http://xxx.lanl.gov/abs/1505.0163}{{\tt arXiv:1505.0163}}.

\bibitem{Dunfield:2005si}
N.~M. Dunfield, S.~Gukov, and J.~Rasmussen, {\it {The Superpolynomial for knot
  homologies}},  \href{http://xxx.lanl.gov/abs/math/0505662}{{\tt
  math/0505662}}.

\bibitem{Cecotti:1992rm}
S.~Cecotti and C.~Vafa, {\it {On classification of N=2 supersymmetric
  theories}},  {\em Commun.Math.Phys.} {\bf 158} (1993) 569--644,
  [\href{http://xxx.lanl.gov/abs/hep-th/9211097}{{\tt hep-th/9211097}}].

\bibitem{Gadde:2013dda}
A.~Gadde and S.~Gukov, {\it {2d Index and Surface operators}},  {\em JHEP} {\bf
  1403} (2014) 080, [\href{http://xxx.lanl.gov/abs/1305.0266}{{\tt
  arXiv:1305.0266}}].

\bibitem{DM11}
T.~{Dyckerhoff} and D.~{Murfet}, {\it {Pushing forward matrix factorisations}},
   \href{http://xxx.lanl.gov/abs/1102.2957}{{\tt arXiv:1102.2957}}.

\bibitem{Behr:2012xg}
N.~Behr and S.~Fredenhagen, {\it {Variable transformation defects}},  {\em
  Proc. Symp. Pure Math.} {\bf 85} (2012) 303--312,
  [\href{http://xxx.lanl.gov/abs/1202.1678}{{\tt arXiv:1202.1678}}].

\bibitem{MY13}
M.~Mackaay and Y.~Yonezawa, {\it sl(N)-Web categories},
  \href{http://xxx.lanl.gov/abs/1306.6242}{{\tt arXiv:1306.6242}}.

\bibitem{Eto:2005fm}
M.~Eto, Y.~Isozumi, M.~Nitta, K.~Ohashi, and N.~Sakai, {\it {Non-Abelian webs
  of walls}},  {\em Phys.Lett.} {\bf B632} (2006) 384--392,
  [\href{http://xxx.lanl.gov/abs/hep-th/0508241}{{\tt hep-th/0508241}}].

\bibitem{Shifman:2003uh}
M.~Shifman and A.~Yung, {\it {Localization of nonAbelian gauge fields on domain
  walls at weak coupling (D-brane prototypes II)}},  {\em Phys.Rev.} {\bf D70}
  (2004) 025013, [\href{http://xxx.lanl.gov/abs/hep-th/0312257}{{\tt
  hep-th/0312257}}].

\bibitem{Eto:2005cp}
M.~Eto, Y.~Isozumi, M.~Nitta, K.~Ohashi, and N.~Sakai, {\it {Webs of walls}},
  {\em Phys.Rev.} {\bf D72} (2005) 085004,
  [\href{http://xxx.lanl.gov/abs/hep-th/0506135}{{\tt hep-th/0506135}}].

\bibitem{eisenbud}
D.~Eisenbud, {\it Homological algebra with an application to group
  representations},  {\em Trans. Amer. Math. Soc.} {\bf 260} (1980) 35--64.

\end{thebibliography}\endgroup

\end{document}